\documentclass[11pt]{article}
\usepackage{a4wide}
\usepackage{amsmath}
\usepackage{axodraw}
\usepackage{graphicx}
\usepackage{longtable}

\newcommand{\NEG}[1]{{#1\hskip-0.7em/\,}}
\newcommand{\unit}[1]{\mbox{#1}}

\begin{document}
\begin{titlepage}
\begin{flushright}
{\large LU TP 03-02\\ January 2003}
\end{flushright}
\vfill
\begin{center}
{\Huge\bf Determination of the\\
Anomalous Chiral Coefficients\\[2mm]
of order \boldmath$p^{6}$}
\\[1cm]
\begin{large}
Master of Science Thesis\\[0.3cm]
\emph{author:}  Olof Strandberg \\[0.3cm] 
\emph{advisor:}  Johan Bijnens
\\[1cm]
Department of Theoretical Physics, Lund University\\*[0pt]
S\"{o}lvegatan 14A, S22362 Lund, Sweden
\end{large}
\end{center}

\vfill

\begin{abstract}
Symmetries affected by the anomaly do not survive quantization and cannot be
understood classically. They are of fundamental importance and offer an
opportunity of expanding the theoretical framework. We examine the theory of
the anomalous sector, starting from lowest order Chiral Perturbation Theory
(ChPT), leading up to the construction of the recently developed Lagrangian
of $\mathcal{O}\left( p^{6}\right) $ describing anomalous processes. This
Lagrangian contains a set of chiral coefficients that must be determined
phenomenologically. Using currently available experimental data, we fit as
many of these coefficients as possible. The results of the ChPT treatment
are then used to test the validity of the two main alternative models
employed in the anomalous sector - Vector Meson Dominance (VMD) and the
constituent Chiral Quark Model (CQM).
\end{abstract}
\vfill
\end{titlepage}

\tableofcontents


\section{Introduction}

In high energy QCD, we have asymptotic freedom and scattering can be
calculated in a power series expansion in the strong coupling constant. The
non-perturbative effects are determined phenomenologically in the form of
structure and fragmentation functions. Predictions take the form of
relations between amplitudes parametrized by $\alpha _{s}$ and the structure
functions.

In low energy QCD, the predictions are relations between amplitudes with a
structure determined by symmetry constraints, parametrized by empirically
determined coefficients. The power series expansion in $\alpha _{s}$ is
replaced by an expansion in the low energy. Heavy fields of the high energy
region can be integrated out and their effect is consequently encoded in the
coefficients of the low energy effective theory. Proceeding up to $\mathcal{O%
}\left( p^{6}\right) $ in ChPT, we are confronted with a large number of
empirical parameters. In principle, QCD should be able to predict these
parameters, as well as the structure functions appearing in the chiral
Lagrangian. No rigorous derivation exists up to date. The combined use of
tentative models and phenomenological knowledge provides us with insight of
the physics leading to the chiral Lagrangian. The aim of this thesis is to
solve for as many low energy chiral coefficients as possible by making use
of the available experimental data in the form of widths, slopes and form
factors.

Classically conserved currents that are affected by the anomaly do not
survive quantization. To evaluate the ensuing effects, it is necessary to
employ a more rigorous analysis, taking quantum corrections into account.
The effect of the anomaly in the chiral Lagrangian framework was first
analyzed by Wess and Zumino \cite{WZ}, who realized that the result could
not be expressed in a local effective Lagrangian. Their result, in the form
of a Taylor series expansion, was later given an elegant geometrical
interpretation by Witten \cite{WZW}.

So far most articles treating anomalous processes quote the leading order
amplitudes and meson one-loop corrections. Higher order contributions are
then estimated using models like Vector Meson Dominance (VMD) or the more
recently developed chiral Constituent Quark Model (CQM). In this thesis we
will instead test the validity of VMD\ and CQM by comparing with the results
of the experimentally fixed chiral $\mathcal{O}\left( p^{6}\right) $
coefficients.\medskip

\section{Symmetries}

\subsection{Chiral Symmetry}

Apart from the more obvious symmetries of the standard model, such as the
gauge symmetries $SU\left( 3\right) _{c}\times SU\left( 2\right) _{L}\times
U\left( 1\right) _{Y},$ there exist global vector symmetries like the
fermion number symmetries, the isospin symmetry, and the less accurate $%
SU\left( 3\right) $ flavor symmetry. The flavor symmetries are valid if the
masses of the quarks included by the symmetry are set equal. Therefore, the
isospin symmetry is broken by the up-down quark mass difference, as well as
by electromagnetic and weak interactions. By imposing the considerably
stricter condition, $m_{q}=0,$ so called chiral symmetries arise. Since
there are no longer any mass terms to couple fields of differing chirality,
the fields are invariant under separate left- and right-handed
transformations. The QCD Lagrangian in the massless limit is given by 
\begin{eqnarray}
\mathcal{L}_{QCD}^{m=0} &=&\sum_{q=u,d,s}\bar{q}\gamma ^{\mu }\left(
i\partial _{\mu }-g_{s}\frac{\lambda _{a}}{2}G_{\mu }^{a}\right) q-\tfrac{1}{%
4}G_{\mu \nu }^{a}G_{a}^{\mu \nu }  \notag \\
&=&\bar{\psi}_{L}\NEG{D}\psi _{L}+\bar{\psi}_{R}\NEG{D}\psi _{R}-\tfrac{1}{4}%
G_{\mu \nu }^{a}G_{a}^{\mu \nu }  \label{qcdlag}
\end{eqnarray}%
where $\psi _{R,L}=\tfrac{1}{2}\left( 1\pm \gamma _{5}\right) \psi
=P_{R,L}\psi $ are the chiral projections of the column vector containing
the relevant quark fields ($u,d$ for chiral $SU\left( 2\right) $ and $u,d,s$
for chiral $SU\left( 3\right) )$. Taking $m_{u}=m_{d}=0,$ the QCD Lagrangian
is invariant under chiral rotations of the fields%
\begin{eqnarray}
\psi _{L} &\rightarrow &e^{-i\vec{\theta}_{L}\cdot \vec{\tau}}\psi
_{L}\equiv g_{L}\psi _{L} \\
\psi _{R} &\rightarrow &e^{-i\vec{\theta}_{R}\cdot \vec{\tau}}\psi
_{R}\equiv g_{R}\psi _{R}
\end{eqnarray}%
where \{$\tau ^{i}=\tfrac{1}{2}\sigma ^{i}$\} $\left( i=1,2,3\right) $ are
the Pauli matrices and \{$\theta _{L,R}^{i}$\} are the components of an
arbitrary constant vector. The chiral $SU\left( 2\right) $ invariance is
referred to as the direct product of the left- and right-handed chiral
transformation groups, $SU\left( 2\right) _{L}\times SU\left( 2\right) _{R}.$
This symmetry is easily extended to include the strange quark, substituting
the Pauli matrices for the Gell-Mann 3$\times $3 $SU\left( 3\right) $
matrices $\tfrac{1}{2}$\{$\lambda ^{a}$\} $\left( a=1,...,8\right) $. In eq.
(\ref{qcdlag}), we also find the vector $U\left( 1\right) _{V}$ and the
axial $U\left( 1\right) _{A}$ invariances

\begin{eqnarray}
\quad \psi &\rightarrow &e^{i\theta }\psi \\
\quad \psi &\rightarrow &e^{i\theta \gamma _{5}}\psi
\end{eqnarray}%
So apart from the normal $SU\left( 3\right) _{c}$ color gauge and discrete
symmetries, the chiral QCD Lagrangian also has the global invariance 
\begin{equation}
U\left( 1\right) _{A}\times U\left( 1\right) _{V}\times SU\left( 3\right)
_{R}\times SU\left( 3\right) _{L}  \label{global symmetries}
\end{equation}%
where the there are now separate invariances under $SU\left( 3\right) _{L}$
and $SU\left( 3\right) _{R},$ for the three massless quarks. The $U\left(
1\right) _{V}$ is conserved, ensuring baryon number conservation. The axial $%
U\left( 1\right) _{A}$ current is not conserved in the full quantum theory,
and is explicitly broken by the Abelian anomaly.

Let $G$ be the direct product group of the left- and right-handed chiral
transformations: $\psi _{L,R}\rightarrow g_{L,R}\psi _{L,R},$ $g_{L,R}\in G$%
. Then (\ref{qcdlag}) is invariant under $G.$ However, the QCD vacuum
structure does not share this invariance$.$ The effect of a non-vanishing
quark condensate, i.e. a non-zero vacuum expectation value $\left\langle
0\right| \bar{\psi}\psi \left| 0\right\rangle \neq 0,$ is the spontaneous
breaking of the axial part of $G$. Spontaneous (dynamic) symmetry breaking
is said to occur whenever the symmetry of the Lagrangian is not shared by
the ground (vacuum) state. For $G$ we have%
\begin{equation}
SU\left( 3\right) _{L}\times SU\left( 3\right) _{R}\rightarrow SU\left(
3\right) _{V=L+R}
\end{equation}%
where $SU\left( 3\right) _{V}\equiv H$ is the remaining unbroken subgroup of
simultaneous transformations of differing chirality. According to the
Goldstone theorem, whenever a global continuous symmetry is broken, massless
so called Goldstone (GS) bosons appear. Since it is the axial part that is
broken, the resulting GS bosons are pseudoscalars. An $SU\left( n\right) $
matrix has $n^{2}-1$ parameters (due to the unitarity and determinant
constraints). The matrix of the coset space $G/H=SU\left( 3\right)
_{L}\times SU\left( 3\right) _{R}\,/\,SU\left( 3\right) _{V},$ used to
parametrize the GS bosons is an $SU\left( 3\right) $ isomorphism, also
represented by a $3\times 3$ matrix, containing 8 pseudoscalars. These are
the light pseudoscalar mesons of the low energy spectrum: $\pi ^{\pm },\pi
^{0},K^{\pm },K^{0},\bar{K}^{0}$ and $\eta $. If we instead would have
chosen to focus on the up and down quark, setting $m_{u}=m_{d}=0$, the
resulting spectrum would have been described by the $2\times 2$ $SU\left(
2\right) $ matrices, containing the $2^{2}-1=3$ pions: $\pi ^{+},\pi ^{-}$
and $\pi ^{0}.$

The only visible symmetry in hadronic states is $U\left( 1\right) _{V}\times
SU\left( 3\right) _{V}.$ The effect of $U\left( 1\right) _{V}$ can only be
seen by including the (non-trivially transforming) baryon fields in the
effective Lagrangian.

The proof of the breaking of the axial symmetry lies in the predictions it
makes, and can be produced by Monte Carlo lattice gauge techniques. The main
motivation for the existence of this mechanism lies in its phenomenological
success and theoretical consistency.

GS bosons are massless, but the physical particles are clearly not. However,
the quark masses are small compared to the breaking scale of the chiral
symmetry ($m_{u,d,s}<\Lambda _{\chi }$ $\sim $ $1GeV$)$.$ This enables us to
treat them as a small perturbation, and we can include the quark masses in
the Lagrangian as external scalar fields. The inclusion of mass terms will
mean the explicit breaking of $G.$ Perturbational treatment is supported by
the relatively small size of the pseudoscalar masses compared to the
hadronic scale ($m_{\pi }^{2}/m_{\rho }^{2}\sim 0.03$ for $SU\left( 2\right) 
$ and $m_{K}^{2}/m_{\rho }^{2}\sim 0.3$ for $SU\left( 3\right) )$.

The task at hand is to construct a chiral invariant low energy effective
quantum theory with the GS\ bosons as dynamic fields.

\subsection{Non-linear Realization of Chiral Symmetry}

Before we can construct an invariant Lagrangian, we must examine the
transformation properties of the Goldstone boson degrees of freedom$.$ $G$
is a compact connected Lie group, that is dynamically broken into the
subgroup $H.$ The coordinates of the coset space $G/H=SU\left( 3\right)
_{L}\times SU\left( 3\right) _{R}/SU\left( 3\right) _{L+R}$ are the
remaining degrees of freedom describing the system, i.e. the GS bosons. The
transformation of an element parametrized by the GS bosons under $G$%
\begin{equation*}
gu\left( \xi _{i}\right) \rightarrow u\left( \xi _{i}^{\prime }\right)
h\left( \xi _{i},g\right) \quad ;g\in G,h\in H
\end{equation*}%
is non-linear in nature since the generators of $H$ and $G/H$ do not
commute. What we want is to construct operators that transform linearly
under $G,$ that is to say are chiral invariant$.$ To this we then add the
explicitly symmetry breaking light quark masses. Linear operators can be
constructed by starting from projections onto the transformation subgroup $%
H. $ As shown by Callan, Coleman, Wess \& Zumino \cite{chintref}, this is
the most general way of constructing operators linear under $G$ in terms of
the GS bosons. A given field $\psi $, transforming linearly under $H,$
transforms under $G$ as $\psi \rightarrow h\left( g,\xi _{i}\right) \psi
h^{-1}\left( g,\xi _{i}\right) .$ This means that all products of type $%
\left( u,u^{\dagger }\right) \cdot \psi \cdot \left( u,u^{\dagger }\right) $
will transform linearly under $G.$ For a more technical treatment of how
this can be done the reader is referred to \cite{chptintro}.

From the derivative of $u$ and $u^{\dagger }$ we can define the Hermitian
operator $u_{\mu }$ and the covariant derivative $\nabla _{\mu },$ both of
which transform linearly under $G.$ 
\begin{eqnarray}
u_{\mu } &=&i\left( u^{\dagger }\partial _{\mu }u-u\partial _{\mu
}u^{\dagger }\right) \\
\nabla _{\mu }\psi &=&\partial _{\mu }\psi -\left[ \Gamma _{\mu },\psi %
\right] \quad ;\Gamma _{\mu }=\tfrac{1}{2}\left( u^{\dagger }\partial _{\mu
}u+u\partial _{\mu }u^{\dagger }\right)
\end{eqnarray}%
These can be used as building blocks to construct the linear operators. To
get the correct behavior in the low energy limit, it is necessary to have
derivative interactions. This is not a problem since the Goldstone fields
can easily be rewritten to accommodate this soft behavior. The invariant
term is realized by taking the trace in flavor space, denoted by $%
\left\langle ...\right\rangle .$

The parametrization of the coset space is not unique. It can be shown that $%
S $-matrix elements are invariant under a field redefinition $\psi =\phi
F\left( \phi \right) ,$ where $F\left( 0\right) =1.$ The power counting
depends on the number of derivatives, which is unchanged by the
redefinition, so the results are representation independent, order by order.
It is advantageous to use an exponential parametrization in $3\times 3$
flavor space. 
\begin{equation}
u^{2}=U=e^{i\frac{\sqrt{2}}{F}M}
\end{equation}%
where $F$ is a dimensional constant of same dimension as $M$ - the Goldstone
boson matrix%
\begin{equation}
M=\frac{1}{\sqrt{2}}\sum_{i=1}^{8}\lambda _{i}\phi ^{i}=\left( 
\begin{array}{ccc}
\frac{\pi ^{0}}{\sqrt{2}}+\frac{\eta _{8}}{\sqrt{6}} & \pi ^{+} & K^{+} \\ 
\pi ^{-} & -\frac{\pi ^{0}}{\sqrt{2}}+\frac{\eta _{8}}{\sqrt{6}} & K^{0} \\ 
K^{-} & \bar{K}^{0} & -\frac{2\eta _{8}}{\sqrt{6}}%
\end{array}%
\right)  \label{gsmatr}
\end{equation}%
where $\lambda _{i}$ are the Gell-Mann $SU\left( 3\right) $ matrices, $\phi
^{i}$ the Goldstone fields and $\eta _{8}$ denotes the octet component of $%
\eta $ (see section on $\eta _{8}$-$\eta _{0}$ mixing).

\section{Effective Quantum Field Theories (EQFT)}

QCD is a gauge theory describing the interactions of quarks and gluons.
Below $\Lambda _{\chi }$ it becomes highly non-perturbative. As a
consequence, we can no longer use the partonic degrees of freedom to
describe it. The effective chiral Lagrangian approach makes use of the very
simple low energy spectrum of light pseudoscalar mesons: $\pi ^{\pm },\pi
^{0},K^{0},\bar{K}^{0},K^{\pm }$ and $\eta ,$ as dynamic fields describing
the theory. Considering the weak nature of interactions amongst and between
the light mesons and nucleons, the perturbative approach can be reinstated
by simply transforming to the relevant degrees of freedom, i.e. the light
pseudoscalar mesons. This is the principle for all effective quantum field
theories. At a given energy, only certain degrees of freedom are relevant in
describing the theory. The non-relevant degrees of freedom can be integrated
out, and their effect is consequently encoded in the coefficients of the
effective theory. All quantum field theories can be regarded as EQFTs.
However, they differ in their degree of renormalizability. Since the theory
is valid only below a given intrinsic scale parameter $\Lambda ,$ we can
expand amplitudes in terms of $E/\Lambda ,$ and require a finite number of
counterterms in order to regularize at any $\mathcal{O}\left( E^{n}/\Lambda
^{n}\right) .$

If we want to be able to study relevant physical processes, the needed
external source fields must enter the Lagrangian. These are conveniently
included in a chiral invariant way. They take the form of external scalar $%
\left( s\right) $, pseudoscalar $\left( p\right) $, right- $\left( r_{\mu
}\right) $ and left-handed $\left( \ell _{\mu }\right) $ $3\times 3$ matrix
source functions. 
\begin{equation}
\mathcal{L}_{QCD}=...-\bar{\psi}\gamma _{\mu }P_{L}\ell ^{\mu }\psi -\bar{%
\psi}\gamma _{\mu }P_{R}r^{\mu }\psi -\bar{\psi}_{L}\left( s+ip\right) \psi
_{R}-\bar{\psi}_{R}\left( s-ip\right) \psi _{L}  \label{sources}
\end{equation}%
Here we see that the quark mass matrix $s=m=diag(m_{u},m_{d},m_{s})$
explicitly breaks the left and right chiral symmetries. The electroweak
gauge fields enter through $\ell _{\mu }$ and $r_{\mu }$.

The low energy effective action for the GS bosons is a functional of
external sources. We obtain the QCD connection by considering the effect of
the sources.%
\begin{equation}
e^{iZ\left( \ell _{\mu },r_{\mu },s,p\right) }=\int [d\bar{\psi}][d\psi
][dA_{\mu }^{a}]e^{i\int d^{4}x\mathcal{L}_{QCD}\left( \psi ,\bar{\psi}%
,A_{\mu }^{a},\ell _{\mu },r_{\mu },s,p\right) }
\end{equation}%
Only the GS bosons are the relevant degrees of freedom at low energies, so
in integrating out the heavy fields (thereby absorbing them into
coefficients) and transforming to the GS system gives 
\begin{equation}
e^{iZ\left( \ell _{\mu },r_{\mu },s,p\right) }=\int [dU]e^{i\int d^{4}x%
\mathcal{L}_{eff}\left( U,\ell _{\mu },r_{\mu },s,p\right) }
\end{equation}

\section{Chiral Perturbation Theory (ChPT)}

Chiral Perturbation Theory is an EQFT describing hadronic interactions in
the low energy limit of the standard model. It is valid below the breaking
scale of chiral symmetry, i.e. for energies $\ll 1GeV\sim \Lambda _{\chi }.$
The chiral theory successfully describes the mesonic sector, and has also
been extended to other fields such as heavy quark and bound state dynamics.
ChPT is the evolved form of Partial Conservation of the Axial current (PCAC)
and current algebra techniques.

\subsection{Lowest Order Effective Lagrangian}

To obtain terms that are invariant under both chiral and Lorenz symmetries,
at least two derivatives of $u$ or $u^{\dagger }$ must be present. In the
absence of external fields there is only one term of $\mathcal{O}\left(
p^{2}\right) $%
\begin{equation}
\mathcal{L}_{2}^{\left( 0\right) }=\frac{F^{2}}{4}\left\langle \partial
_{\mu }U\partial ^{\mu }U^{\dagger }\right\rangle  \label{L20}
\end{equation}%
where the coupling is set to reproduce the correct kinetic term. Adding
source functions coupled to their associated currents (following Gasser \&
Leutwyler \cite{gasleut}), as in eq.\thinspace (\ref{sources}), we find the
global symmetries (\ref{global symmetries}) implying the following
invariances:%
\begin{eqnarray}
&\psi &\overset{\smallskip }{\underset{U\left( 1\right) _{V}}{%
\longrightarrow }}e^{i\varepsilon }\psi \\
\,\, &\psi &\overset{\smallskip }{\underset{U\left( 1\right) _{A}}{%
\longrightarrow }}e^{i\gamma _{5}\varepsilon }\psi \\
&\psi &\underset{G}{\longrightarrow }\left( g_{L}P_{L}+g_{R}P_{R}\right)
\psi \quad ;g_{R,L}=e^{-i\vec{\theta}_{L,R}\cdot \vec{\lambda}/2}\in G \\
\ell _{\mu } &=&v_{\mu }-a_{\mu }\underset{G}{\longrightarrow }g_{L}\ell
_{\mu }g_{L}^{-1} \\
r_{\mu } &=&v_{\mu }+a_{\mu }\underset{G}{\longrightarrow }g_{R}r_{\mu
}g_{R}^{-1} \\
&&s\pm ip\underset{G}{\longrightarrow }g_{R,L}\left( s\pm ip\right)
g_{L,R}^{-1}
\end{eqnarray}%
These can all be made local by adding terms to the transformation of the
vector fields $\ell _{\mu },r_{\mu }.$%
\begin{equation}
\left. 
\begin{array}{c}
\ell _{\mu } \\ 
r_{\mu }%
\end{array}%
\right\} \,\,\,\,\,\,%
\begin{array}{l}
\underset{U\left( 1\right) _{V}}{\longrightarrow }\left\{ 
\begin{array}{c}
\ell _{\mu }-\partial _{\mu }\varepsilon \\ 
r_{\mu }-\partial _{\mu }\varepsilon%
\end{array}%
\right. \smallskip \\ 
\underset{U\left( 1\right) _{A}}{\longrightarrow }\left\{ 
\begin{array}{c}
\ell _{\mu }+\partial _{\mu }\varepsilon \\ 
r_{\mu }-\partial _{\mu }\varepsilon%
\end{array}%
\right. \smallskip \\ 
\,\,\,\underset{G}{\longrightarrow }\left\{ 
\begin{array}{c}
g_{L}\left( \ell _{\mu }+i\partial _{\mu }\right) g_{L}^{-1} \\ 
g_{R}\left( r_{\mu }+i\partial _{\mu }\right) g_{R}^{-1}%
\end{array}%
\right.%
\end{array}%
\end{equation}%
The pseudoscalar source is not relevant for the processes considered here
(it appears for example in the Higgs-sector), and the scalar field is set to
the light quark mass matrix$.$ (12) shows that the $s$ field transforms
non-trivially under $G$, explicitly breaking the chiral symmetry. But
because the light quark masses are much smaller than the chiral breaking
scale, we still have approximate chiral symmetry.

Comparing with the SM\ Lagrangian we can make the identifications 
\begin{eqnarray}
v_{\mu } &\rightarrow &-eQA_{\mu }-\frac{g}{2\sqrt{2}}\left( T_{+}W_{\mu
}^{+}+T_{-}W_{\mu }^{-}\right) \\
a_{\mu } &\rightarrow &\frac{g}{2\sqrt{2}}\left( T_{+}W_{\mu
}^{+}+T_{-}W_{\mu }^{-}\right)
\end{eqnarray}%
where the neutral weak gauge fields have been omitted, since their
contribution will be strongly suppressed by the heavy $Z^{0}$ mass. $Q$ is
the electric charge matrix $Q=\tfrac{1}{3}diag\left( 2,-1,-1\right) $ and 
\begin{equation}
T_{+}=\left( 
\begin{array}{ccc}
0 & V_{ud} & V_{us} \\ 
0 & 0 & 0 \\ 
0 & 0 & 0%
\end{array}%
\right)
\end{equation}%
and its Hermitian conjugate $T_{-}$ contain elements from the weak mixing
matrix.

The non-abelian field strength tensors are given by%
\begin{eqnarray}
F_{L}^{\mu \nu } &=&\partial ^{\mu }\ell ^{\nu }-\partial ^{\nu }\ell ^{\mu
}-i\left[ \ell ^{\mu },\ell ^{\nu }\right] \\
F_{R}^{\mu \nu } &=&\partial ^{\mu }r^{\nu }-\partial ^{\nu }r^{\mu }-i\left[
r^{\mu },r^{\nu }\right]
\end{eqnarray}%
Local invariance is maintained by replacing $\partial _{\mu }$ in (\ref{L20}%
) by the covariant derivative $D_{\mu }.$%
\begin{equation}
\partial _{\mu }U\rightarrow D_{\mu }U=\partial _{\mu }U-ir_{\mu }U+iU\ell
_{\mu }
\end{equation}

\subsection{Power Counting}

To organize our results we must examine how powers in the chiral expansion
are counted, so that we can assign them to the Lagrangian of appropriate
order. It is the order expansion that makes the whole approach useful. If we
assign $\partial _{\mu }u$ and $a_{\mu },v_{\mu }$ the same power counting, $%
D_{\mu }U$ becomes a first order homogeneous term in the derivative
expansion. We then have $U\sim \mathcal{O}\left( p^{0}\right) ,$ $a_{\mu
},v_{\mu },u_{\mu }\sim \mathcal{O}\left( p^{1}\right) ,$ $s,p,F_{L,R}^{\mu
\nu }\sim \mathcal{O}\left( p^{2}\right) .$ The lowest order Lagrangian
including external sources is given by:%
\begin{equation}
\mathcal{L}_{2}=\frac{F^{2}}{4}\left\langle D_{\mu }UD^{\mu }U^{\dagger
}+\chi U^{\dagger }+U\chi ^{\dagger }\right\rangle  \label{L2}
\end{equation}%
where $\chi =2B_{0}\left( s+ip\right) $, and $B_{0}$ is a constant related
to the vacuum expectation value $\left\langle \bar{\psi}\psi \right\rangle
_{0}$. The coupling $F$ is can be identified with the pion decay constant $%
F_{\pi }.$

\subsection{Vacuum Expectation Values and Masses}

The axial symmetry is a hidden symmetry, meaning that it is dynamically
broken - the invariance of the Lagrangian is not shared by the vacuum state.
It can be seen that the vacuum state does not transform separately under
left and right chiral transformations, as it couples $\psi _{L}$ with $\psi
_{R}$.%
\begin{eqnarray}
\left\langle 0\right| \bar{\psi}\psi \left| 0\right\rangle &=&\left\langle
0\right| \bar{\psi}\left( P_{L}+P_{R}\right) ^{2}\psi \left| 0\right\rangle 
\notag \\
&=&\left\langle 0\right| \bar{\psi}_{L}\psi _{R}\left| 0\right\rangle
+\left\langle 0\right| \bar{\psi}_{R}\psi _{L}\left| 0\right\rangle
\label{vacexp}
\end{eqnarray}%
where we have used the projection operator properties and 
\begin{equation}
\bar{\psi}_{L,R}=\left( P_{L,R}\psi \right) ^{\dagger }\gamma ^{0}=\psi
^{\dagger }P_{L,R}\gamma ^{0}=\bar{\psi}P_{R,L}
\end{equation}%
(by the commutation relations for the gamma matrices). In a world of
massless quarks, the cost of producing a quark-antiquark pair with total
angular momentum and momentum zero, is small. The vacuum can be seen as
containing a condensate of $q\bar{q}$ pairs with strong attractive
interactions. Mixing of chirality (\ref{vacexp}) implies that up and down
quarks can acquire an effective (constituent) mass by moving through and
interacting with the vacuum (see for example \cite{qft1}).

We can evaluate current matrix elements by differentiating the classical
action $S_{2}=\int d^{4}x\mathcal{L}_{2}$ with respect to external sources.
This gives the lowest order result: 
\begin{eqnarray}
\left\langle 0\right| \bar{d}\gamma ^{\mu }\gamma _{5}u\left| \pi ^{+}\left(
p\right) \right\rangle &=&\left\langle 0\right| \frac{\delta S_{2}}{\delta
a_{\mu }}\left| \pi ^{+}\left( p\right) \right\rangle =i\sqrt{2}Fp^{\mu }
\label{exp1} \\
\left\langle 0\right| \bar{\psi}\psi \left| 0\right\rangle &=&-\left\langle
0\right| \frac{\delta S_{2}}{\delta s}\left| 0\right\rangle =-F^{2}B_{0}
\label{exp2}
\end{eqnarray}%
From the definition of the pion decay constant, 
\begin{equation*}
i\sqrt{2}F_{\pi }p^{\mu }=\left\langle 0\right| \bar{d}\gamma ^{\mu }\gamma
_{5}u\left| \pi ^{+}\left( p\right) \right\rangle
\end{equation*}%
we see that to lowest order we can make the identification $F=F_{\pi }.$
Relations (\ref{exp1}) \& (\ref{exp2}) are only valid in the chiral limit,
and are subject to corrections of order $\mathcal{O}\left( m_{q}\right) .$ (%
\ref{exp2}) relates $B_{0}$ to the vacuum expectation value.

The combination $B_{0}m_{q}$ is experimentally determinable. The reason for
identifying $\mathcal{O}\left( m_{q}\right) \sim \mathcal{O}\left(
p^{2}\right) $ becomes clear if we expand $(\ref{L2})$ to second order in
meson fields. We then get the relations%
\begin{eqnarray}
M_{\pi ^{+}}^{2} &=&B_{0}\left( m_{u}+m_{d}\right) \\
M_{K^{+}}^{2} &=&B_{0}\left( m_{u}+m_{s}\right) \\
M_{K^{0}}^{2} &=&B_{0}\left( m_{d}+m_{s}\right) \\
M_{\eta _{8}}^{2} &=&\tfrac{1}{3}B_{0}\left( m_{u}+m_{d}+4m_{s}\right)
\end{eqnarray}%
By elimination we obtain the Gell-Mann-Okubo consistency relation for the GS
bosons $3M_{\eta _{8}}^{2}=4M_{K}^{2}-M_{\pi }^{2}$ \cite{GeOr}$.$ This is
reasonably\footnote{%
Putting in the numbers we find that we're off by approximately 20 MeV, but
if we instead use linear relations, the numbers get worse ($\sim 70$ MeV).}
well satisfied using $m_{u}\simeq m_{d}$ and $M_{\eta _{8}}\simeq M_{\eta }$
(see $\eta $ mixing). For higher mass resonances both linear and quadratic
mass formulas give acceptable relations \cite{dynsm}. This is because to
first order in symmetry breaking we have $\delta \left( m^{2}\right) =\left(
m_{0}+\delta m\right) ^{2}-m_{0}^{2}=2m_{0}\delta m+...$ But when expanding
around a massless limit $m$ and $m^{2}$ distinction becomes important. In
the normalized effective theory, the pion mass prediction is%
\begin{equation}
m_{\pi }^{2}=\left( m_{u}+m_{d}\right) B_{0}+\left( m_{u}+m_{d}\right)
^{2}C_{0}+...
\end{equation}%
However, there is no symmetry constraint to force the renormalized parameter 
$B_{0}$ to zero, so the squared pion mass is (mostly) linear in the symmetry
breaking quark masses.

\subsection{The $\mathcal{O}\left( p^{4}\right) $ effective action}

Listing all possible operators invariant under discrete and continuous
symmetries, transforming linearly, we can construct the tree-level effective
chiral action of order $p^{4}.$ The number of terms can be reduced to a
minimum using the equations of motion for $\mathcal{O}\left( p^{2}\right) ,$
partial integration, the unitarity of $U$ and $SU\left( n\right) $ $n$%
-depending trace identities. $\mathcal{L}_{4}$ for $SU\left( 3\right) $ was
first determined by Gasser and Leutwyler \cite{gasleut} and is given by%
\begin{eqnarray}
\mathcal{L}_{4} &=&L_{1}\left\langle D_{\mu }UD^{\mu }U^{\dagger
}\right\rangle ^{2}+L_{2}\left\langle D_{\mu }UD_{\nu }U^{\dagger
}\right\rangle ^{2}+L_{3}\left\langle D_{\mu }UD^{\mu }U^{\dagger }D_{\nu
}UD^{\nu }U^{\dagger }\right\rangle  \notag \\
&&+L_{4}\left\langle D_{\mu }UD^{\mu }U^{\dagger }\right\rangle \left\langle
\chi U^{\dagger }+U\chi ^{\dagger }\right\rangle +L_{5}\left\langle \left(
D_{\mu }UD^{\mu }U^{\dagger }\right) \left( \chi U^{\dagger }+U\chi
^{\dagger }\right) \right\rangle  \notag \\
&&+L_{6}\left\langle \chi U^{\dagger }+U\chi ^{\dagger }\right\rangle
^{2}+L_{7}\left\langle \chi U^{\dagger }-U\chi ^{\dagger }\right\rangle
^{2}+L_{8}\left\langle \chi U^{\dagger }\chi U^{\dagger }+\chi ^{\dagger
}U\chi ^{\dagger }U\right\rangle  \notag \\
&&+iL_{9}\left\langle L_{\mu \nu }D^{\mu }UD^{\nu }U^{\dagger }+R_{\mu \nu
}D^{\mu }U^{\dagger }D^{\nu }U\right\rangle +L_{10}\left\langle
L_{\mu \nu }UR^{\mu \nu }U^{\dagger }\right\rangle  \label{L4}
\end{eqnarray}%
\newline
where $L_{i}$ are the expansion coefficients that must be determined
phenomenologically. Both $\mathcal{L}_{2}$ and $\mathcal{L}_{4}$ are
invariant under%
\begin{equation}
U\leftrightarrow U^{\dagger },\quad \chi \leftrightarrow \chi ^{\dagger
},\quad \ell _{\mu }\leftrightarrow r_{\mu }  \label{ipar}
\end{equation}%
since the trace is invariant under cyclic shifting of matrices. This the
so-called intrinsic parity operation (originally introduced by Witten \cite%
{WZW}), which operates on the function, but not on the space-time
coordinates. If $\ell _{\mu }=r_{\mu }$, then we can have terms containing
an odd or even number of pseudoscalars, but no transition between the two,
as this would violate intrinsic parity conservation. $\ell _{\mu }$ is equal
to $r_{\mu }$ if we have only EM interactions or direct meson interaction
(no external fields). Including the $W$ fields, the two sectors of differing
intrinsic parity are coupled and intrinsic parity can be violated. Hence,
both $\mathcal{L}_{2}$ and $\mathcal{L}_{4}$ are unable to describe
anomalous processes like $3\pi 2K$ or $\pi ^{0}\gamma \gamma $. For further
details see \cite{u1v}.

What have we missed? There is the possibility of terms that transform
non-trivially under $G,$ but still preserve chiral symmetry if their
variation under $G$ is a total derivative \cite{WZW}. This type of term is
anomalous.

In order to describe anomalous processes we need terms that take the axial
anomaly into account. The first terms that contribute are $\mathcal{O}\left(
p^{4}\right) $ and make up the so-called Wess-Zumino-Witten anomalous action.

\subsection{Anomalous Processes}

In this thesis we consider a number of anomalous processes. First we have
the pseudoscalar to $\gamma \gamma ^{\star }$ decays: $\pi ^{0}/\eta
\rightarrow \gamma \gamma ^{\star },$ where $\gamma ^{\star }$ is on-shell
or off-shell going to an $e^{+}e^{-}$ pair. We also consider the
semileptonic weak decays $\pi ^{+}/K^{+}\rightarrow \gamma e^{+}\nu _{e},$ $%
K^{+}\rightarrow \pi ^{+}\pi ^{-}e^{+}\nu ,\,K^{+}\rightarrow \pi ^{0}\pi
^{0}e^{+}\nu $ and the three-pseudoscalar-photon interactions $\gamma \pi
^{-}\rightarrow \pi ^{-}\pi ^{0}$ and $\eta \rightarrow \pi ^{+}\pi
^{-}\gamma .$ The $\eta $ calculations were performed for both the octet $%
\eta _{8}$ and the singlet $\eta _{0}$ component$,$ but only the octet is
relevant at the level we're working at. The Feynman diagrams for the
processes are depicted in figure 1.

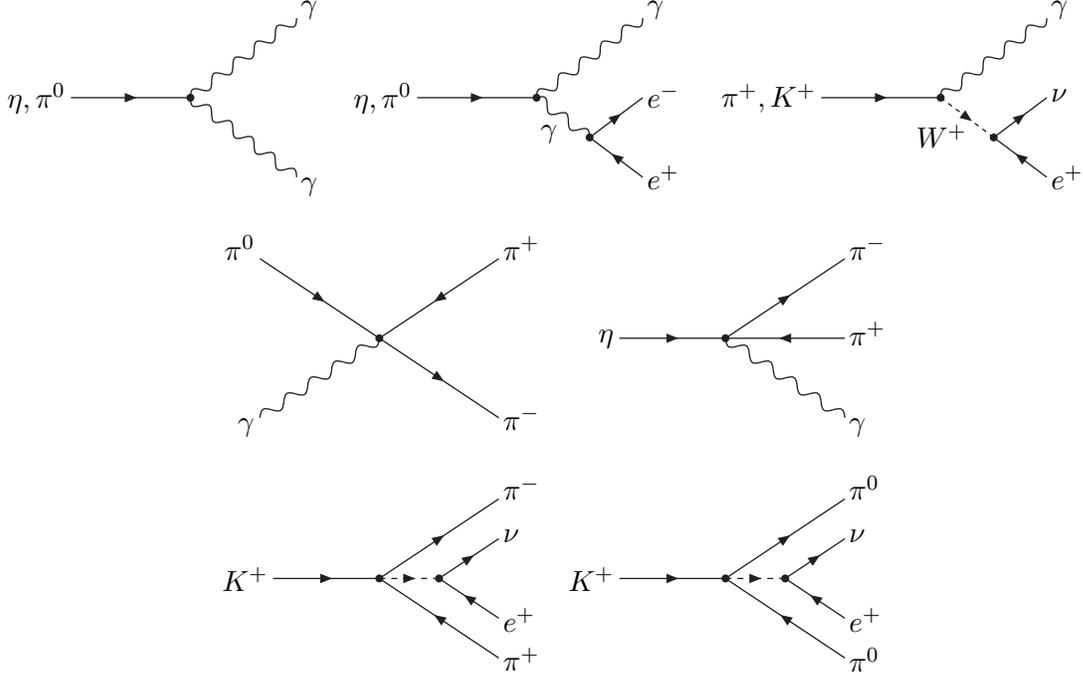
\begin{figure}
\begin{center}
\begin{picture}(120,90)(0,0)
\ArrowLine(15,50)(60,50)
\Vertex(60,50){1.5}
\Photon(60,50)(100,80){2}{5}
\Photon(60,50)(100,20){-2}{5}
\Text(13,50)[r]{$\eta, \pi ^{0}$}
\Text(102,20)[tl]{$\gamma$}
\Text(102,80)[bl]{$\gamma$}
\end{picture}$\quad $%
\begin{picture}(120,90)(0,0)
\ArrowLine(15,50)(60,50)
\Vertex(60,50){1.5}
\Photon(60,50)(100,80){2}{5}
\Photon(60,50)(80,35){2}{2.5}
\Vertex(80,35){1.5}
\Text(13,50)[r]{$\eta, \pi ^0$}
\Text(102,20)[l]{$e ^{+}$}
\Text(102,51)[l]{$e ^{-}$}
\Text(102,80)[bl]{$\gamma$}
\Text(68,40)[tr]{$\gamma$}
\ArrowLine(80,35)(100,50)
\ArrowLine(100,20)(80,35)
\end{picture}$\quad \quad \quad $%
\begin{picture}(120,90)(0,0)
\ArrowLine(15,50)(60,50)
\Vertex(60,50){1.5}
\Photon(60,50)(100,80){2}{5}
\DashArrowLine(60,50)(80,35){2}
\Vertex(80,35){1.5}
\Text(13,50)[r]{$\pi^{+},K^{+}$}
\Text(102,20)[l]{$e ^{+}$}
\Text(102,51)[l]{$\nu$}
\Text(102,80)[bl]{$\gamma$}
\Text(70,40)[tr]{$W ^{+}$}
\ArrowLine(80,35)(100,50)
\ArrowLine(100,20)(80,35)
\end{picture}

\begin{picture}(120,90)(0,0)
\Photon(10,20)(55,50){2}{5}
\Vertex(55,50){1.5}
\ArrowLine(10,80)(55,50)
\Text(8,20)[rt]{$\gamma$}
\Text(8,80)[rb]{$\pi ^{0}$}
\Text(102,20)[l]{$\pi ^{-}$}
\Text(102,80)[bl]{$\pi ^{+}$}
\ArrowLine(55,50)(100,20)
\ArrowLine(100,80)(55,50)
\end{picture}$\quad $%
\begin{picture}(120,90)(0,0)
\ArrowLine(15,50)(55,50)
\Vertex(55,50){1.5}
\Photon(55,50)(100,20){-2}{5}
\Text(102,20)[lt]{$\gamma$}
\Text(13,50)[r]{$\eta$}
\Text(102,52)[l]{$\pi ^{+}$}
\Text(102,80)[lb]{$\pi ^{-}$}
\ArrowLine(100,50)(55,50)
\ArrowLine(55,50)(100,80)
\end{picture}

\begin{picture}(120,90)(0,0)
\ArrowLine(15,50)(55,50)
\Vertex(55,50){1.5}
\ArrowLine(100,20)(55,50)
\Text(13,50)[r]{$K^{+}$}
\Text(102,24)[lt]{$\pi ^{+}$}
\Text(102,80)[lb]{$\pi ^{-}$}
\DashArrowLine(55,50)(77.5,50){2}
\Vertex(77.5,50){1.5}
\ArrowLine(77.5,50)(100,65)
\ArrowLine(100,35)(77.5,50)
\Text(102,64)[lb]{$\nu$}
\Text(102,39)[lt]{$e^{+}$}
\ArrowLine(55,50)(100,80)
\end{picture}$\quad $%
\begin{picture}(120,90)(0,0)
\ArrowLine(15,50)(55,50)
\Vertex(55,50){1.5}
\ArrowLine(100,20)(55,50)
\Text(13,50)[r]{$K^{+}$}
\Text(102,24)[lt]{$\pi ^{0}$}
\Text(102,80)[lb]{$\pi ^{0}$}
\DashArrowLine(55,50)(77.5,50){2}
\Vertex(77.5,50){1.5}
\ArrowLine(77.5,50)(100,65)
\ArrowLine(100,35)(77.5,50)
\Text(102,64)[lb]{$\nu$}
\Text(102,39)[lt]{$e^{+}$}
\ArrowLine(55,50)(100,80)
\end{picture}
\end{center}
\caption{
\emph{ Feynman diagrams for the anomalous processes.}}
\end{figure}

\section{$\protect\eta ^{\prime }\left( 958\right) $}

Invoking the quark model, we see that the quantum numbers of $\pi ^{+},\pi
^{-},\pi ^{0},K^{+},K^{-},K^{0}$,$\bar{K}^{0}$ and $\eta _{8}$ are the same
as for $u\bar{d},d\bar{u},(u\bar{u}-d\bar{d}),u\bar{s},s\bar{u},d\bar{s},s%
\bar{d}$ and ($u\bar{u}+d\bar{d}-2s\bar{s}$). Extending this logic, we find
one more state $u\bar{u}+d\bar{d}+s\bar{s}$, identifiable with the next
lightest pseudoscalar meson in turn, the $\eta ^{\prime }\left( 958\right) .$
The $\eta ^{\prime }$ has an abnormally large mass compared to the other
pseudoscalars. This is because it receives a mass contribution from the
axial $U\left( 1\right) _{A}$ anomaly \cite{dynsm}. By Noether's theorem the
axial $SU\left( 3\right) $ singlet current is $J_{5\mu }^{\left( 0\right)
}=\sum_{q=u,d,s}\bar{q}\gamma _{\mu }\gamma _{5}q,$ the divergence of which
receives an anomalous contribution%
\begin{equation}
\partial ^{\mu }J_{5\mu }^{\left( 0\right) }=\tfrac{3\alpha _{s}}{8\pi }%
G_{\mu \nu }^{a}\tilde{G}^{a\mu \nu }+2im_{q}\sum_{q=u,d,s}\bar{q}\gamma
_{5}q\quad ;\tilde{G}^{a\mu \nu }\equiv \varepsilon ^{\mu \nu \alpha \beta
}G_{\alpha \beta }^{a}
\end{equation}%
where $G_{\mu \nu }$ is the field strength tensor for the strong interaction
and $\varepsilon ^{\mu \nu \alpha \beta }$ the antisymmetric Levi-Civita
tensor. Taking the divergence of the vacuum to $\eta _{0}$ matrix element,
we get%
\begin{eqnarray}
\left\langle 0\right| J_{5\mu }^{\left( 0\right) }\left| \eta _{0}\left( 
\mathbf{p}\right) \right\rangle &=&iF_{\eta _{0}}p_{\mu }e^{-ip\cdot
x}\Rightarrow \\
\left\langle 0\right| \partial ^{\mu }J_{5\mu }^{\left( 0\right) }\left|
\eta _{0}\left( \mathbf{p}\right) \right\rangle &=&F_{\eta _{0}}m_{\eta
_{0}}^{2}\Rightarrow \\
\underset{m_{q}\rightarrow 0}{\lim }m_{\eta _{0}}^{2} &=&\tfrac{1}{F_{\eta
_{0}}}\tfrac{3\alpha _{s}}{8\pi }\left\langle 0\right| G_{\mu \nu }^{a}%
\tilde{G}^{a\mu \nu }\left| \eta _{0}\left( \vec{p}\right) \right\rangle
\end{eqnarray}%
which tells us that the $\eta _{0}$ mass is non-vanishing in the chiral
limit. If it were not for the anomalous $G\tilde{G}$-term the $U\left(
1\right) _{A}$ symmetry would be approximately conserved and break
dynamically along with the chiral $SU\left( 3\right) ,$ producing a nonet of
GS bosons.

\section{$\protect\eta _{8}-\protect\eta _{0}$ mixing}

$SU\left( 3\right) $ breaking in the quark mass matrix causes the singlet $%
\eta _{0}$ and octet $\eta _{8}$ components to mix, yielding the physical
states $\eta $ and $\eta ^{\prime }.$%
\begin{equation}
\left( 
\begin{array}{c}
\left| \eta \right\rangle \\ 
\left| \eta ^{\prime }\right\rangle%
\end{array}%
\right) =\left( 
\begin{array}{cc}
\cos \theta & -\sin \theta \\ 
\sin \theta & \cos \theta%
\end{array}%
\right) \left( 
\begin{array}{c}
\left| \eta _{8}\right\rangle \\ 
\left| \eta _{0}\right\rangle%
\end{array}%
\right) \quad ;\left\{ 
\begin{array}{l}
\left| \eta _{8}\right\rangle =\tfrac{1}{\sqrt{6}}\left( u\bar{u}+d\bar{d}-2s%
\bar{s}\right) \smallskip \\ 
\left| \eta _{0}\right\rangle =\tfrac{1}{\sqrt{3}}\left( u\bar{u}+d\bar{d}%
+\,s\bar{s}\right)%
\end{array}%
\right.
\end{equation}%
Quantum mechanically, other states with $I=0$ can also enter the mix. These
we assume to be too heavy to be of significance. Another crucial assumption
is that the mixing does not depend on the energy of the state, which allows
for this simple phenomenologically motivated model \cite{etamixmod}. Making
use of the $\pi ^{0},\eta ,\eta ^{\prime }$ to $\gamma \gamma $ data, the
mixing angle $\theta $ is determined to be approximately $-20^{\text{o}}.$
With this angle we find the quark content%
\begin{eqnarray}
\eta &\sim &0.58\left( u\bar{u}+d\bar{d}\right) -0.57s\bar{s} \\
\eta ^{\prime } &\sim &0.40\left( u\bar{u}+d\bar{d}\right) +0.82s\bar{s}
\end{eqnarray}%
The $s\bar{s}$ is decreased for the $\eta $, and increased for the $\eta
^{\prime },$ accounting for the large mass difference. Since the up quark
has a greater charge magnitude, the $\eta \rightarrow \gamma \gamma $
amplitude is boosted. Heuristically, this can be seen by%
\begin{eqnarray}
A &\sim &e_{u}^{2}\left( \frac{\cos \theta }{\sqrt{6}}-\frac{\sin \theta }{%
\sqrt{3}}\right) +e_{d}^{2}\left( \frac{\cos \theta }{\sqrt{6}}-\frac{\sin
\theta }{\sqrt{3}}\right) +e_{s}^{2}\left( -\frac{2\cos \theta }{\sqrt{6}}-%
\frac{\sin \theta }{\sqrt{3}}\right)  \notag \\
&\propto &\left\{ 
\begin{array}{ll}
\frac{1}{\sqrt{6}}\smallskip & ;\theta =0\smallskip \\ 
\frac{1}{\sqrt{3}} & ;\theta \simeq 20^{\text{o}}%
\end{array}%
\right.
\end{eqnarray}%
so that there is a factor 2 difference in width if we take the mixing into
account. This mixing model can also be directly encoded into the chiral
coefficients, but offers an explanation for lowest order
experimental-theoretical discrepancies.

\section{Anomalies}

Anomalies are said to appear whenever a classical symmetry is violated by
the existence of quantum corrections. They are crucial in expanding the
theoretical framework, as they signal new physics in the standard model. In
this section we examine the origin of the anomaly.

\subsection{Classical vs. Quantum Symmetries}

Let's compare the classical Noether current with the one obtained from the
full quantum theory using path integrals. A more complete treatment of the
subject can be found in \cite{dynsm}. The infinitesimal transformation 
\begin{equation}
\varphi _{i}\rightarrow \varphi _{i}^{\prime }=\varphi _{i}+\varepsilon
\left( x\right) f_{i}\left( \varphi \right) ,  \label{redef}
\end{equation}%
where $\varepsilon \left( x\right) $ is the infinitesimal parameter (which
is temporarily given a coordinate dependence for the purposes of Noether
current construction) and $f_{i}$ is an arbitrary function of the fields,
implies a Noether current and an invariance condition:%
\begin{equation}
J^{\mu }\left( x\right) =\frac{\partial \mathcal{L}^{\prime }}{\partial
\left( \partial _{\mu }\varepsilon \right) }\Rightarrow \mathcal{\mathcal{L}}%
\left( \varphi ^{\prime },\partial \varphi ^{\prime }\right) =\mathcal{L}%
\left( \varphi ,\partial \varphi \right) +J^{\mu }\partial _{\mu }\varepsilon
\label{class}
\end{equation}%
So that the Lagrangian is invariant if $\varepsilon $ is a constant. (\ref%
{class}) represents the classical symmetry.

In the path integral formalism, all matrix elements can be obtained via the
generating functional $W[j],$ which is a functional of the sources. 
\begin{equation}
W\left[ j\right] =e^{iZ\left[ j\right] }=\int \left[ d\varphi \right]
e^{i\int d^{4}x\left( \mathcal{L}\left( \varphi ,\partial \varphi \right)
-j\varphi \right) }  \label{genfunc}
\end{equation}%
where the classical source field $j\left( x\right) $ is used to probe the
theory, and where $\left[ d\varphi \right] $ stands for integration over all
possible values of $\varphi \left( x\right) $ at each space-time point.
Matrix elements describing physical processes are obtained by taking the
functional derivative\footnote{%
Functional differentiation is defined by%
\begin{equation*}
j\left( t\right) =\int dt^{\prime }\delta \left( t-t^{\prime }\right)
j\left( t^{\prime }\right) \Rightarrow \frac{\delta j\left( t\right) }{%
\delta j\left( t^{\prime }\right) }=\delta \left( t-t^{\prime }\right)
\end{equation*}%
} of the logarithm of the path integral.%
\begin{equation}
\left\langle 0\right| T\left( \varphi \left( x_{k}\right) ...\varphi \left(
x_{p}\right) \right) \left| 0\right\rangle =\left( i\right) ^{n}\frac{\delta
^{n}\ln W\left[ j\right] }{\delta j\left( x_{k}\right) ...\delta j\left(
x_{p}\right) }  \label{matrelm}
\end{equation}%
We can study the classical Noether current $J^{\mu }\left( x\right) $ by
coupling it to a classical source field $v_{\mu }$ and inserting it in the
generating functional (\ref{genfunc}).%
\begin{equation}
W\left[ v_{\mu }\right] =\int \left[ d\varphi \right] e^{i\int d^{4}x\left( 
\mathcal{L}\left( \varphi ,\partial \varphi \right) -v_{\mu }J^{\mu }\right)
}
\end{equation}%
(\ref{matrelm}) lets us take current matrix elements (denoted by bar),%
\begin{eqnarray}
\bar{J}^{\mu }\left( x\right) &=&i\frac{\delta \ln W\left[ v_{\nu }\right] }{%
\delta v_{\mu }\left( x\right) }\Rightarrow \\
\delta \ln W\left[ v_{\mu }\right] &=&\ln W\left[ v_{\mu }+\delta v_{\mu }%
\right] -\ln W\left[ v_{\mu }\right] \equiv -i\int d^{4}x\bar{J}^{\mu
}\left( x\right) \delta v_{\mu }\left( x\right)
\end{eqnarray}%
If we choose $\delta v_{\mu }=-\partial _{\mu }\varepsilon \left( x\right) ,$
then%
\begin{eqnarray}
\delta _{\varepsilon }\ln W\left[ v_{\mu }\right] &=&\ln W\left[ v_{\mu
}-\partial _{\mu }\varepsilon \right] -\ln W\left[ v_{\mu }\right] \\
&=&i\int d^{4}x\bar{J}^{\mu }\left( x\right) \partial _{\mu }\varepsilon
\left( x\right) =-i\int d^{4}x\varepsilon \left( x\right) \partial _{\mu }%
\bar{J}^{\mu }\left( x\right)
\end{eqnarray}%
where in the last step partial integration has been used. From this follows
that if $\delta _{\varepsilon }\ln W\left[ v_{\mu }\right] =0$ then all
matrix elements are divergenceless, i.e. $\partial ^{\mu }\bar{J}_{\mu
}\left( x\right) =0.$

Since we are integrating over all values of $\varphi \left( x\right) $ at
each point in space-time, one can argue that it should make no difference if
we shift the origin of integration at each point $x$ by a constant,
redefining $\varphi _{i}\left( x\right) \equiv \varphi _{i}^{\prime }\left(
x\right) -\varepsilon \left( x\right) f_{i}\left( \varphi \right) $, as in (%
\ref{redef}), with an accompanying Jacobian $\mathcal{J}=1.$ 
\begin{eqnarray}
\ln W\left[ v_{\mu }-\partial _{\mu }\varepsilon \right] &=&\int \left[
d\varphi _{i}\right] e^{i\int d^{4}x\left( \mathcal{L}\left( \varphi
,\partial \varphi \right) -\left( v_{\mu }-\partial _{\mu }\varepsilon
\right) J^{\mu }\right) } \\
&=&\int \left[ d\varphi _{i}^{\prime }\right] e^{i\int d^{4}x\left( \mathcal{%
L}\left( \varphi ^{\prime },\partial \varphi ^{\prime }\right) -v_{\mu
}J^{\mu }\right) }=\ln W\left[ v_{\mu }\right]  \label{redefeps}
\end{eqnarray}%
This would imply $\partial ^{\mu }\bar{J}_{\mu }\left( x\right) =0,$ in
accordance with classical current conservation. However, as was first shown
by Fujikawa \cite{fuji}, shifting the origin like this is not always
allowed. The transformation (\ref{redef}) can have a Jacobian $\mathcal{J}%
\neq 1,$ so that the path integral measure is changed. If the change of
variables is non-trivial then $\partial ^{\mu }\bar{J}_{\mu }\left( x\right)
\neq 0,$ and we have encountered an anomaly.

\subsection{The $U\left( 1\right) _{A}$ Axial Anomaly}

In the chiral limit$,$ the QCD Lagrangian is invariant under the global $%
U\left( 1\right) _{A}$ axial transformation $\psi \rightarrow e^{-i\theta
\gamma _{5}}\psi .$ Using Noether's theorem we get the classically conserved
singlet current $J_{5\mu }^{\left( 0\right) }=\sum \bar{q}\gamma _{\mu
}\gamma _{5}q,$ with $\partial ^{\mu }J_{5\mu }^{\left( 0\right) }=0.$
However, in going through the full quantum theory analysis, it is found that
the divergence of the matrix elements of the axial vector singlet current is
proportional an anomaly,%
\begin{equation}
\partial ^{\mu }J_{5\mu }^{\left( 0\right) }=\tfrac{3\alpha _{s}}{8\pi }%
G_{\mu \nu }^{a}\tilde{G}^{a\mu \nu }\quad ;\tilde{G}_{\mu \nu }^{a}\equiv
\varepsilon ^{\mu \nu \alpha \beta }G_{\alpha \beta }^{a}
\label{anom singlet}
\end{equation}%
Important consequences of the anomaly involves a large anomalous
contribution to the $\pi ^{0}\rightarrow \gamma \gamma $ decay rate, the
prevention of $\eta ^{\prime }$ becoming a Goldstone boson (thus separating
the chiral $SU\left( 3\right) $ spectra into one octet and one singlet part).

There are two main approaches one can take in examining the anomaly: 1) a
direct calculation via the $U\left( 1\right) _{A}\rightarrow gg$ triangle
diagram, or 2) the path integral analysis. The direct approach was
originally taken by the discoverers of the $U\left( 1\right) _{A}$ anomaly%
\footnote{%
A.k.a. the Adler-Bell-Jackiw anomaly.}: Adler \cite{adler}, Bell and Jackiw 
\cite{belljackiw}. They calculated the matrix element for the $U\left(
1\right) _{A}$ current going to two gluons via the Feynman diagram in figure
2. In the resulting four-momentum integral we can make a change of
variables, yielding an expression that is compatible with conservation of
the vector color current or the $U\left( 1\right) _{A}$ current, but not
both. Phenomenologically, we know that only the vector color current is
conserved. In the presence of the $U\left( 1\right) _{A}$ anomaly, the axial
symmetry is not even approximately conserved. The anomaly is also present in
the $U\left( 1\right) _{A}\rightarrow \gamma \gamma $ diagram, producing
equation ($\ref{anom singlet})$ with $\alpha _{s}\rightarrow \alpha _{EM}$
and $G_{\mu \nu }\rightarrow F_{\mu \nu }.$

\begin{figure}
\begin{center}
$\qquad \quad 
\begin{picture}(250,70)(0,0)  
\SetScale{0.8} 
\ArrowLine(5,50)(60,80) 
\ArrowLine(60,80)(60,20) 
\ArrowLine(60,20)(5,50) 
\Vertex(5,50){1.5} 
\Vertex(60,80){1.5} 
\Vertex(60,20){1.5} 
\Gluon(60,80)(110,80){4}{5} 
\Gluon(60,20)(110,20){4}{5} 
\ArrowLine(130,50)(185,80) 
\ArrowLine(185,80)(185,20) 
\ArrowLine(185,20)(130,50) 
\Vertex(130,50){1.5} 
\Vertex(185,80){1.5} 
\Vertex(185,20){1.5} 
\Gluon(185,80)(245,20){4}{8} 
\Gluon(185,20)(245,80){4}{8} 
\end{picture}$

\end{center}
\caption{
$U\left( 1\right) _{A}\rightarrow gg$ \emph{diagrams
leading to an axial vector anomaly.}}

\end{figure}
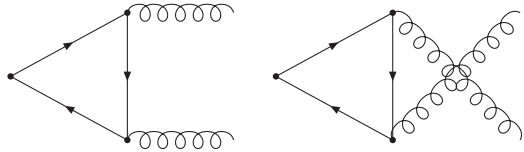

The path integral approach was taken by Fujikawa \cite{fuji}, and clarifies
the origin of the anomaly. In the path integral formalism we introduce the
singlet axial current coupled to an axial current source $a_{\mu }$ into a
functional of the gluon field $A_{\mu }^{a}.$%
\begin{equation}
W[a_{\mu },A_{\mu }^{a}]=\int [d\bar{\psi}][d\psi ]e^{i\int d^{4}x\left( 
\mathcal{L}_{QCD}\left( \psi ,\bar{\psi},A_{\lambda }^{c}\right) -a_{\mu
}J_{5}^{\left( 0\right) \mu }\right) }
\end{equation}%
The crucial point is that in redefining the fermion fields to absorb the $%
\varepsilon $ term of eq. (\ref{redefeps}), the path integral measure is
changed:%
\begin{equation}
\int [d\bar{\psi}][d\psi ]=\mathcal{J}\int [d\bar{\psi}^{\prime }][d\psi
^{\prime }]
\end{equation}%
where the Jacobian $\mathcal{J}$ is independent of the fermion fields. The
Jacobian is divergent, but can be regularized (as was done by Fujikawa) by
removing the high energy eigenmodes of the Dirac field in a gauge invariant
way. The freedom to change integration variables in the direct approach,
corresponds to the freedom in choice of regulator in the path integral
approach.

\section{The Wess-Zumino-Witten (WZW) Anomaly Action}

In section 4.4 we saw that if we want to describe anomalous processes, we
must construct an effective Lagrangian that violates intrinsic parity. This
can be ensured by always including the totally antisymmetric Levi-Civita
tensor $\varepsilon ^{\mu \nu \alpha \beta }$, the presence of which will
preserve normal parity while at the same time violating intrinsic parity.
The origin of this symmetry breaking lies in the axial anomaly. It affects
photonic processes like $\pi ^{0}\rightarrow \gamma \gamma ,$ but also
hadronic ones like $K\bar{K}\rightarrow \pi ^{0}\pi ^{+}\pi ^{-}.$ For
reactions like these we need to construct a Lagrangian, which takes the
axial anomaly into account. This section contains a brief sketch of how this
can be done. The derivation follows the sigma model approach employed by
Donoghue et al \cite{dynsm}.

The history of the anomalous $\mathcal{O}\left( p^{4}\right) $ WZW action
follows a somewhat crooked path. The first effective action analysis of the
anomaly was made by Wess and Zumino, who arrived at their expression by
integrating the anomalous Ward identities \cite{WZ}. Their Lagrangian was
given as a Taylor expansion, and was later given a geometric interpretation
and rewritten on a compact form as a five-dimensional integral with
four-dimensional space-time boundaries by Witten \cite{WZW} (his form did
not conserve parity, but was later corrected).

\subsection{Anomalous Terms}

We emulate QCD behavior by introducing a color quantum number (with $N_{c}$
colors), and we are using the $u,d,s$ quarks so the number of fermions is
set to three (all with mass $M)$. All that is needed to derive the anomalous
action are the correct symmetries. A convenient starting point is the sigma
model, which incorporates the essential features of spontaneous symmetry
breaking and chiral invariance. Making use of the representation
independence theorem one can then proceed with the non-linear sigma model
using exponential parametrization. This facilitates the extension of the
formalism from 2 to 3 flavors. Dropping all terms irrelevant to the anomaly
and imposing the unitary change of variables $\psi _{L}\rightarrow \xi
^{\dagger }\psi _{L},$ $\psi _{R}\rightarrow \xi \psi _{R}$, one arrives at
a Lagrangian for fermions of mass $M$ coupled to axial and vector sources.%
\begin{equation}
\mathcal{L}=\bar{\psi}\left( i\mathcal{\NEG{D}}-M\right) \psi
\end{equation}%
where%
\begin{eqnarray}
D_{\mu }=\partial _{\mu }+i\bar{V}_{\mu }+i\bar{A}_{\mu }\gamma _{5}\quad &;&%
\bar{V}_{\mu }=-\tfrac{i}{2}\left( \xi ^{\dagger }\partial _{\mu }\xi +\xi
\partial _{\mu }\xi ^{\dagger }\right)  \label{curdef} \\
&;&\bar{A}_{\mu }=-\tfrac{i}{2}\left( \xi ^{\dagger }\partial _{\mu }\xi
-\xi \partial _{\mu }\xi ^{\dagger }\right)
\end{eqnarray}%
Our model contains no gluons, but according to the Adler-Bardeen theorem 
\cite{adbardeen} this will not modify the result. Keeping in mind the
section on anomalies, we see that the change of variables induces a change
in the path integral measure, and the Jacobian must enter into the effective
action. 
\begin{equation}
e^{i\Gamma \left( U\right) }=\int \left[ d\psi \right] [d\bar{\psi}]\mathcal{%
J}e^{i\int d^{4}x\bar{\psi}\left( i\mathcal{\NEG{D}-M}\right) \psi }=e^{\ln 
\mathcal{J}}e^{\mathrm{t\mathrm{r}}\ln \left( i\mathcal{\NEG{D}-M}\right) }
\end{equation}%
The second exponent cannot produce $\varepsilon ^{\mu \nu \alpha \beta }$ at 
$\mathcal{O}\left( p^{4}\right) $, so the anomalous effect must be lodged in
the Jacobian. Having the distinct goal of calculating the Jacobian in mind,
we introduce a continuous parameter dependence via the transformation $\xi
\rightarrow \xi _{\tau }=\exp (i\tau \vec{\lambda}\cdot \vec{\varphi}%
/2F_{\pi })\equiv \exp \left( i\tau \bar{\varphi}\right) $. Transforming
infinitesimally in $\tau $ induces a change $\delta \mathcal{J}$ in the
Jacobian$,$ 
\begin{eqnarray}
\psi &\rightarrow &\psi ^{\prime }=\xi _{\delta \tau }^{\dagger }\psi
_{L}+\xi _{\delta \tau }\psi _{R}\Rightarrow \int \left[ d\psi \right] [d%
\bar{\psi}]=\int [d\psi ^{\prime }][d\bar{\psi}^{\prime }]e^{\ln \delta 
\mathcal{J}} \\
&\Rightarrow &\delta \mathcal{J}=e^{-2i\delta \tau \text{\textrm{t}}\mathrm{%
\mathrm{r}}\bar{\varphi}\gamma _{5}}\Rightarrow \left. \frac{d\ln \mathcal{J}%
}{d\tau }\right| _{\tau =0}=-2i\text{\textrm{t}}\mathrm{r}\bar{\varphi}%
\gamma _{5}
\end{eqnarray}%
Due to the anomaly, the Jacobian is divergent and must be regularized.
Making use of Fujikawas method of removing high energy eigenmodes in a gauge
invariant way we take the limit%
\begin{equation}
\text{\textrm{t}}\mathrm{r}\bar{\varphi}\gamma _{5}=\underset{\varepsilon
\rightarrow 0}{\lim }\text{\textrm{t}}\mathrm{r}\left( \bar{\varphi}\gamma
_{5}e^{-\varepsilon \mathcal{\NEG{D}}_{\tau }\mathcal{\NEG{D}}_{\tau
}}\right) \quad ;\mathcal{\NEG{D}}_{\tau }\equiv \partial ^{\mu }+i\bar{V}%
_{\tau }^{\mu }+i\bar{A}_{\tau }^{\mu }\gamma _{5}
\end{equation}%
From the definition of vector and axial-vector currents (\ref{curdef}) with $%
\xi \rightarrow \xi _{\tau }$, we have that 
\begin{eqnarray}
\mathcal{\NEG{D}}_{\tau }\mathcal{\NEG{D}}_{\tau }=d_{\mu }d^{\mu }+\sigma
\quad &;&d_{\mu }=\partial _{\mu }+i\bar{V}_{\tau \mu }+\sigma _{\mu \nu }%
\bar{A}_{\tau }^{\nu }\gamma _{5}=\partial _{\mu }+\Gamma _{\tau \mu } 
\notag \\
&;&\sigma =-2\bar{A}_{\tau \mu }\bar{A}_{\tau }^{\mu }+i\left[ \partial
_{\mu }+i\bar{V}_{\tau \mu },\bar{A}_{\tau }^{\mu }\right] \gamma _{5}
\end{eqnarray}%
We now make use of the heat kernel expansion from thermodynamics and take
the limit $\varepsilon \rightarrow 0,$ to obtain an expression for the
regulated anomalous action $\Gamma \left( \bar{\varphi}\right) .$%
\begin{eqnarray}
\text{\textrm{t}}\mathrm{r}\bar{\varphi}\gamma _{5} &=&i\int d^{4}x\mathrm{Tr%
}\left( \frac{\bar{\varphi}\gamma _{5}}{\left( 4\pi \varepsilon \right) ^{2}}%
\sum_{n}\varepsilon ^{n}a_{n}\right) \underset{\varepsilon \rightarrow 0}{%
\longrightarrow }\frac{i}{16\pi ^{2}}\int d^{4}x\mathrm{Tr}\left( \bar{%
\varphi}\gamma _{5}a_{2}\right) \Rightarrow \\
\Gamma \left( \bar{\varphi}\right) &=&-i\ln \mathcal{J}+\text{...}=\frac{%
N_{c}}{4\pi ^{2}}\int_{0}^{1}d\tau \int d^{4}x\mathrm{Tr}\left( \tfrac{8}{3}%
\bar{\varphi}\varepsilon _{\mu \nu \alpha \beta }\bar{A}_{\tau }^{\mu }\bar{A%
}_{\tau }^{\nu }\bar{A}_{\tau }^{\alpha }\bar{A}_{\tau }^{\beta }\right) +%
\text{ }...  \label{wzim}
\end{eqnarray}%
where $N_{c}$ comes from the color sum and the ellipses signify terms
without correct symmetry $\left( \varepsilon _{\mu \nu \alpha \beta }\right)
.$ The only way to integrate this expression in a closed form, is to Taylor
expand each axial-vector current around $\tau =0$, and then integrate to
obtain a series of local Lagrangians. What Witten did was to treat the $\tau 
$ as a time-like fifth dimension $x_{5},$ with $\tau =1$ corresponding to
normal space-time. By expansion one can prove that the final the result
depends only on normal space-time. To describe processes other than direct
meson interaction, we must include external electroweak gauge fields via $%
\ell _{\mu }$ and $r_{\mu }$. This will alter the covariant derivative,
which in turn affects the Jacobian. The result is a gauge invariant
tree-level action describing the effect of the anomaly at $\mathcal{O}\left(
p^{4}\right) $ in the chiral expansion.

The WZW action can be written on a form more suitable for calculations, by
performing the $\tau $ integration where possible and changing variables 
\cite{gasleut}.%
\begin{eqnarray}
S[U,\ell ,r]_{WZW} &=&-\frac{iN_{c}}{240\pi ^{2}}\int d\sigma
^{ijklm}\left\langle \Sigma _{i}^{L}\Sigma _{j}^{L}\Sigma _{k}^{L}\Sigma
_{l}^{L}\Sigma _{m}^{L}\right\rangle  \notag \\
&&-\frac{iN_{c}}{48\pi ^{2}}\int d^{4}x\varepsilon _{\mu \nu \alpha \beta
}\left( W\left( U,\ell ,r\right) ^{\mu \nu \alpha \beta }-W\left( \mathbf{1}%
,\ell ,r\right) \right)  \label{wzwa}
\end{eqnarray}%
where

\begin{eqnarray}
W\left( U,\ell ,r\right) _{\mu \nu \alpha \beta } &=&\langle U\ell _{\mu
}\ell _{\nu }\ell _{\alpha }U^{\dagger }r_{\beta }+\tfrac{1}{4}U\ell _{\mu
}U^{\dagger }r_{\nu }U\ell _{\alpha }U^{\dagger }r_{\beta }  \notag \\
&&+iU\partial _{\mu }\ell _{\nu }\ell _{\alpha }U^{\dagger }r_{\beta
}+i\partial _{\mu }r_{\nu }U\ell _{\alpha }U^{\dagger }r_{\beta }-i\Sigma
_{\mu }^{L}\ell _{\nu }U^{\dagger }r_{\alpha }U\ell _{\beta }  \notag \\
&&+\Sigma _{\mu }^{L}U^{\dagger }\partial _{\nu }r_{\alpha }U\ell _{\beta
}-\Sigma _{\mu }^{L}\Sigma _{\nu }^{L}U^{\dagger }r_{\alpha }U\ell _{\beta
}+\Sigma _{\mu }^{L}\ell _{\nu }\partial _{\alpha }\ell _{\beta }+\Sigma
_{\mu }^{L}\partial _{\nu }\ell _{\alpha }\ell _{\beta }  \notag \\
&&-i\Sigma _{\mu }^{L}\ell _{\nu }\ell _{\alpha }\ell _{\beta }+\tfrac{1}{2}%
\Sigma _{\mu }^{L}\ell _{\nu }\Sigma _{\alpha }^{L}\ell _{\beta }-i\Sigma
_{\mu }^{L}\Sigma _{\nu }^{L}\Sigma _{\alpha }^{L}\ell _{\beta }\rangle
-\left( L\leftrightarrow R\right)
\end{eqnarray}%
where%
\begin{equation}
\left\{ 
\begin{array}{l}
\Sigma _{\mu }^{L}=U^{\dagger }\partial _{\mu }U\smallskip \\ 
\Sigma _{\nu }^{R}=U\partial _{\mu }U^{\dagger }%
\end{array}%
\right.
\end{equation}%
and%
\begin{equation}
U=\exp \left[ i\frac{\sqrt{2}}{F}\left( 
\begin{array}{ccc}
\frac{\pi ^{0}}{\sqrt{2}}+\frac{\eta }{\sqrt{6}} & \pi ^{+} & K^{+}\smallskip
\\ 
\pi ^{-} & -\frac{\pi ^{0}}{\sqrt{2}}+\frac{\eta }{\sqrt{6}} & 
K^{0}\smallskip \\ 
K^{-} & \bar{K}^{0} & -\frac{2}{\sqrt{6}}\eta%
\end{array}%
\right) \right]
\end{equation}%
$L\leftrightarrow R$ stands for%
\begin{equation}
\left\{ 
\begin{array}{l}
U\leftrightarrow U^{\dagger } \\ 
\ell _{\mu }\leftrightarrow r_{\mu } \\ 
\Sigma _{\mu }^{L}\leftrightarrow \Sigma _{\mu }^{R}%
\end{array}%
\right.
\end{equation}

\subsection{Example of Calculation}

As an example of how to use (\ref{wzwa}) we calculate the amplitude for $\pi
^{0}\rightarrow \gamma e^{+}e^{-}.$ The five dimensional term does not
contribute as it contains too many fields. The second term in the normal
space-time integral is just there for mathematical consistency, which leaves
us with $W_{\mu \nu \alpha \beta }.$ It's easy to see that the $%
L\leftrightarrow R$ operation leads to the same result (since $\ell _{\mu
}=r_{\mu }).$ So we can calculate just the $L$ terms and then multiply by 2.
Below we have used partial integration and the antisymmetry of $\varepsilon
^{\mu \nu \alpha \beta }.$

\begin{eqnarray}
W_{\mu \nu \alpha \beta } &=&2\left\langle \Sigma _{\mu }^{L}U^{\dagger
}\partial _{\nu }r_{\alpha }U\ell _{\beta }+\Sigma _{\mu }^{L}\ell _{\nu
}\partial _{\alpha }\ell _{\beta }+\Sigma _{\mu }^{L}\partial _{\nu }\ell
_{\alpha }\ell _{\beta }\right\rangle  \notag \\
&=&i\frac{2\sqrt{2}}{F}e^{2}\left\langle \partial _{\mu }M\partial _{\nu
}A_{\alpha }A_{\beta }Q^{2}+\partial _{\mu }MA_{\nu }\partial _{\alpha
}A_{\beta }Q^{2}+\partial _{\mu }M\partial _{\nu }A_{\alpha }A_{\beta
}Q^{2}\right\rangle  \notag \\
&=&-i\frac{e^{2}}{F}6\sqrt{2}\left\langle MQ^{2}\right\rangle \partial _{\mu
}A_{\nu }\partial _{\alpha }A_{\beta }=-i2\frac{e^{2}}{F}\partial _{\mu
}A_{\nu }\partial _{\alpha }A_{\beta }\Rightarrow  \notag \\
A\left( \pi ^{0}\rightarrow \gamma \gamma ^{\star }\right) &=&i\frac{1}{4\pi
^{2}}\frac{e}{F}\varepsilon ^{\mu \nu \alpha \beta }k_{\mu }\varepsilon
_{\nu }k_{\alpha }^{\star }\varepsilon _{\beta }^{\star }  \label{mel2}
\end{eqnarray}%
where the $\pi ^{0}\rightarrow \gamma e^{+}e^{-}$ amplitude can easily be
obtained by applying the Feynman rules for QED. $\varepsilon $ is the photon
polarization vector and $k$ the photon 4-momentum. The $\pi ^{0}\rightarrow
\gamma \gamma $ decay is historically important as it paved the way for the
theory of the anomalous sector. This is where Steinberger \cite{steinberger}
in 1949 first observed the effects of the anomaly.

In contrast with the effective Lagrangians, the only parameter appearing in
the WZW action is $N_{c}\footnote{%
One can also introduce the number of fermions $N_{f},$ which we here set to
3.}.$ This is because it is a prediction of the QCD anomaly structure. In
accordance with Adler and Bardeen \cite{adbardeen} there are no radiative
corrections. One can show that $N_{c}$ is an integer by making use of the
fact that the five-dimensional integral can only depend on normal
four-dimensional space-time.

Squaring the matrix element, summing over photon polarizations and inserting
one half for identical particles gives%
\begin{eqnarray*}
\left| A_{WZW}^{\pi ^{0}\gamma \gamma }\right| ^{2} &=&\frac{N_{c}^{2}\alpha
^{2}}{9\pi ^{2}F_{\pi }^{2}}\tfrac{1}{2}\sum_{Pol}\varepsilon ^{\mu \nu
\alpha \beta }\varepsilon _{\mu }k_{\nu }\varepsilon _{\alpha }^{\star
}k_{\beta }^{\star }\varepsilon _{\mu ^{\prime }\nu ^{\prime }\alpha
^{\prime }\beta ^{\prime }}\varepsilon ^{\mu ^{\prime }}k^{\nu ^{\prime
}}\varepsilon ^{\alpha ^{\prime }\star }k^{\beta ^{\prime }\star }\quad
;\left\{ 
\begin{array}{l}
\varepsilon /\varepsilon ^{\star }=\left( 0,1,0,0\right) \\ 
\varepsilon ^{\star }/\varepsilon =\left( 0,0,1,0\right) \\ 
k/k^{\star }=\tfrac{m_{\pi }}{2}\left( 1,0,0,\pm 1\right)%
\end{array}%
\right. \\
\sum_{Pol}\cdot \cdot \cdot &=&-2\times \left( 2013\right) \left(
2310\right) -2\times \left( 1023\right) \left( 1320\right) \quad ;\left(
2013\right) =\varepsilon _{2}k_{0}\varepsilon _{1}^{\star }k_{3}^{\star
}\,\,etc. \\
&{}&+\left( 2013\right) ^{2}+\left( 2310\right) ^{2}+\left( 1023\right)
^{2}+\left( 1320\right) ^{2}=\tfrac{1}{2}m_{\pi }^{4}\Rightarrow \\
\Gamma _{\gamma \gamma } &=&\frac{1}{16\pi }\frac{\left| A_{WZW}\right| ^{2}%
}{m_{\pi }}=\frac{1}{16\pi m_{\pi }}\frac{N_{c}^{2}\alpha ^{2}}{9\pi
^{2}F_{\pi }^{2}}\tfrac{1}{2}\tfrac{1}{2}m_{\pi }^{4}\underset{N_{c}=3}{%
\longrightarrow }\frac{\alpha ^{2}}{64\pi ^{3}F_{\pi }^{2}}m_{\pi
}^{3}\simeq 7.73eV
\end{eqnarray*}%
In excellent agreement with the experimental value $\Gamma _{\gamma \gamma
}=7.7\pm 0.5\pm 0.5eV$ \cite{pdg1}. This is an important test for the number
of colors as well as the anomaly structure and chiral symmetries.

In section 10 we follow through with the $\mathcal{O}\left( p^{6}\right) $
contribution to $\pi ^{0}\rightarrow \gamma \gamma ^{\star }$, allowing us
to solve for the chiral coefficients.

\section{Loops and Renormalization}

The anomalous sector is subject to meson loop corrections. QFT loop
corrections to the Born amplitude lead to ultra-violet divergences in the
form of polynomials in masses or external momenta. All non-analytical
divergences must cancel. The UV\ divergences are removed by introducing a
finite number of counterterms at a given order. Poles from one loops in
dimensional regularization can be obtained by considering quantum
fluctuations around the classical solution of the EOM.

\subsection{Power Counting}

In this section we consider the order at which a general Feynman diagram
contributes. The distinction should be clear from section 4.2 where we are
asking which tree-level terms will contribute at a given order. Here we
follow a purely diagrammatic approach (as in \cite{dynsm}) to arrive at
Weinberg's power counting theorem \cite{weinberg}. We are specifically
interested in the order at which loop integrals contribute.

Consider a general diagram with $N_{V}$ vertices and $N_{n}$ vertices from $%
\mathcal{L}_{n},$ so that $N_{V}=\sum_{n}N_{n}.$ The dimensionality of the
couplings is $M^{N_{C}},$ where $N_{C}=\sum_{n}N_{n}\left( 4-n\right) $ and $%
M$ is the characteristic mass scale. For $N_{I}$ internal lines and $N_{E}$
external lines we get $M^{2N_{I}-N_{E}}.$ There is a relation between
internal lines, vertices and loops;%
\begin{equation}
N_{I}=N_{L}+N_{V}-1=N_{L}+\sum_{n}N_{n}-1
\end{equation}%
Remaining factors must composed of a power of energy times the logarithm of
the dimensionless $E^{2}/\mu ^{2}$ (where $\mu $ is the renormalization
scale)$.$ Putting all this together, we get the matrix element energy
dimensionality%
\begin{equation}
\mathcal{M}\sim \frac{M^{\Sigma _{n}N_{n}\left( 4-n\right) }}{%
M^{N_{E}+2N_{L}+2\Sigma _{n}N_{n}-2}}E^{D}F\left( E/\mu \right) \Rightarrow
D=2+\sum_{n}N_{n}\left( n-2\right) +2N_{L}
\end{equation}%
So that a diagram with $N_{L}$ loops contributes at $E^{2N_{L}}$ higher than
the tree-level used in the calculation. This simplifies calculations
considerably, since at low energies, only a few loops need to be taken into
account. Loop divergences are handled in the usual way, and can be removed
by renormalizing the parameters of the theory. The general effective
Lagrangian, compatible with the symmetry conditions, must have enough
parameters to absorb the divergences.

The lowest order action for anomalous processes is already $\mathcal{O}%
\left( p^{4}\right) $ (the WZW action). At $\mathcal{O}\left( p^{4}\right) $
we have the power counting: tree level diagrams from $S_{WZW}$ with one
vertex, and the rest from $\mathcal{L}_{2}$. In this thesis we will proceed
up to the $\mathcal{O}\left( p^{6}\right) $ anomalous action. At $\mathcal{O}%
\left( p^{6}\right) $ we have: tree level diagrams with one vertex from $%
S_{WZW},$ one from $\mathcal{L}_{4}$ and the rest from $\mathcal{L}_{2},$
tree level diagrams with one vertex from the anomalous $\mathcal{O}\left(
p^{6}\right) $ action and the rest from $\mathcal{L}_{2},$ one-loop diagrams
with one vertex from $S_{WZW}$ and the rest from $\mathcal{L}_{2}.$

\subsection{Infinite parts}

Starting from a chiral-invariant effective Lagrangian, the divergences from
the loop integrals are constrained by chiral symmetry. In the previous
section we saw that one-loops contribute at $\mathcal{O}\left( p^{2}\right) $
higher than the order of the Lagrangian from which they where calculated.
The anomalous power counting laws tell us that chiral invariant counterterms
in the anomalous Lagrangian of $\mathcal{O}\left( p^{6}\right) $ will absorb
the one-loop divergences, leaving behind the renormalized coefficients. To
calculate the divergent parts of the $\mathcal{L}_{2}$ one-loops, we can
expand around the classical solution to the EOMs of our chiral theory, i.e.
the GS boson matrix. 
\begin{eqnarray*}
\frac{\delta S_{2}}{\delta U} &=&0\Rightarrow \bar{U}=e^{i\frac{\sqrt{2}}{%
F_{\pi }}M} \\
U &=&e^{i\frac{\sqrt{2}}{F}\left( M+\xi ^{\prime }\right) }=e^{i\frac{M}{%
\sqrt{2}F}}\,e^{i\xi }e^{i\frac{M}{\sqrt{2}F}}=u\,e^{i\xi }u
\end{eqnarray*}%
where $\xi $ is an Hermitian matrix that preserves unitarity. The one-loop
divergences are obtained by expanding $\mathcal{L}_{2}$ (eq. (\ref{L2})) to $%
\mathcal{O}\left( \xi ^{2}\right) :$%
\begin{equation}
\int d^{4}x\mathcal{L}_{2}=\int d^{4}x\mathcal{\bar{L}}_{2}+\tfrac{1}{2}\int
d^{4}x\xi ^{i}\Delta _{ij}\xi ^{j}+\mathcal{O}\left( \xi ^{3}\right)
\label{l2ex}
\end{equation}%
The divergent parts of the new term in (\ref{l2ex}) can be pinpointed by
calculating the second variation of $S_{2}$ and identifying the relevant
operators. The power counting of section 9.1 told us that the WZW action can
contribute a vertex to one-loop diagrams. Including the WZW action, the
equations of motion and the variation of second order in $\xi $ now receives
new contributions, and the corresponding anomalous quantities are calculated
through the second variation of $S_{WZ}.$ The calculation of divergent terms
is very technical and the reader is referred to \cite{u1v} for a more
in-depth analysis.

\subsection{Meson One-Loop Corrections}

We begin by considering the $PS\rightarrow \gamma \gamma ^{\star }$ decays.
If both photons are on-shell there are no infinite parts coming from the
meson one-loops. This means that we should find no counterterms in the
chiral coefficients describing the decay. If however, the photon goes to an $%
e^{+}e^{-}$ pair, coefficients containing parts designated to cancel the
infinities will appear. The meson one-loop corrections were calculated by
Bijnens et al. \cite{pggloop} via the Feynman diagrams in figure 3. To fully
describe situation we must include the results from the wavefunction
renormalization and the decay constant corrections. For the semileptonic $%
\pi ^{+}/K^{+}\rightarrow \gamma e^{+}\nu $ decays, the one-loops
corrections have been calculated in \cite{semileploop}. These have Feynman
diagrams similar to those in figure 3.

We are also considering the experimentally well charted
3-pseudoscalar-photon interactions $\eta \pi \pi \gamma $ \& $\pi \pi \pi
\gamma $. The one-loop corrections were calculated in \cite{vtxloop} via the
Feynman diagrams in figure 4. Only on-shell photons are considered here. For 
$\pi \pi \pi \gamma $, $k_{\gamma }^{2}$ is very small $\left( \ll m_{\pi
}^{2}\right) $, and for $\eta \pi \pi \gamma $ $k_{\gamma }^{2}=0$ unless $%
e^{+}e^{-}\rightarrow \gamma ^{\star }\rightarrow 3\pi ,$ which we do not
consider. The decays $K^{+}\rightarrow \pi ^{+}\pi ^{-}e^{+}\nu $ and $%
K^{0}\rightarrow \pi ^{0}\pi ^{-}e^{+}\nu $ are subject to similar loop
corrections (as calculated in \cite{semileploop}).

In the loop corrections (listed in appendix A) we find the so-called chiral
logarithms of the form $m_{PS}^{2}\log m_{PS}^{2}/\mu ^{2},$ where $\mu
=m_{\rho }$ is taken as the arbitrary renormalization scale. We also find
the divergent terms proportional to $\lambda $ (see eq. (\ref{ms})), which
will be exactly canceled by the counterterms at $\mathcal{O}\left(
p^{6}\right) .$ These are either proportional to pseudoscalar masses or
particle four-momenta. The latter can be seen to vanish in the soft (low
energy) limit. Most loop corrections were calculated using $m_{u}=m_{d}\neq
m_{s},$ and the Gell-Mann-Okubo relation can be used to eliminate the $\eta $
mass.

\begin{figure}
\begin{center}
\begin{picture}(240,90)(0,0)
\SetScale{0.85}
\SetOffset(20,0)
\Line(15,50)(60,50)
\Vertex(60,50){1.5}
\Oval(59.5,65)(15,7.5)(0)
\Photon(60,50)(105,60){2}{5}
\Photon(60,50)(100,20){-2}{5}
\Text(88,20)[tl]{$\gamma$}
\Text(91,46)[bl]{$\gamma^{\star }$}
\Line(125,50)(170,50)
\Vertex(170,50){1.5}
\Oval(181,61)(15,7.5)(-48)
\Vertex(192,71){1.5}
\Photon(192,71)(213,88){2}{2.5}
\Photon(170,50)(210,20){-2}{5}
\Text(182,20)[tl]{$\gamma$}
\Text(183,73)[bl]{$\gamma^{\star }$}
\end{picture}
\end{center}
\caption{
 \emph{One-loop diagrams for PS} $\rightarrow \gamma \gamma
^{\star }\medskip $}
\end{figure}
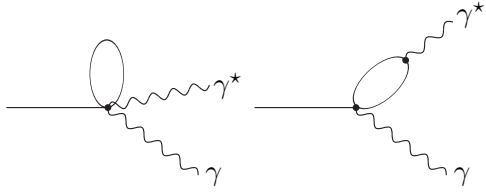

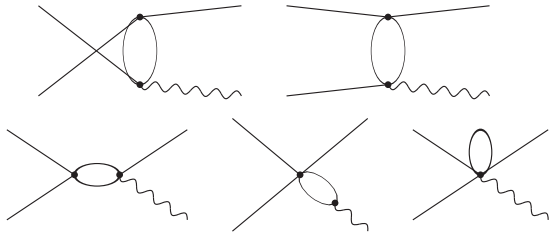
\begin{figure}
\begin{center}
$%
\begin{picture}(300,100)(0,90)
\SetOffset(50,15)
\SetScale{0.85}
\Line(15,190)(60,155)
\Oval(60,170)(15,7.5)(0)
\Vertex(60,155){1.5}
\Vertex(60,185){1.5}
\Line(15,150)(60,185)
\Photon(60,155)(105,150){-2}{5}
\Line(60,185)(105,190)

\Line(125,190)(170,185)
\Vertex(170,155){1.5}
\Vertex(170,185){1.5}
\Oval(170,170)(15,7.5)(0)
\Line(125,150)(170,155)
\Photon(170,155)(215,150){-2}{5}
\Line(170,185)(215,190)

\Line(1,135)(31,115)
\Line(1,95)(31,115)
\Vertex(31,115){1.5}
\Oval(41,115)(5,10)(0)
\Vertex(51,115){1.5}
\Line(51,115)(81,135)
\Photon(51,115)(81,95){-2}{4}

\Line(101,140)(131,115)
\Line(101,90)(161,140)
\Vertex(131,115){1.5}
\Oval(139,109)(10,5)(-128)
\Photon(146.5,102.5)(161,90){-2}{2}
\Vertex(146.5,102.5){1.5}

\Line(181,135)(211,115)
\Line(181,95)(211,115)
\Vertex(211,115){1.5}
\Oval(211,125)(5,10)(90)
\Line(211,115)(241,135)
\Photon(211,115)(241,95){-2}{4}

\end{picture}\medskip $
\end{center}
\caption{
 \emph{One-loop diagrams for} $\gamma \pi ^{0}\rightarrow \pi
^{+}\pi ^{-},\eta \rightarrow \pi ^{+}\pi ^{-}\gamma $, $K^{+}\rightarrow
\pi ^{+}\pi ^{-}e^{+}\nu $ \emph{\&} $K^{0}\rightarrow \pi ^{0}\pi
^{0}e^{+}\nu $}
\end{figure}

\section{Anomalous Lagrangian of $\mathcal{O}\left( p^{6}\right) $}

The construction of the anomalous $\mathcal{O}\left( p^{6}\right) $ is a
formidable task and was first completed for an arbitrary number of flavors ($%
N_{f}$) in 2001 by Bijnens et al. \cite{bijp6}. The only symmetries needed
to construct the desired EQFT, are those of the initial and broken subgroup.
One starts by listing all operators of a given order in the power expansion,
compliant with all continuous and discrete symmetries. If we want to keep
track of the phenomenologically relevant terms, it is imperative to reduce
the list of operators to a minimum using all available constraints and
identities. The regularizing terms that cancel the infinite parts of the
ultra-violet loop graphs contributing at $\mathcal{O}\left( p^{6}\right) $
must also be calculated.

\subsection{Construction of the Effective Action}

The following building blocks can be used in constructing monomials of $%
\mathcal{O}\left( p^{6}\right) $ in the chiral expansion: $u_{\mu },$ $%
h_{\mu \nu }\equiv \nabla _{\mu }u_{\nu }+\nabla _{\nu }u_{\mu },$ $f_{\pm
\mu \nu },$ $\chi _{\pm }\equiv u^{\dagger }\chi u^{\dagger }\pm u\chi u.$
The motivation for using these is that they appear in the calculation of the
divergent parts. Also, the number of terms built of these can easily be
reduced to a minimum. QCD symmetry conditions of parity, charge conjugation
and hermicity must then be imposed on the list of operators. Since we are
dealing with the anomalous sector the antisymmetric $\varepsilon ^{\mu \nu
\alpha \beta }$ enters, imposing further conditions.

Use of partial integration, antisymmetry conditions, the mathematical
Bianchi \& Schouten identities and the EOMs reduces the number of monomials
to 24 for the general $N_{f}$ case. In the three flavor case we use $%
SU\left( 3\right) $ matrices obeying the Cayley-Hamilton relation, yielding
an equation allowing for the removal of one more monomial. The Lagrangian
density for $N_{f}=3$ is then given by%
\begin{equation}
\mathcal{L}_{6}^{W}=\sum_{i=1}^{23}C_{i}^{W}O_{i}^{W}
\end{equation}%
where the monomials $O_{i}^{W}$ are listed in table 1. The infinities are
removed using the $\overline{MS}$ dimensional regularization scheme
subtracting both infinite and omnipresent terms:%
\begin{equation}
C_{i}^{W}=C_{i}^{Wr}+\eta _{i}\frac{\mu ^{d-4}}{32\pi ^{2}}\lambda \quad
;\lambda =\frac{2}{d-4}-\ln \left( 4\pi \right) +\gamma _{E}-1  \label{ms}
\end{equation}%
where $C_{i}^{Wr}$ are the renormalized coefficients, $\gamma _{E}$ is the
Euler constant, $\mu $ the arbitrary renormalization scale and $d$ is the
number of dimensions $\left( \rightarrow 4\right) $. The coefficients $\eta
_{i}$ can be deduced from table 1. Once the infinite parts have canceled, we
can solve for the renormalized coefficients $C_{i}^{Wr}$ by experimental
comparison. The triple $\left( \text{anti}\right) $commutators in table 1
are defined as%
\begin{eqnarray*}
\left[ a,b,c\right] &=&abc-cba \\
\{a,b,c\} &=&abc+cba
\end{eqnarray*}%
We put the non-abelian field strengths in the operator $f_{\pm }^{\mu \nu
}=u\,F_{L}^{\mu \nu }u^{\dagger }\pm u^{\dagger }F_{R}^{\mu \nu }u.$

\begin{table}
\begin{center}
$%
\begin{tabular}{lll|lll}
$i$ & Monomial $O_{i}^{W}$ & 384$\pi ^{2}F^{2}\eta _{i}$ & $i$ & Monomial $%
O_{i}^{W}$ & 384$\pi ^{2}F^{2}\eta _{i}$ \\ \hline
1 & $i\varepsilon ^{\mu \nu \alpha \beta }\left\langle \chi _{-}u_{\mu
}u_{\nu }u_{\alpha }u_{\beta }\right\rangle $ & \multicolumn{1}{c|}{12} & 13
& $i\varepsilon ^{\mu \nu \alpha \beta }\left\langle h_{\gamma \mu
}\{f_{+\gamma \nu },u_{\alpha }u_{\beta }\}\right\rangle $ & 
\multicolumn{1}{c}{-30} \\ 
2 & $\varepsilon ^{\mu \nu \alpha \beta }\left\langle \chi _{+}\left[
f_{-\mu \nu },u_{\alpha }u_{\beta }\right] \right\rangle $ & 
\multicolumn{1}{c|}{-7} & 14 & $i\varepsilon ^{\mu \nu \alpha \beta
}\left\langle h_{\gamma \mu }\left[ f_{+\nu \alpha },u_{\gamma },u_{\beta }%
\right] \right\rangle $ & \multicolumn{1}{c}{-9} \\ 
3 & $\varepsilon ^{\mu \nu \alpha \beta }\left\langle \chi _{+}u_{\mu
}\right\rangle \left\langle u_{\nu }f_{-\alpha \beta }\right\rangle $ & 
\multicolumn{1}{c|}{-6} & 15 & $i\varepsilon ^{\mu \nu \alpha \beta
}\left\langle h_{\gamma \mu }\left[ u_{\gamma },f_{+\nu \alpha },u_{\beta }%
\right] \right\rangle $ & \multicolumn{1}{c}{3} \\ 
4 & $\varepsilon ^{\mu \nu \alpha \beta }\left\langle \chi _{-}\{f_{+\mu \nu
},u_{\alpha }u_{\beta }\}\right\rangle $ & \multicolumn{1}{c|}{-6} & 16 & $%
\varepsilon ^{\mu \nu \alpha \beta }\left\langle f_{-\gamma \mu }\left[
u_{\gamma },u_{\nu }u_{\alpha }u_{\beta }\right] \right\rangle $ & 
\multicolumn{1}{c}{18} \\ 
5 & $\varepsilon ^{\mu \nu \alpha \beta }\left\langle \chi _{-}u_{\mu
}f_{+\nu \alpha }u_{\beta }\right\rangle $ & \multicolumn{1}{c|}{12} & 17 & $%
\varepsilon ^{\mu \nu \alpha \beta }\left\langle f_{-\mu \nu }\left[
u_{\gamma }u_{\gamma },u_{\alpha }u_{\beta }\right] \right\rangle $ & 
\multicolumn{1}{c}{15} \\ 
6 & $\varepsilon ^{\mu \nu \alpha \beta }\left\langle \chi _{-}\right\rangle
\left\langle f_{+\mu \nu }u_{\alpha }u_{\beta }\right\rangle $ & 
\multicolumn{1}{c|}{8} & 18 & $\varepsilon ^{\mu \nu \alpha \beta
}\left\langle f_{-\mu \nu }u_{\alpha }\right\rangle \left\langle u_{\gamma
}u_{\gamma }u_{\beta }\right\rangle $ & \multicolumn{1}{c}{18} \\ 
7 & $i\varepsilon ^{\mu \nu \alpha \beta }\left\langle \chi _{-}f_{+\mu \nu
}f_{+\alpha \beta }\right\rangle $ & \multicolumn{1}{c|}{0} & 19 & $%
i\varepsilon ^{\mu \nu \alpha \beta }\left\langle f_{+\gamma \mu
}\{f_{-\gamma \nu },u_{\alpha }u_{\beta }\}\right\rangle $ & 
\multicolumn{1}{c}{-12} \\ 
8 & $i\varepsilon ^{\mu \nu \alpha \beta }\left\langle \chi
_{-}\right\rangle \left\langle f_{+\mu \nu }f_{+\alpha \beta }\right\rangle $
& \multicolumn{1}{c|}{0} & 20 & $i\varepsilon ^{\mu \nu \alpha \beta
}\left\langle f_{+\gamma \mu }\left[ f_{-\nu \alpha },u_{\gamma },u_{\beta }%
\right] \right\rangle $ & \multicolumn{1}{c}{-3} \\ 
9 & $i\varepsilon ^{\mu \nu \alpha \beta }\left\langle \chi _{-}f_{-\mu \nu
}f_{-\alpha \beta }\right\rangle $ & \multicolumn{1}{c|}{0} & 21 & $%
i\varepsilon ^{\mu \nu \alpha \beta }\left\langle f_{+\gamma \mu }\left[
u_{\beta },f_{-\nu \alpha },u_{\gamma }\right] \right\rangle $ & 
\multicolumn{1}{c}{15} \\ 
10 & $i\varepsilon ^{\mu \nu \alpha \beta }\left\langle \chi
_{-}\right\rangle \left\langle f_{-\mu \nu }f_{-\alpha \beta }\right\rangle $
& \multicolumn{1}{c|}{0} & 22 & $\varepsilon ^{\mu \nu \alpha \beta
}\left\langle u_{\mu }\{\bigtriangledown _{\gamma }f_{+\gamma \nu
},f_{+\alpha \beta }\}\right\rangle $ & \multicolumn{1}{c}{12} \\ 
11 & $i\varepsilon ^{\mu \nu \alpha \beta }\left\langle \chi _{+}\left[
f_{+\mu \nu },f_{-\alpha \beta }\right] \right\rangle $ & 
\multicolumn{1}{c|}{-$\tfrac{5}{2}$} & 23 & $\varepsilon ^{\mu \nu \alpha
\beta }\left\langle u_{\mu }\{\bigtriangledown _{\gamma }f_{-\gamma \nu
},f_{-\alpha \beta }\}\right\rangle $ & \multicolumn{1}{c}{0} \\ 
12 & $\varepsilon ^{\mu \nu \alpha \beta }\left\langle h_{\gamma \mu }\left[
u_{\gamma },u_{\nu }u_{\alpha }u_{\beta }\right] \right\rangle $ & 
\multicolumn{1}{c|}{-6} &  &  & \multicolumn{1}{c}{}%
\end{tabular}%
\medskip $
\end{center}
\caption{
 $\mathcal{O}\left( p^{6}\right) $ \emph{monomials \&
renormalization coefficients}}
\end{table}

\subsection{Infinite parts}

Calculating the second variation of the $WZW$ $\&$ $S_{2}$ action ($\mathcal{%
O}\left( \xi ^{2}\right) $ terms), and proceeding with dimensional
regularization one obtains an expression for the divergent one-loop parts in
arbitrary number of flavors $\left( N_{f}\right) $ and colors $\left(
N_{c}\right) .$%
\begin{eqnarray}
Z_{1-loop}^{WZ\infty } &=&-\frac{1}{16\pi ^{2}\left( d-4\right) }\frac{%
N_{c}N_{f}}{1152\pi ^{2}F^{2}}\{4O_{1}^{W}+\left( -3+\frac{6}{N_{f}^{2}}%
\right) O_{2}^{W}-\frac{6}{N_{f}}O_{3}^{W}-2O_{4}^{W}  \notag \\
&&+4O_{5}^{W}+\frac{8}{N_{f}}O_{6}^{W}+\left( -\frac{5}{6}+\frac{6}{N_{f}^{2}%
}\right) O_{11}^{W}-2O_{12}^{W}-10O_{13}^{W}-3O_{14}^{W}+O_{15}^{W}  \notag
\\
&&+2O_{16}^{W}+O_{17}^{W}+\frac{6}{N_{f}}%
O_{18}^{W}-4O_{19}^{W}-O_{20}^{W}+5O_{21}^{W}+4O_{22}^{W}-\frac{6}{N_{f}}%
O_{24}^{W}\}
\end{eqnarray}%
where the $O_{i}^{W}$ are the monomials listed in table 1.

\subsection{Example of Calculation}

We continue with the $\pi ^{0}\rightarrow \gamma e^{+}e^{-}$ decay. The
fastest way to work yourself through the list is by counting the minimum
number of fields a monomial must contribute. Then many of them can quickly
be discarded. We are looking for terms that can accommodate one meson and
two photons. By rewriting operators in terms of vector $\left( v_{\mu
}\right) $ and axial vector $\left( a_{\mu }\right) $ sources, we can
quickly see which ones will contribute (since only $v_{\mu }$ contains the
EM field). The most useful tools in simplifying expressions is partial
integration and the anti-symmetry of $\varepsilon ^{\mu \nu \alpha \beta }.$ 
\begin{eqnarray*}
O_{7}^{W} &=&i\varepsilon ^{\mu \nu \alpha \beta }\left\langle \chi
_{-}f_{+\mu \nu }f_{+\alpha \beta }\right\rangle =64\sqrt{2}\frac{B_{0}e^{2}%
}{F}\varepsilon ^{\mu \nu \alpha \beta }\left\langle sMQ^{2}\right\rangle
\partial _{\mu }A_{\nu }\partial _{\alpha }A_{\beta } \\
&=&\frac{64}{9}\frac{B_{0}e^{2}}{F}\left( 4m_{u}-m_{d}\right) \varepsilon
^{\mu \nu \alpha \beta }\pi ^{0}\partial _{\mu }A_{\nu }\partial _{\alpha
}A_{\beta } \\
O_{8}^{W} &=&i\varepsilon ^{\mu \nu \alpha \beta }\left\langle \chi
_{-}\right\rangle \left\langle f_{+\mu \nu }f_{+\alpha \beta }\right\rangle
=64\sqrt{2}\frac{B_{0}e^{2}}{F}\varepsilon ^{\mu \nu \alpha \beta }\frac{1}{%
\sqrt{2}}\left( m_{u}-m_{d}\right) \pi ^{0}\left\langle Q^{2}\right\rangle
\partial _{\mu }A_{\nu }\partial _{\alpha }A_{\beta } \\
&=&\frac{128}{3}\frac{B_{0}e^{2}}{F}\left( m_{u}-m_{d}\right) \varepsilon
^{\mu \nu \alpha \beta }\pi ^{0}\partial _{\mu }A_{\nu }\partial _{\alpha
}A_{\beta }
\end{eqnarray*}%
$O_{22}^{W}$ and $O_{23}^{W}$ may contribute since the photon going to $%
e^{+}e^{-}$ is has $k_{\gamma ^{\star }}^{2}\neq 0.$ However,%
\begin{equation*}
f_{\pm }^{\mu \nu }\simeq F_{L}^{\mu \nu }\pm F_{R}^{\mu \nu }-\left[
F_{L}^{\mu \nu },m\right] \pm \left[ F_{R}^{\mu \nu },m\right] \equiv F_{\pm
}^{\mu \nu }-\frac{i}{\sqrt{2}F}\left[ F_{\mp }^{\mu \nu },M\right]
\end{equation*}%
to next-to-leading order in expanding the exponential, and $F_{+\mu \nu
}\simeq 4\partial _{\mu }v_{\nu }$, $F_{-\mu \nu }\simeq -4\partial _{\mu
}a_{\nu },$ so we can exclude $O_{23}^{W}$. For $O_{22}^{W}$ we get

\begin{eqnarray*}
O_{22}^{W} &=&\varepsilon ^{\mu \nu \alpha \beta }\left\langle u_{\mu
}\{\bigtriangledown _{\gamma },f_{+\gamma \nu }f_{+\alpha \beta
}\}\right\rangle \\
&=&-\frac{4\sqrt{2}}{F}\varepsilon ^{\mu \nu \alpha \beta }\langle \partial
_{\mu }M(\partial _{\gamma }F_{L\gamma \nu }F_{L\alpha \beta }+F_{L\alpha
\beta }\partial _{\gamma }F_{L\gamma \nu })\rangle \\
&=&16\sqrt{2}\frac{e^{2}}{F}\varepsilon ^{\mu \nu \alpha \beta }\langle
MQ^{2}\rangle \partial _{\gamma }^{2}\partial _{\mu }A_{\nu }\partial
_{\alpha }A_{\beta } \\
&=&\frac{16}{3}\frac{e^{2}}{F}\varepsilon ^{\mu \nu \alpha \beta }\pi
^{0}\partial _{\gamma }^{2}\partial _{\mu }A_{\nu }\partial _{\alpha
}A_{\beta }
\end{eqnarray*}%
Adding up the contributing monomials, we get the amplitude%
\begin{eqnarray}
A\left( \pi ^{0}\gamma e^{+}e^{-}\right) &=&\frac{16}{9}\frac{e^{3}}{F}%
\varepsilon ^{\mu \nu \alpha \beta }k_{\mu }\varepsilon _{\nu }k_{\alpha
}^{\star }\frac{\bar{e}\gamma _{\beta }e}{k_{\alpha }^{\star 2}}\times 
\notag \\
&&\left( 8B_{0}\left( 4m_{u}-m_{d}\right) C_{7}^{W}+48B_{0}\left(
m_{u}-m_{d}\right) C_{8}^{W}-3k^{\star 2}C_{22}^{W}\right)  \label{op6gee}
\end{eqnarray}%
The relations in section 4.3 can be used to rewrite the quark masses in
terms of the pseudoscalar masses.

\section{Models for the Chiral Coefficients}

\subsection{Vector Meson Dominance (VMD)}

VMD is a phenomenologically very successful model that has not yet been
derived from the standard model. It is based on the fact that higher mass
resonances always enter the theory through virtual effects. VMD makes a
dramatic appearance in the pion form factor, where the Breit-Wigner shape of
the $\rho $ meson can be seen.

The exchange of higher mass resonances can be used to estimate the finite
part of the chiral coefficients. Roughly speaking, VMD states that for most
processes, the main dynamic effect below the chiral breaking scale comes
from the exchange of a vector meson. Using the hidden symmetry formulation
by Bando et al. \cite{bando}, and extending the formalism to include the
anomalous sector J. Bijnens \cite{u1v} has constructed an anomalous vector
meson Lagrangian. Vector meson exchange can be described by including the
ideally mixed $\rho $-nonet%
\begin{equation}
\rho _{\mu }=\frac{1}{\sqrt{2}}\rho _{\mu }^{a}\lambda ^{a}+\frac{1}{\sqrt{3}%
}\rho _{\mu }^{1}
\end{equation}%
in a chiral invariant way. Operator strings with the correct symmetry
properties can be constructed using the covariant and hermitian building
blocks $\rho _{\mu \nu }=\partial _{\mu }\rho _{\nu }-\partial _{\nu }\rho
_{\mu }+ig[\rho _{\mu },\rho _{\nu }],$ $\xi ^{\dagger }\ell _{\mu \nu }\xi $
$,$ $\xi r_{\mu \nu }\xi ,$ where $\xi =\exp iM/F\sqrt{2}$. The coupling $g$
can be determined from the $\rho \rightarrow \pi \pi $ width.

The heavy $\rho $-meson nonet, consisting of $\rho ^{\pm ,0},K^{\star \pm
,0},\bar{K}^{\star 0},\omega $ \& $\phi $, will only enter virtually and can
be integrated out, assuming that their masses are much greater than the
momenta. The resulting expression will contain parameters that can be
constrained by comparing with the radiative decay widths of vector mesons.
We will not list the full anomalous vector Lagrangian here - the reader is
referred to \cite{u1v} for further details. As an example, coefficients can
be chosen such that one obtains a Lagrangian describing direct $\rho \rho M$
interaction (where $M$ stands for a pseudoscalar meson). 
\begin{equation}
\mathcal{L}\left( \rho _{\mu }\rightarrow \rho _{\mu }M\right) =\frac{3}{%
4\pi ^{2}}\frac{g^{2}}{\sqrt{2}F}\varepsilon ^{\mu \nu \alpha \beta }\text{%
\textrm{t}}\mathrm{\mathrm{r}}\left( \partial _{\mu }\rho _{\nu }\partial
_{\alpha }\rho _{\beta }M\right)
\end{equation}%
This Lagrangian can be used to calculate the $M\rightarrow \gamma \gamma
^{\star }$ amplitudes via the $\rho \rho $ resonance: $M\rightarrow \rho
\rho \rightarrow \gamma \gamma ^{\star }$. To obtain the connection with $%
\ell _{\mu }$ and $r_{\mu }$ one must first integrate out the $\rho $
resonance. It should be noted that it is also possible to calculate the $%
\rho \rho M$ amplitude using an ordinary Feynman diagram approach.
Integrating out the vector mesons one then obtains the effective Lagrangian
for $M\rightarrow \gamma \gamma ^{\star }$ processes:%
\begin{eqnarray*}
\mathcal{L}_{6}^{M\rightarrow \gamma \gamma ^{\star }} &=&...-i\frac{e^{2}}{%
4M_{\rho }^{2}}\frac{3}{8\pi ^{2}}\varepsilon ^{\mu \nu \alpha \beta
}F_{\alpha \beta }\partial ^{\lambda }F_{\lambda \nu } \\
&&\times \left\langle Q^{2}\Sigma ^{\dagger }\partial _{\mu }\Sigma
-Q^{2}\Sigma \partial _{\mu }\Sigma ^{\dagger }+Q\Sigma ^{\dagger }Q\partial
_{\mu }\Sigma -Q\Sigma Q\partial _{\mu }\Sigma ^{\dagger }\right\rangle
\end{eqnarray*}%
where $F_{\mu \nu }$ is the EM field strength tensor, $Q=\tfrac{1}{3}%
diag(2,-1,-1)$ and $\Sigma =\exp (i\sqrt{2}M/F)$. Similar $\mathcal{O}\left(
p^{6}\right) $ VMD expressions can be obtained for the other anomalous
processes.

\subsection{Chiral Constituent Quark Model (CQM)}

The constituent quark model is based on the assumption that vertices has to
come from constituent quark loops. This effect is merged with the chiral
formalism by including the constituent quarks in a chiral invariant way.
Here follows a brief introduction to the model, for a more in-depth and
technical analysis the reader is referred to Bijnens \cite{u1v} or Ball \cite%
{Ball}. The euclidean space Lagrangian density can be written 
\begin{equation}
\mathcal{L}=\mathcal{L}_{QCD}+\mathcal{L}_{M}=\bar{q}\mathcal{D}q
\label{cqmlag}
\end{equation}%
where 
\begin{eqnarray}
\mathcal{D}=\gamma _{\mu }D_{\mu }+\mathcal{M}\quad &;&D_{\mu }=\partial
_{\mu }+ig_{S}G_{\mu }+i\ell _{\mu }P_{L}+ir_{\mu }P_{R}  \label{cqm} \\
&;&\mathcal{M}=-m_{Q}\left( U^{\dagger }P_{L}+UP_{R}\right)  \notag
\end{eqnarray}%
where $m_{Q}$ is the constituent quark mass. We neglect gluonic corrections
and will consequently drop $G_{\mu }$ in (\ref{cqm}). This is a large
distance QCD approximation that will reproduce the anomaly correctly and is
expected to work reasonably well. A connection with the effective approach
in terms of the GS bosons can be obtained by integrating out the quarks. 
\begin{eqnarray}
e^{\Gamma \left( U,\ell _{\mu },r_{\mu }\right) } &=&\int [d\bar{q}%
][dq]e^{\int d^{4}x\mathcal{L}_{QCD}}\Rightarrow \\
\Gamma &=&\log \det \mathcal{D}=\text{\textrm{t}}\mathrm{\mathrm{r}}\log 
\mathcal{D}  \label{effac}
\end{eqnarray}%
and similarly for $\Gamma ^{\star }$ replacing $\mathcal{D}$ with $\mathcal{%
D^{\dagger }}$. The determinant corresponds to a trace over color, flavor,
Dirac indices and space-time. With $\Gamma ^{\star }$ and $\Gamma $ we can
obtain a real part $\Gamma ^{+}$ and imaginary part $\Gamma ^{-}.$ It should
be emphasized that this is only possible since we're working in euclidean
space. In the physical Minkowski space-time imaginary parts are not allowed.
Since we are ultimately interested in the anomalous sector, we evaluate the
effect of the intrinsic parity operation, and find that $\Gamma ^{\pm
}\rightarrow \pm \Gamma ^{\pm }.$ This means that $\Gamma ^{+}$ contains an
even and $\Gamma ^{-}$ an odd number of $\varepsilon ^{\mu \nu \alpha \beta
}.$ In order to preserve the ordinary and break the intrinsic parity
symmetry we know that we must have an odd power of $\varepsilon ^{\mu \nu
\alpha \beta }.$ (\ref{effac}) gives%
\begin{equation}
\Gamma ^{-}=-\log \det \mathcal{D^{\dagger }}  \label{anpart}
\end{equation}%
This can be manipulated to obtain the correct expression describing the
anomalous sector. The process involves rewriting (\ref{anpart}) on a
five-dimensional integral with four-dimensional space-time boundaries,
singling out $\varepsilon ^{\mu \nu \alpha \beta }$-terms and expanding in $%
m_{Q}.$ All but the WZW terms will be suppressed by the constituent quark
mass. In order to perform the integration over the extra time-like dimension
(c.f. $\tau $ in WZW section), one can then make a Seeley-DeWitt expansion,
singling out the contributing terms with SDW coefficients $a_{i}$.

In this thesis we want to test the CQM $\mathcal{O}\left( p^{6}\right) $
predictions by comparing them to the chiral $\mathcal{O}\left( p^{6}\right) $
coefficients fixed by experiment. We will thus focus on the $\mathcal{O}%
\left( p^{6}\right) $ abnormal intrinsic parity effective action. The $%
\Gamma ^{+}$ describing the normal parity sector, can also produce anomalous
terms. This is because the EOMs have been used in rewriting $\Gamma ^{+}$,
and these contain both anomalous and non-anomalous parts. However, this type
of terms can be shown to be of $\mathcal{O}\left( p^{8}\right) $ and are not
of interest here. The final result involves the contribution of terms with
Seely-DeWitt coefficients $a_{3}$ (describing one external field and 3 PS
mesons)$,a_{4}$ (2 external vector fields and one PS meson) and $a_{5}$ (5
PS interaction). The explicit form of $\sum_{i=3,4,5}\Gamma ^{-}\left(
a_{i}\right) $ is very lengthy and can be found in appendix B.

\section{Results}

\subsection{Combining Amplitudes}

We are now ready to combine all the amplitudes: the WZW $\mathcal{O}\left(
p^{4}\right) ,$ the one-loops and the $\mathcal{O}\left( p^{6}\right) $
terms. Following through with the example of the $\pi ^{0}\rightarrow \gamma
e^{+}e^{-}$ decay, we can verify that the divergent part of the loops are
canceled by the parameters of $\mathcal{O}\left( p^{6}\right) .$ This serves
a test for consistent calculations. Combining (\ref{mel2}) \& (\ref{op6gee})
with the loop amplitude in appendix A we get%
\begin{eqnarray*}
A^{an} &=&A_{WZW}+A_{1\ell }+A_{6}=\frac{1}{4\pi ^{2}}\frac{e^{3}}{F_{\pi }}%
\varepsilon ^{\mu \nu \alpha \beta }\varepsilon _{\mu }k_{\nu }\frac{\bar{e}%
\gamma _{\alpha }e}{k^{\star 2}}k_{\beta }^{\star }\times \\
&&[1+\frac{1}{32\pi ^{2}F^{2}}\{\tfrac{2}{3}\lambda k^{\star 2}-\tfrac{1}{3}%
k^{\star 2}(\log \frac{m_{K}^{2}}{\mu ^{2}}+\log \frac{m_{\pi }^{2}}{\mu ^{2}%
})+\tfrac{10}{9}k^{\star 2} \\
&&+\tfrac{4}{3}[F\left( k^{\star 2},m_{\pi }^{2}\right) +F\left( k^{\star
2},m_{K}^{2}\right) ]\}-\frac{512}{9}B_{0}\left( 4m_{u}-m_{d}\right) \pi
^{2}C_{7}^{Wr} \\
&&-\frac{1024}{3}B_{0}\left( m_{u}-m_{d}\right) \pi ^{2}C_{8}^{Wr}+k_{\gamma
}^{\star 2}\left( \frac{64}{3}\pi ^{2}C_{22}^{Wr}-\frac{1}{48\pi ^{2}F^{2}}%
\mu ^{d-4}\lambda \right) ]
\end{eqnarray*}%
We see that the infinite parts cancel: $A^{\lambda }=\frac{1}{32\pi ^{2}F^{2}%
}\tfrac{2}{3}\lambda k^{\star 2}-\frac{1}{48\pi ^{2}F^{2}}\mu ^{d-4}\lambda
k^{\star 2}=0$. The function $F$ comes from the evaluation of the loop
integrals and can be found in appendix A.

\subsection{Theoretical Quantum Field Calculations}

The results of the theoretical calculations are displayed in table 2, where
the first term in the brackets is due to the WZW action. For the sake of
completeness the one-loop contributions $A_{1\ell }$ are listed in appendix
A.

\begin{longtable}{||l|l||}
\hline\hline
\textsc{Process:} & \textsc{Amplitude:} \\ \hline
$\pi ^{0}\rightarrow \gamma \gamma $ & $%
\begin{array}{l}
\\ 
\frac{\alpha }{\pi F_{\pi }}\varepsilon ^{\mu \nu \alpha \beta }\varepsilon
_{\mu }k_{\nu }\varepsilon _{\alpha }^{\star }k_{\beta }^{\star }\times
\smallskip \\ 
\lbrack 1-\tfrac{128}{9}\pi ^{2}\left( 20m_{K}^{2}+m_{\pi }^{2}-15m_{\eta
}^{2}\right) C_{7}^{Wr}
\smallskip\\
-12\tfrac{128}{9}\pi ^{2}\left( 4m_{K}^{2}-m_{\pi
}^{2}-3m_{\eta }^{2}\right) C_{8}^{Wr}]\smallskip \\ 
\smallskip%
\end{array}%
$ \\ \hline
$\eta \rightarrow \gamma \gamma $ & $%
\begin{array}{l}
\\ 
\frac{\alpha }{\sqrt{3}\pi F_{\eta }}\varepsilon ^{\mu \nu \alpha \beta
}\varepsilon _{\mu }k_{\nu }\varepsilon _{\alpha }^{\star }k_{\beta }^{\star
}\times \smallskip \\ 
\lbrack 1-\frac{128}{3}\left( 4m_{K}^{2}-5m_{\eta }^{2}+3m_{\pi }^{2}\right)
\pi ^{2}C_{7}^{W}+512\left( m_{\pi }^{2}-m_{\eta }^{2}\right) \pi
^{2}C_{8}^{W}]\smallskip \\ 
\smallskip%
\end{array}%
$ \\ \hline
$\pi ^{0}\rightarrow \gamma e^{+}e^{-}$ & $%
\begin{array}{l}
\\ 
\frac{1}{4\pi ^{2}}\frac{e^{3}}{F_{\pi }}\varepsilon ^{\mu \nu \alpha \beta
}\varepsilon _{\mu }k_{\nu }\frac{\bar{e}\gamma _{\alpha }e}{k^{\star 2}}%
k_{\beta }^{\star }\times \smallskip \\ 
\lbrack 1-\frac{256}{3}\pi ^{2}m_{\pi }^{2}C_{7}^{Wr}+\frac{64}{3}\pi
^{2}k^{\star 2}C_{22}^{Wr}]+A_{\pi ^{0}\gamma e^{+}e^{-}}^{1\ell }\smallskip
\\ 
\smallskip%
\end{array}%
$ \\ \hline
$\eta \rightarrow \gamma e^{+}e^{-}$ & $%
\begin{array}{l}
\\ 
\frac{e^{3}}{4\sqrt{3}\pi ^{2}F_{\eta }}\varepsilon ^{\mu \nu \alpha \beta
}\varepsilon _{\mu }k_{\nu }\frac{\bar{e}\gamma _{\alpha }e}{k^{\star 2}}%
k_{\beta }^{\star }\times \smallskip \\ 
\lbrack 1-\tfrac{128}{9}\pi ^{2}\left( 12m_{K}^{2}-15m_{\eta }^{2}+9m_{\pi
}^{2}\right) C_{7}^{Wr}-36\tfrac{128}{9}\pi ^{2}\left( m_{\pi }^{2}-m_{\eta
}^{2}\right) C_{8}^{Wr}\smallskip \\ 
+\tfrac{192}{9}\pi ^{2}k^{\star 2}C_{22}^{Wr}]+A_{\eta \gamma
e^{+}e^{-}}^{1\ell }\smallskip \\ 
\smallskip%
\end{array}%
$ \\ \hline
$\pi ^{+}\rightarrow \gamma e^{+}\nu $ & $%
\begin{array}{l}
\\ 
\frac{eG_{F}\cos \theta }{8\pi ^{2}F_{\pi }}\varepsilon ^{\mu \nu \alpha
\beta }l_{\mu }q_{\nu }\varepsilon _{\alpha }k_{\beta }\times \smallskip \\ 
\lbrack 1-\frac{256}{3}\pi ^{2}m_{\pi }^{2}C_{7}^{Wr}+\frac{64}{3}\pi
^{2}\left( q^{2}+k^{2}\right) C_{22}^{Wr}]\smallskip +A_{\pi ^{+}\gamma
e^{+}\nu ^{-}}^{1\ell } \\ 
\smallskip%
\end{array}%
$ \\ \hline
$K^{+}\rightarrow \gamma e^{+}\nu $ & $%
\begin{array}{l}
\\ 
\frac{eG_{F}\sin \theta }{8\pi ^{2}F_{K}}\varepsilon ^{\mu \nu \alpha \beta
}l_{\mu }q_{\alpha }\varepsilon _{\alpha }k_{\beta }\times \smallskip \\ 
\lbrack 1-\frac{256}{3}\pi ^{2}m_{K}^{2}C_{7}^{Wr}+256\pi ^{2}\left(
m_{K}^{2}-m_{\pi }^{2}\right) C_{11}^{Wr}+\frac{64}{3}\pi ^{2}\left(
q^{2}+k^{2}\right) C_{22}^{Wr}]\smallskip \\ 
+A_{K^{+}\gamma e^{+}\nu }^{1\ell }\smallskip \\ 
\smallskip%
\end{array}%
$ \\ \hline
$\gamma \pi ^{0}\rightarrow \pi ^{+}\pi ^{-}$ & $%
\begin{array}{l}
\\ 
\frac{1}{4\pi ^{2}}\frac{e}{F_{\pi }^{3}}\varepsilon ^{\mu \nu \alpha \beta
}\varepsilon _{\mu }p_{\nu }p_{\alpha }p_{\beta }\times \smallskip \\ 
\lbrack 1+64\pi ^{2}m_{\pi }^{2}\left(
2C_{4}^{Wr}+C_{5}^{Wr}-C_{14}^{Wr}-C_{15}^{Wr}\right) \smallskip \\ 
+\frac{64}{3}\pi ^{2}k^{2}\left( C_{14}^{Wr}+C_{15}^{Wr}-C_{13}^{Wr}\right)
]+A_{\gamma \pi ^{0}\pi ^{+}\pi ^{-}}^{1\ell }\smallskip \\ 
\smallskip%
\end{array}%
$ \\ \hline
$\eta \rightarrow \gamma \pi ^{+}\pi ^{-}$ & $%
\begin{array}{l}
\\ 
\frac{1}{4\pi ^{2}\sqrt{3}}\frac{e}{F_{\pi }^{2}F_{\eta _{8}}}\varepsilon
^{\mu \nu \alpha \beta }\varepsilon _{\mu }p_{\nu }^{\eta }p_{\alpha }^{\pi
^{+}}p_{\beta }^{\pi ^{-}}\times \smallskip \\ 
\lbrack 1+192\pi ^{2}\left( m_{\pi }^{2}-m_{\eta }^{2}\right) \left(
C_{6}^{Wr}-C_{3}^{Wr}\right) \smallskip \\ 
+\frac{32}{3}\pi ^{2}\left( 4m_{K}^{2}-3m_{\eta }^{2}+5m_{\pi }^{2}\right)
\left( C_{5}^{Wr}+2C_{4}^{Wr}\right) \smallskip \\ 
+128\pi ^{2}p_{+}p_{-}C_{15}^{Wr}-128\pi ^{2}[p_{\eta }p_{+}+p_{\eta
}p_{-}+p_{+}p_{-}]C_{14}^{Wr}\smallskip \\ 
+64\pi ^{2}[p_{\eta }p_{+}+p_{\eta }p_{-}-m_{\eta
}^{2}]C_{13}^{Wr}]+A_{\gamma \eta _{8}\pi ^{+}\pi ^{-}}^{1\ell } \\ 
\smallskip%
\end{array}%
$ \\\hline
$ K^{+}\rightarrow \pi ^{+}\pi ^{-}e^{+}\nu$ & $%
\begin{array}{l}
\\ 
\frac{G_{F}\sin \theta }{4\pi ^{2}F_{\pi }^{2}F_{K}}\varepsilon ^{\mu \nu
\alpha \beta }l_{\mu }q_{\nu }p_{\alpha }^{+}p_{\beta }^{-}\times \smallskip
\\ 
\lbrack 1+64\pi ^{2}\left( m_{K}^{2}-m_{\pi }^{2}\right) C_{2}^{Wr}+32\pi
^{2}\left( 3m_{\pi }^{2}-2m_{K}^{2}\right) C_{4}^{Wr}\smallskip \\ 
+64\pi ^{2}m_{\pi }^{2}C_{5}^{Wr}-32\pi ^{2}q\left( q+p_{+}\right)
C_{13}^{Wr}-128\pi ^{2}p_{-}p_{K}C_{14}^{Wr}\smallskip \\ 
-64\pi ^{2}p_{+}\left( q+p_{+}\right) C_{15}^{Wr}]+A_{K^{+}\pi ^{+}\pi
^{-}e^{+}\nu }^{1\ell } \\ 
\smallskip%
\end{array}%
$ \\ \hline
$K^{+}\rightarrow \pi ^{0}\pi ^{0}e^{+}\nu $ & $%
\begin{array}{l}
\\ 
\frac{G_{F}\sin \theta }{4\pi ^{2}F_{K}F_{\pi }^{2}}\varepsilon ^{\mu \nu
\alpha \beta }l_{\mu }q_{\nu }p_{\alpha }p_{\beta }\times \smallskip \\ 
(p_{\pi _{1}^{0}}-p_{\pi _{2}^{0}})[-16\pi ^{2}qC_{13}^{Wr}+64\pi
^{2}p_{K}C_{14}^{Wr}+32\pi ^{2}p_{K}C_{15}^{Wr}]\smallskip \\ 
+A_{K^{+}\pi ^{0}\pi ^{0}e^{+}\nu }^{1\ell } \\ 
\smallskip%
\end{array}%
$ \\ \hline\hline

\caption[Nothing]{
 \emph{Results of theoretical quantum field
calculations.} \rule{0cm}{2em} }
 \end{longtable}

$k$ stands for the photon and $q$ for the di-lepton four-momentum. The other
momenta have been labeled as needed. The $W_{\mu }$ has been replaced with
the leptonic current%
\begin{equation*}
W_{\mu }\rightarrow \frac{g}{2\sqrt{2}M_{W}^{2}}l_{\mu }\equiv \frac{g}{2%
\sqrt{2}M_{W}^{2}}\bar{u}_{\nu }\gamma _{\mu }\left( 1-\gamma _{5}\right)
v_{e}
\end{equation*}%
with $G_{F}/\sqrt{2}=g^{2}/8M_{W}^{2}.$ In the $M\rightarrow \gamma \gamma
^{\star }$ decays$,$ note the absence of loop corrections in the case of
real photons. It is interesting to see that the kinetic part of the
electroweak fields in the decays with one pseudoscalar and two external
fields are all connected to the $C_{22}^{Wr}$ coefficient. From experiment
we expect this chiral coefficient to be rather large with respect to the
others in the decay amplitude. $C_{22}^{Wr}$ obviously becomes increasingly
significant at higher energies, so it is important to extract a good value.
In the cases where they are active, $C_{13}^{Wr},C_{14}^{Wr}$ and $%
C_{15}^{Wr}$ play a similar role - they are also connected to kinematical
factors, but do not vanish in the soft limit. Inspecting the corresponding
monomials in table 1, we can trace the kinematical dependence to the
presence of an extra $\partial _{\gamma }$ 4-derivative. Other monomials
have the same property but forcably contain a minimum of fields that exceeds
the number allowed by the process.

The Gell-Mann-Okubo relation has been used to eliminate the $\eta _{8}$ mass
in all processes except for those involving an $\eta .$ Terms connected to $%
C_{8}^{Wr}$ in $\pi ^{0}\gamma e^{+}e^{-}$ and to $C_{11}^{Wr}$ in $\pi
^{0}\gamma e^{+}\nu $ are proportional to $m_{u}-m_{d}$, and do not appear
in table 2 since we are working in the isospin limit. The loop corrections
must be recalculated if we wish evaluate the amplitudes away from the
isospin limit.

In all processes containing an on-shell photon, the term proportional to $%
k^{2}$ has been kept, even though this will be set to zero when comparing
with the experimental data. Such terms will become useful when we are
comparing with the VMD \& CQM predictions.

\subsection{Experimental Comparison}

Experimental data in the form of slopes, form factors and decay rates will
allow us to extract numerical values for the chiral coefficients. The slope
parameter $b$ is defined as%
\begin{equation}
b=\frac{1}{A\left( M\rightarrow \gamma \gamma \right) }\frac{d}{dk^{\star 2}}%
\left. A\left( M\rightarrow \gamma \gamma ^{\star }\right) \right|
_{k^{\star }=0}  \label{slope}
\end{equation}%
i.e. the factor in front of the off-shell photon squared four-momentum,
normalized by the on-shell amplitude. To calculate the width we make use of
the standard formula for two body decays, except in the case of $\gamma \eta
\pi ^{+}\pi ^{-}$ where it's necessary to perform three body phase space
integration. The form factors allow for easy comparison, as they can be
directly related to the matrix element. Below, all cases are treated
individually - the results are summarized in table 3. The coefficients have
been solved for using the least square solver in the MINUIT program (part of
the CERN programming resource library). The errors are mainly due to
experimental uncertainty, but all error limits have a contribution coming
from the decay constants, of which we still have relatively poor knowledge.
Measurements on charged pion decays give $F_{\pi }=92.4\pm 0.33\unit{MeV}$.
Next-to-leading order values for the other decay constants can be extracted
through wavefunction renormalization as in \cite{u1v}. Many sources use an
alternative convention and quote $f_{M}=\sqrt{2}F_{M}.$

\begin{table}[t]
\begin{center}
\begin{tabular}{||l|l|l||}
\hline\hline
\emph{Process:} & \emph{Experimental input:} & \emph{Solved Coefficients }[$%
10^{-9}$MeV$^{-2}]$ \\ \hline
$%
\begin{array}{l}
\\ 
\pi ^{0}\rightarrow \gamma \gamma \text{ }\&\smallskip \\ 
\eta \rightarrow \gamma \gamma \smallskip \smallskip \smallskip%
\end{array}%
$ & $%
\begin{array}{l}
\\ 
\text{Width \cite{pdg1}}\smallskip \\ 
\text{Width \cite{pdg1}}\smallskip \smallskip \smallskip%
\end{array}%
$ & $%
\begin{array}{l}
\\ 
C_{7}^{Wr}\simeq 0.013\pm 1.17 \\ 
C_{8}^{Wr}\simeq 0.76\pm 0.18\smallskip \smallskip \smallskip%
\end{array}%
$ \\ \hline
$%
\begin{array}{l}
\\ 
\pi ^{0}\rightarrow \gamma e^{+}e^{-}\smallskip \smallskip \smallskip%
\end{array}%
$ & $%
\begin{array}{l}
\\ 
\text{Slope parameter \cite{cleo}}\smallskip \smallskip \smallskip%
\end{array}%
$ & $%
\begin{array}{l}
\\ 
C_{22}^{Wr}\simeq 6.52\pm 0.78\smallskip \smallskip \smallskip%
\end{array}%
$ \\ \hline
$%
\begin{array}{l}
\\ 
\eta \rightarrow \gamma e^{+}e^{-}\smallskip \smallskip \smallskip%
\end{array}%
$ & $%
\begin{array}{l}
\\ 
\text{Slope parameter \cite{cleo}}\smallskip \smallskip \smallskip%
\end{array}%
$ & $%
\begin{array}{l}
\\ 
C_{22}^{Wr}\simeq 5.07\pm 0.71\smallskip \smallskip \smallskip%
\end{array}%
$ \\ \hline
$%
\begin{array}{l}
\\ 
\pi ^{+}\rightarrow \gamma e^{+}\nu \smallskip \smallskip \smallskip%
\end{array}%
$ & $%
\begin{array}{l}
\\ 
\text{Form factor \cite{pdg1}}\smallskip \smallskip \smallskip%
\end{array}%
$ & $%
\begin{array}{l}
\\ 
C_{7}^{Wr}\simeq 20.3\pm 18.7\smallskip \smallskip \smallskip%
\end{array}%
$ \\ \hline
$%
\begin{array}{l}
\\ 
K^{+}\rightarrow \gamma e^{+}\nu \smallskip \smallskip \smallskip%
\end{array}%
$ & $%
\begin{array}{l}
\\ 
\text{Form factor \cite{pdg1}}\smallskip \smallskip \smallskip%
\end{array}%
$ & $%
\begin{array}{l}
\\ 
C_{11}^{Wr}\simeq -6.37\pm 4.54\smallskip \smallskip \smallskip%
\end{array}%
$ \\ \hline
$%
\begin{array}{l}
\\ 
\\ 
\gamma \pi ^{0}\pi ^{+}\pi ^{-}\,\,\& \\ 
K^{+}\rightarrow \pi ^{+}\pi ^{-}e^{+}\nu \smallskip \\ 
\\ 
\\ 
\smallskip \smallskip \smallskip%
\end{array}%
$ & $%
\begin{array}{l}
\\ 
\\ 
\text{Form factor \cite{area51}} \\ 
\text{Form factors \cite{kl4}}\smallskip \\ 
\\ 
\\ 
\smallskip \smallskip \smallskip%
\end{array}%
$ & $%
\begin{array}{l}
\\ 
C_{2}^{Wr}\simeq -0.32\pm 10.4\smallskip \\ 
\smallskip C_{4}^{Wr}\simeq 0.28\pm 9.19 \\ 
C_{5}^{Wr}\simeq 28.50\pm 28.83\smallskip \\ 
C_{13}^{Wr}\simeq -74.09\pm 55.89\smallskip \\ 
C_{14}^{Wr}\simeq 29.99\pm 11.14\smallskip \\ 
C_{15}^{Wr}\simeq -25.30\pm 23.93\smallskip \smallskip \smallskip%
\end{array}%
$ \\ \hline
$%
\begin{array}{l}
\\ 
\eta \gamma \pi ^{+}\pi ^{-}\smallskip \smallskip \smallskip%
\end{array}%
$ & $%
\begin{array}{l}
\\ 
\smallskip \smallskip \smallskip \text{Width }\cite{pdg1}%
\end{array}%
$ & $%
\begin{array}{l}
\\ 
\smallskip \smallskip \smallskip C_{3}^{Wr}-C_{6}^{Wr}\simeq 21.67\pm 17.41%
\end{array}%
$ \\ \hline
$%
\begin{array}{l}
\\ 
K^{+}\rightarrow \pi ^{0}\pi ^{0}e^{+}\nu \smallskip \smallskip \smallskip%
\end{array}%
$ & $%
\begin{array}{l}
\\ 
NA\smallskip \smallskip \smallskip%
\end{array}%
$ &  \\ \hline\hline
\end{tabular}%
\end{center}

\caption{
 \emph{Solved coefficients}}
\end{table}

\subsubsection{$\protect\pi ^{0}/\protect\eta \rightarrow \protect\gamma 
\protect\gamma ^{\star }$}

For real photons we form the decay rate by squaring the matrix element and
tagging it with the appropriate factors. The squaring involves inserting one
half for identical photons and then summing over all polarizations as in
section 8.2. Width data is quoted in \cite{pdg1} as a weighted average of
several measurements, yielding $\Gamma _{\pi ^{0}\gamma \gamma
}=7.\,\allowbreak 836\pm 0.523$ eV and $\Gamma _{\eta \gamma \gamma }=465\pm
44$ eV respectively. Note that the calculation was performed with the
Goldstone boson matrix (\ref{gsmatr}), i.e. using $\eta _{8}.$ Instead of
explicitly implementing the mixing model of section 6, we conveniently let
the chiral coefficients supply the needed extra factor. Proceeding in this
way we are left with two equations, allowing us to solve for $C_{7}^{Wr}$
and $C_{8}^{Wr}$.

For on-shell photons we can extract $C_{22}^{Wr}$ using (\ref{slope}). To
this end, we use the results of the CLEO II detector differential cross
section measurements \cite{cleo}. Fitting the form factor data they arrive
at the pole parameters $\Lambda _{\pi ^{0}}=776\pm 38\unit{MeV}$ $\left(
=b^{-1/2}\right) $ and $\Lambda _{\eta }=774\pm 49\unit{MeV}.$ The similar
slopes indicate that the two meson wavefunctions are nearly identical.
Looking at table 3 we see that the two extracted values for $C_{22}^{Wr}$
agree within the error limits. Also note that $C_{22}^{Wr}$ is an order of
magnitude larger than $C_{7}^{Wr}$ and $C_{8}^{Wr}.$

\subsubsection{$\protect\pi ^{+}/K^{+}\rightarrow \protect\gamma e^{+}%
\protect\nu $}

For these decays the Particle Data Group \cite{pdg1} quotes constant vector
form factors ($F_{V})$, as no momentum dependence can be seen in the region
of the experiments. The axial current is not of interest here as is belongs
to the non-anomalous sector. Experimental vector form factor measurements
yield the matrix element%
\begin{equation}
M\left( SD_{V}\right) =\frac{eG_{F}V_{qq^{\prime }}}{\sqrt{2}m_{P}}%
\varepsilon ^{\mu }l^{\nu }F_{V}^{\pi }\varepsilon _{\mu \nu \sigma \tau
}k^{\sigma }q^{\tau }\quad ;F_{V}^{\pi }=\left( 0.017\pm 0.008\right)
\end{equation}%
Comparing with the pion amplitude in table 2 we find%
\begin{eqnarray}
F_{V}^{\pi } &=&\frac{m_{\pi ^{+}}^{2}}{4\sqrt{2}\pi ^{2}F_{\pi }}\times
\lbrack 1+\frac{1}{32\pi ^{2}F_{\pi }^{2}}\{-4m_{\pi }^{2}\ln \frac{m_{\pi
}^{2}}{\mu ^{2}}-4m_{K}^{2}\ln \frac{m_{K}^{2}}{\mu ^{2}} \\
&&+4I\left( q^{2},m_{\pi }^{2},m_{\pi }^{2}\right) +4I\left( k^{2},m_{\pi
}^{2},m_{K}^{2}\right) \}-\frac{256}{3}\pi ^{2}m_{\pi }^{2}C_{7}^{Wr}+\frac{%
64}{3}\pi ^{2}\left( q^{2}+k^{2}\right) C_{22}^{Wr}]  \notag
\end{eqnarray}%
where the one-loop corrections from appendix A have been inserted. Setting
the photon and dilepton four-momenta squared ($k^{2}$ and $q^{2}$
respectively) to zero, allows solving for $C_{7}^{Wr}.$ The measurements are
rather imprecise, so the extracted value for $C_{7}^{Wr}$ should simply be
regarded as an upper limit. The $M\rightarrow \gamma \gamma ^{\star }$ value
is closer to the truth. As for $C_{11}^{Wr}$, this coefficient can only be
extracted from the $K^{+}$ decay, or alternatively from the $\pi ^{+}$ decay
recalculated away from the isospin limit. Considering the poor data
presently available, the recalculation hardly seems worth the effort.
Repeating the above procedure for the kaon, with $F_{V}^{K}=\left( 0.204\pm
0.070\right) ,$ we can extract a value for $C_{11}^{Wr}.$ The fact that this
values is negative is not a problem, since there is nothing in the theory
which contradicts negative values for the chiral coefficients.

\subsubsection{$\protect\gamma \protect\pi ^{0}\protect\pi ^{+}\protect\pi %
^{-}$}

Experimental data \cite{area51} does not reveal any kinematical dependence.
The amplitude is expressible in terms of the $\gamma 3\pi $ coupling
constant $F^{3\pi }=12.9\pm 1.4\,$GeV$^{-3}.$ Parsing the amplitude from
table 2 and the loop corrections from appendix A, we set $k^{2}=0$ and the
invariant $p_{ij}^{2}=\left( p_{i}+p_{j}\right) ^{2}$ to their average value
- one third into their respective kinematical range.%
\begin{equation*}
\begin{array}{l}
-3.5m_{\pi }^{2}\leq p_{01}^{2}\equiv \left( p_{\pi ^{0}}-p_{\pi
^{-}}\right) ^{2}\leq 0\Rightarrow \left\langle p_{01}^{2}\right\rangle
\simeq -\tfrac{7}{6}m_{\pi }^{2}\smallskip \\ 
\,\,\,\,\,\,\,\,\,4m_{\pi }^{2}\leq p_{02}^{2}\equiv \left( p_{\pi
^{0}}-p_{\pi ^{+}}\right) ^{2}\leq 13m_{\pi }^{2}\Rightarrow \left\langle
p_{02}^{2}\right\rangle \simeq 7m_{\pi }^{2}\smallskip \\ 
\sum p_{ij}^{2}=\sum m_{i}^{2}\Rightarrow \left\langle
p_{12}^{2}\right\rangle =-\tfrac{17}{6}m_{\pi }^{2}%
\end{array}%
\end{equation*}%
This gives%
\begin{eqnarray}
F^{3\pi } &\simeq &\sqrt{\frac{\alpha }{4\pi }}\frac{1}{\pi F_{\pi }^{3}}%
\times \lbrack 1+\frac{1}{96\pi ^{2}F^{2}}\{-3m_{\pi }^{2}\log \frac{m_{\pi
}^{2}}{\mu ^{2}}+5m_{\pi }^{2} \\
&&+4F\left( m_{\pi }^{2},-\tfrac{17}{6}m_{\pi }^{2}\right) +4F\left( m_{\pi
}^{2},7m_{\pi }^{2}\right) +4F\left( m_{\pi }^{2},-\tfrac{7}{6}m_{\pi
}^{2}\right) \}  \notag \\
&&-64\pi ^{2}m_{\pi }^{2}\left[
C_{14}^{Wr}+C_{15}^{Wr}-2C_{4}^{Wr}-C_{5}^{Wr}\right] ]  \notag
\end{eqnarray}%
producing one equation with the unknown combination $\left(
C_{14}^{Wr}+C_{15}^{Wr}-2C_{4}^{Wr}-C_{5}^{Wr}\right) .$ What we have done
is clarified in figure 5, where the theoretical form factor and $F^{3\pi }$
have been plotted as functions of $p_{01}^{2}$ and $p_{12}^{2}$ in units of $%
m_{\pi }^{2}.$ We have simply aligned the form factor average with the
experimentally observed, constant plane $F^{3\pi }$. Note that the error on $%
F^{3\pi }$ transcends the maximum deviation of the theoretical surface from
the plane.

\begin{figure}
\begin{center}
\includegraphics{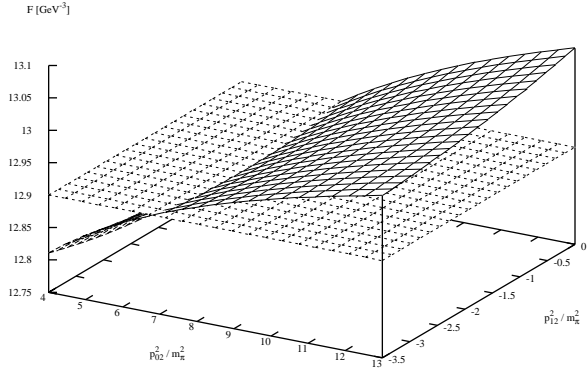}
\end{center}
\caption{
 \emph{Experimental (dashed) \& theoretical (solid) form
factor.}}
\end{figure}

\subsubsection{$\protect\eta \protect\gamma \protect\pi ^{+}\protect\pi ^{-}$%
}

To compare with available data in the form of the decay rate $\Gamma _{\eta
\gamma \pi \pi }=55.2\pm 6.6$ eV \cite{pdg1}, we must perform three-body
phase space integration. We make use of the formula%
\begin{equation}
d\Gamma =\frac{1}{\left( 2\pi \right) ^{3}}\frac{1}{32m_{\eta }^{3}}\left| 
\mathcal{M}\right| ^{2}dp_{12}^{2}dp_{23}^{2}
\end{equation}%
by rewriting the amplitude in terms of the invariants $p_{12}^{2}\equiv
\left( p_{+}+p_{-}\right) ^{2}$ and $p_{23}^{2}\equiv \left( p_{-}+k\right)
^{2}$ and evaluating the integral numerically. The final result is an
equation with chiral coefficients 4-6 and 13-14 as unknowns. This equation
contains a term proportional to $C_{3}^{Wr}-C_{6}^{Wr}.$ Since these
coefficients only appear in $\eta \gamma \pi \pi ,$ they cannot be solved
for individually, and all we can do is to extract a value for the difference
between them.

\subsubsection{$K^{+}\rightarrow \protect\pi ^{+}\protect\pi ^{-}e^{+}%
\protect\nu $}

\noindent This decay provides the best experimental data in the form of an
energy dependent vector form factor. Data from 4$\times 10^{5}$ events where
recently (2001) collected at the Brookhaven Alternating Gradient Synchrotron 
\cite{kl4} over a broad kinematical range. The matrix element is quoted as%
\begin{equation}
M=\frac{G_{F}}{\sqrt{2}}V_{us}^{\star }\bar{u}\left( p_{\nu }\right) \gamma
_{\mu }\left( 1-\gamma _{5}\right) v\left( p_{e}\right) \left( V^{\mu
}-A^{\mu }\right)  \label{mex}
\end{equation}%
where we only concern ourselves with the hadronic vector contribution $%
V^{\mu }=H\varepsilon ^{\mu \nu \rho \sigma }L_{\nu }P_{\rho }Q_{\sigma }$,
with $P=p_{+}+p_{-},$ $Q=p_{+}-p_{-},$ $L=p_{e}+p_{\nu }$ in units of $m_{K}$
and where the dimensionless $H$ is a function of the invariant dipion mass $%
M_{\pi \pi }=\left| p_{+}+p_{-}\right| .$ 6 datapoints for $H$ are quoted in
the range $280\unit{MeV}\leq M_{\pi \pi }\leq 380\unit{MeV}.$ No angular
dependence was detected in this energy range. Comparing (\ref{mex}) with the
amplitude in table 3, we can extract 6 equations with $H(M_{\pi \pi
}^{2},\left( p_{K}-p_{+}\right) ^{2},\left( p_{K}-p_{-}\right) ^{2}).$ This
requires rewriting the matrix element using relativistic kinematics and
verifying that the dependence on $\theta _{\pi }$ (the polar angle of the $%
\pi ^{+}$ with respect to the dipion in the kaon rf) is small. Let the $z$
axis be parallel to the dipion flight direction. In the kaon restframe we
have 
\begin{eqnarray*}
p_{K} &=&(m_{K},\mathbf{0}) \\
p_{2\pi } &=&(\sqrt{M_{\pi \pi }^{2}+\left| \mathbf{p}_{2\pi }\right| ^{2}}%
,0,0,\left| \mathbf{p}_{2\pi }\right| )
\end{eqnarray*}%
where $\left| \mathbf{p}_{2\pi }\right| \simeq \left( m_{K}^{2}-M_{\pi \pi
}^{2}\right) ^{2}/m_{K}$ with $q^{2}\simeq 0$. In the dipion restframe we
have%
\begin{eqnarray*}
p_{2\pi } &=&\left( E_{2\pi },0,0,0\right) \\
p_{+} &=&\left( \tfrac{1}{2}E_{2\pi },0,p_{\pi }\sin \theta _{\pi },p_{\pi
}\cos \theta _{\pi }\right) \quad ;p_{\pi }=\sqrt{\tfrac{1}{4}M_{\pi \pi
}^{2}-m_{\pi }^{2}} \\
p_{-} &=&\left( \tfrac{1}{2}E_{2\pi },0,-p_{\pi }\sin \theta _{\pi },-p_{\pi
}\cos \theta _{\pi }\right) \\
p_{K} &=&\left( E_{K},0,0,-\left| \mathbf{p}_{K}\right| \right)
\end{eqnarray*}%
where we can extract $E_{K}$ and $\left| \mathbf{p}_{K}\right| $ using the
frame invariance of scalars: $p_{K}^{Krf}p_{2\pi }^{Krf}=p_{K}^{2\pi
rf}p_{2\pi }^{2\pi rf}.$ We can now recast the amplitude in the appropriate
form and verify that the angular dependence is small. For example:%
\begin{equation}
p_{K}p_{-}=\tfrac{1}{4}\left( m_{K}^{2}+M_{\pi \pi }^{2}\right) -\frac{%
\left( m_{K}^{2}-M_{\pi \pi }^{2}\right) }{2M_{\pi \pi }}\sqrt{\tfrac{1}{4}%
M_{\pi \pi }^{2}-m_{\pi }^{2}}\cos \theta _{\pi }
\end{equation}%
The second term becomes very small when $M_{\pi \pi }\simeq m_{K}$ or $%
4m_{\pi }.$ In addition, the term is suppressed by $M_{\pi \pi }$ and the
cos factor.

The 6 extracted equations, involving coefficients 2, 4, 5 and 13-15, are now
merged with the equation from $\gamma 3\pi ,$ giving a total of 7 equations
with 6 unknowns. This system is overdetermined, which will help to reduce
the errors.

The data series has been plotted in figure 6, along with the prediction of
the fitted chiral parameters. Evidently the data points at the high end of
the spectrum deviates significantly from the prediction. It is unclear
whether or not this is just a statistical fluke, or if it originates
somewhere in the experimental setup. The dominating accidental background
was from $K\pi ^{0}\pi ^{+}\pi ^{-}$, with a $\pi ^{+}\pi ^{-}$ pair
detection along with an $e^{+}$ from the beam or coincident decay. However,
this was reduced to $2.4\pm 1.2\%$ using a likelihood method.

\begin{figure}
\begin{center}
\includegraphics{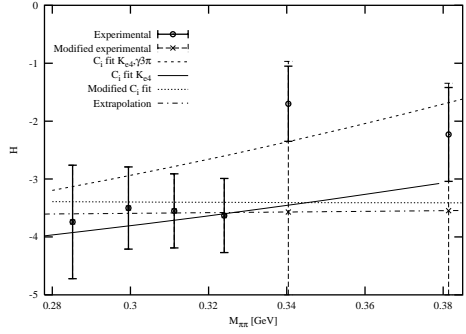}
\end{center}
\caption{
\emph{Experimental \& chiral predictions of }$H.$}
\end{figure}

If we hypothesize that the deviating points are indeed due to a statistical
fluke, this can be compensated for by extrapolating a line using the first
four data points to the high energy region. Also, we exclude the $\gamma
3\pi $ equation as the measurement is not very good. Proceeding in this way,
we can refit the chiral coefficients, producing a radically altered slope
(see figure 6) and the $C_{i}s$ in table 4. The errors have been
overestimated, reflecting the uncertainty in the extrapolation procedure.
This is why the errors in table 4 have been inflated.

\begin{table}
\begin{center}
\begin{tabular}{||l|l||}
\hline\hline
\emph{Process:} & $C_{i}^{Wr}$ [10$^{-9}$ MeV$^{-2}$] \\ \hline
$K^{+}\rightarrow \pi ^{+}\pi ^{-}e^{+}\nu $ & $%
\begin{array}{l}
\\ 
C_{2}^{Wr}\simeq 0.78\pm 12.7\smallskip \\ 
C_{4}^{Wr}\simeq 0.67\pm 10.9\smallskip \\ 
C_{5}^{Wr}\simeq 9.38\pm 152.2\smallskip \\ 
C_{13}^{Wr}\simeq -8.44\pm 69.9\smallskip \\ 
C_{14}^{Wr}\simeq 0.72\pm 15.3\smallskip \\ 
C_{15}^{Wr}\simeq -3.10\pm 28.6\smallskip \smallskip \smallskip%
\end{array}%
$ \\ \hline
$\eta \rightarrow \gamma \pi ^{+}\pi ^{-}$ & $%
\begin{array}{l}
\\ 
\smallskip \smallskip \smallskip C_{3}^{Wr}-C_{6}^{Wr}\simeq 4.6\pm 26.6%
\end{array}%
$ \\ \hline\hline
\end{tabular}%
\end{center}
\caption{
 \emph{Coefficient values as a result of extrapolation.}}
\end{table}

\subsection{VMD Comparison}

VMD is unable to predict the mass terms of the chiral theory, as the
Lagrangian contains no explicit mass parameters. Table 5 shows the $\mathcal{%
O}\left( p^{6}\right) $ VMD expressions that should be compared with the
chiral predictions. The prefactors are the same as for the $\mathcal{O}%
\left( p^{6}\right) $ chiral expressions, allowing for direct comparison
with the terms proportional to kinematical factors.

\begin{table}
\begin{center}
$%
\begin{tabular}[b]{lll}
$\emph{process:}$ & \emph{Amplitude:\smallskip } & $\emph{source:}$ \\ 
$\pi ^{0}\left( \eta \right) \rightarrow \gamma e^{+}e^{-}\smallskip $ & $%
\frac{k^{2}\smallskip }{M_{\rho }^{2}}$ & $\cite{u1v}$ \\ 
$\smallskip \pi ^{+}\left( K^{+}\right) \rightarrow \gamma e^{+}\nu $ & $%
\frac{1}{M_{\rho }^{2}}\left( k^{2}+q^{2}\smallskip \right) $ & $\cite%
{semileploop}$ \\ 
$\smallskip \gamma \pi ^{0}\rightarrow \pi ^{+}\pi ^{-}$ & $\tfrac{1}{%
2M_{\rho }^{2}}\left( p_{01}^{2}+p_{02}^{2}+p_{12}^{2}+3k^{2}\right) $ & $%
\cite{vmd}$ \\ 
$\smallskip \eta _{8}\gamma \rightarrow \pi ^{+}\pi ^{-}$ & $\tfrac{3}{%
2M_{\rho }^{2}}\left( p_{12}^{2}+k^{2}\right) $ & $\cite{vmd}$ \\ 
$\smallskip K^{+}\pi ^{+}\pi ^{-}e^{+}\nu $ & $\tfrac{3}{4M_{\rho }^{2}}%
\left( 2q^{2}+\left( p_{+}+p_{-}\right) ^{2}+\left( q+p_{-}\right)
^{2}\right) $ & $\cite{semileploop}\smallskip $%
\end{tabular}%
\medskip $
\end{center}
\caption{
 \emph{VMD\ }$\mathcal{O}\left( p^{6}\right) $\emph{\
predictions.}}
\end{table}

The $M\rightarrow \gamma \gamma ^{\star }$ contains no kinematical
dependence for on-shell photons. For off-shell photons $\pi ^{0}$ and $\eta $
give the same result$:$%
\begin{equation*}
C_{22}^{Wr}=\frac{3}{64M_{\rho }^{2}\pi ^{2}}\simeq 8.01\times 10^{-9}\unit{%
MeV}^{-2}
\end{equation*}%
in excellent agreement with the result in table 3. $\pi ^{+}\left(
K^{+}\right) \gamma e^{+}\nu $ offers no new constraints. Two equations can
be extracted from the $\gamma 3\pi $ amplitude: 
\begin{eqnarray}
-C_{13}^{Wr}+C_{14}^{Wr}+C_{15}^{Wr} &=&\frac{12}{128M_{\rho }^{2}\pi ^{2}}
\label{g3pivmd} \\
C_{14}^{Wr}+C_{15}^{Wr} &=&-\frac{3}{128m_{\rho }^{2}\pi ^{2}}
\label{g32vmd}
\end{eqnarray}%
From $\eta \gamma \pi \pi $ we get:%
\begin{eqnarray}
2C_{15}^{Wr}-4C_{14}^{Wr}+C_{13}^{Wr} &=&\frac{3}{64M_{\rho }^{2}\pi ^{2}}
\label{eta1} \\
2C_{14}^{Wr}-C_{13}^{Wr} &=&\frac{3}{64M_{\rho }^{2}\pi ^{2}}  \label{eta2}
\end{eqnarray}%
Exactly the same two equations follow from $K_{e4}.$ There are three
independent equations since (\ref{eta1}) \& (\ref{eta2}) can be combined to
give (\ref{g3pivmd}). In fact, the situation is worse - (\ref{g32vmd})
originates in mass terms produced by kinematical factors. In the derived VMD
Lagrangian, approximations have been made removing some of these terms. This
means that (\ref{g32vmd}) is subject to corrections that will alter the
numerical value somewhat. Only (\ref{eta1}) \& (\ref{eta2}) are exact
predictions of the VMD model.

$3/64M_{\rho }^{2}\pi ^{2}\approx 8.\,\allowbreak 0\times 10^{-9}\allowbreak 
$ MeV$^{-2}$ and evaluating the lefthand side of (\ref{eta1}) \& (\ref{eta2}%
) using the values in table 3, we obtain $%
2C_{15}^{Wr}-4C_{14}^{Wr}+C_{13}^{Wr}\approx \left( -244.\,\allowbreak 7\pm
\allowbreak 148.\,\allowbreak 4\right) \times 10^{-9}$ MeV$^{-2}$ and $%
2C_{14}^{Wr}-C_{13}^{Wr}\approx \left( 134.\,\allowbreak 1\pm
78.\,\allowbreak 17\right) \times 10^{-9}$ MeV$^{-2}.$ Using the results of
extrapolation in table 4 gives $\left( -17.\allowbreak 52\pm 188.\allowbreak
3\right) $ $\times 10^{-9}$ MeV$^{-2}$ and $\left( 9.\,\allowbreak 88\pm
100.\allowbreak 5\right) $ $\times 10^{-9}$ MeV$^{-2}$ respectively - in
much better agreement with the VMD prediction.

Including (\ref{g32vmd}), we can solve for coefficients 13-15. The
coefficient values that can be extracted using VMD comparison and the
respective chiral predictions are displayed in table 6 in units of $10^{-9}$
MeV$^{-2}$.

\begin{table}
\begin{center}
$\smallskip 
\begin{tabular}{||l|l|l|l||}
\hline\hline
& $%
\begin{array}{l}
\mathbf{VMD}%
\end{array}%
$ & $%
\begin{array}{l}
\mathbf{ChPT\smallskip }%
\end{array}%
$ & $%
\begin{array}{l}
\mathbf{ChPT\smallskip }\text{ (extrapolated)}%
\end{array}%
$ \\ \hline
$C_{22}^{Wr}$ & $%
\begin{array}{l}
\\ 
\frac{3}{64M_{\rho }^{2}\pi ^{2}}\simeq 8.01\smallskip \smallskip \smallskip%
\end{array}%
$ & $%
\begin{array}{l}
\\ 
\left\{ 
\begin{array}{l}
6.52\pm 0.78 \\ 
5.07\pm 0.71%
\end{array}%
\right. \smallskip \smallskip \smallskip%
\end{array}%
$ &  \\ \hline
$C_{13}^{Wr}$ & $%
\begin{array}{l}
\\ 
-\frac{15}{128M_{\rho }^{2}\pi ^{2}}\simeq -20.0\smallskip \smallskip
\smallskip%
\end{array}%
$ & $%
\begin{array}{l}
\\ 
\smallskip \smallskip \smallskip -74.09\pm 55.89%
\end{array}%
$ & $%
\begin{array}{l}
\\ 
\smallskip \smallskip \smallskip -8.44\pm 69.9%
\end{array}%
$ \\ \hline
$C_{14}^{Wr}$ & $%
\begin{array}{l}
\\ 
-\tfrac{9}{2}\frac{1}{128M_{\rho }^{2}\pi ^{2}}\simeq -6.01%
\end{array}%
$ & $%
\begin{array}{l}
\\ 
29.99\pm 11.14\smallskip \smallskip \smallskip%
\end{array}%
$ & $%
\begin{array}{l}
\\ 
0.72\pm 15.3\smallskip \smallskip \smallskip%
\end{array}%
$ \\ \hline
$C_{15}^{Wr}$ & $%
\begin{array}{l}
\\ 
\tfrac{3}{2}\frac{1}{128M_{\rho }^{2}\pi ^{2}}\simeq 2.00\smallskip
\smallskip \smallskip%
\end{array}%
$ & $%
\begin{array}{l}
\\ 
-25.30\pm 23.93\smallskip \smallskip \smallskip%
\end{array}%
$ & $%
\begin{array}{l}
\\ 
-3.10\pm 28.6\smallskip \smallskip \smallskip%
\end{array}%
$ \\ \hline\hline
\end{tabular}%
$
\end{center}

\caption{
\emph{VMD \& chiral predictions in MeV}$^{-2}$.}
\end{table}

The values for $C_{13}^{Wr},C_{14}^{Wr}$ \& C$_{15}^{Wr}$ are not great, but
we have to remember that coefficients 13-14 also has a part proportional to
mass terms that VMD cannot predict. By contrast, $C_{22}^{Wr}$ is only
proportional to $k^{2}$ and there VMD does fine. The values obtained by
extrapolation are compatible with the VMD\ predictions.

Of course, the larger the momenta is in the experiment, the more dominant
the terms with coefficients connected to the kinematical dependence will be,
making the VMD parts increasingly significant.

The shortcomings of the VMD\ model is illustrated by the plots in figure 7,
showing the respective form factors normalized to 1 (with the exception of $%
K_{e4}$). The top left plot is for the $\pi ^{+}\gamma e^{+}\nu $ decay,
where the upper plane represents the VMD prediction and the lower one the
result from ChPT. Here the influence from the mass term proportional to $%
C_{7}^{Wr}$ is very small and $C_{22}$ dominates, so the VMD prediction is
well within the error limits.

In the top right plot (for the $K^{+}\gamma e^{+}\nu ),$ the effect of mass
terms is more dramatic. Both coefficients $C_{7}^{Wr}$ \& $C_{11}^{Wr}$ are
active and bring down the ChPT prediction considerably. The middle plane
shows the effect of removing these terms, aligning the ChPT with the VMD
prediction.

In the lower left graph (showing $\pi ^{0}\left( \eta \right) \gamma \gamma
^{\star })$ we see that the effect of mass terms is hardly noticeable for
the pion decay. Removing the $C_{7}^{Wr}$ term from the $\pi ^{+}\gamma
\gamma ^{\star }$ amplitude does not produce a visibly different result. In
the $\eta \gamma \gamma ^{\star }$ decay, both $C_{7}^{Wr}$ and $C_{8}^{Wr}$
are active, and become augmented by the eta mass. Removing these shifts the
curve considerably towards the VMD prediction. The error bars represent the
error due to the kinematical term. From looking at the graphs, we can
conclude that the pion processes $\pi ^{+}\gamma e\nu $ \& $\pi ^{0}\gamma
\gamma ^{\star }$ are well predicted by VMD, as they are relatively
unaffected by the mass terms. Similarly for $\gamma 3\pi $ \& $\eta \gamma
2\pi ,$ VMD does better when only pions are involved. This is because the
presence of the $\eta $ (or $K^{+}$ for that matter) induces a greater mass
term contribution.

The lower right graph is for the $K_{e4}$ process, where the chiral
prediction (solid line) has been extracted using equations from $K_{e4}$ \& $%
\gamma 3\pi .$ $K_{e4}$ produces 6 equations and is the major contributor to
the predicted form factor. Removing mass terms shifts the ChPT towards the
VMD prediction. The solutions in table 3, using all available data, does not
produce the same slope as VMD. The possible statistical fluke in the high
end of the spectrum is responsible for this, and can only be eliminated by
future experiments. Ignoring the fluke by extrapolation, produces a slope
similar to VMDs. And if we then remove the mass terms, we get almost perfect
correspondence.

\begin{figure}[t]
\begin{center}
\includegraphics[width=0.49\textwidth]{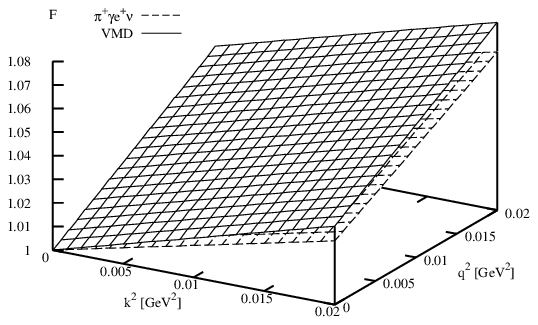}
\includegraphics[width=0.49\textwidth]{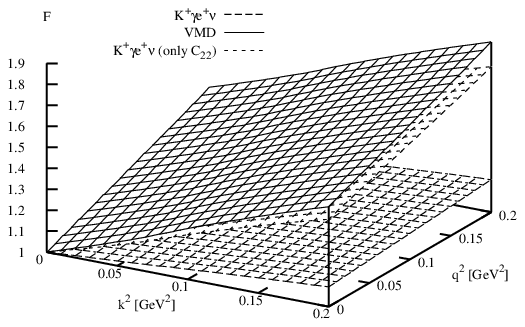}

\includegraphics[width=0.49\textwidth]{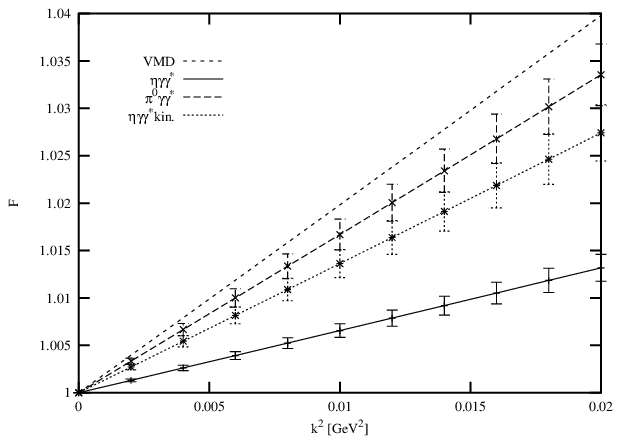}
\includegraphics[width=0.49\textwidth]{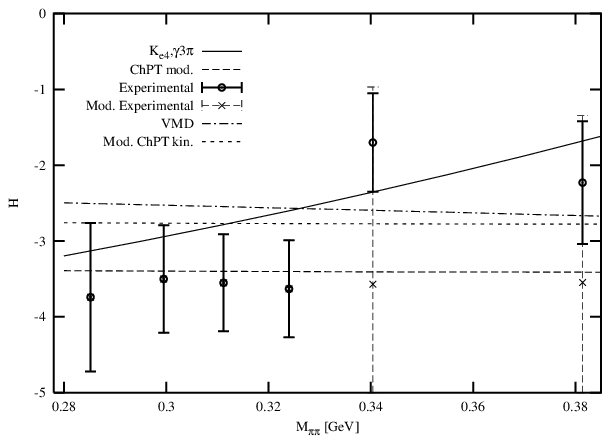}
\end{center}
\caption{
\emph{Form factors. }}
\end{figure}

\subsection{CQM Comparison}

With the Lagrangians of appendix B, the corresponding $\mathcal{O}\left(
p^{6}\right) $ results from the chiral constituent quark model can be
calculated and subsequently compared with the ChPT predictions of table 3.
First of all, we need to extract a value for the constituent quark mass - $%
m_{Q}.$ We proceed as in \cite{u1v}, making use of the $M\rightarrow \gamma
\gamma $ prediction and experimental slope parameters \cite{cleo}. Setting $%
R_{\mu }=L_{\mu }=eQA_{\mu }$ in $\Gamma ^{-}\left( a_{3}\right) ,$ then
performing the calculation and adding lowest order (WZW) amplitude gives:%
\begin{equation}
A_{M\rightarrow \gamma \gamma }^{CQM}=\frac{ie^{2}}{4\pi ^{2}F_{M}}%
C_{M}\varepsilon ^{\mu \nu \alpha \beta }\varepsilon _{\mu }k_{\nu
}\varepsilon _{\alpha }^{\star }k_{\beta }^{\star }\left( 1+\frac{k^{\star 2}%
}{12m_{Q}^{2}}\right)  \label{cqmmgg}
\end{equation}%
where $C_{\pi }=1$ \& $C_{\eta _{8}}=1/\sqrt{3}.$ Taking an average of the
pole parameters from \cite{cleo} and using (\ref{slope}) to calculate the
slope, we get%
\begin{eqnarray*}
\left\{ 
\begin{array}{c}
\Lambda _{\pi ^{0}}=776\pm 38\,\text{MeV} \\ 
\Lambda _{\eta }=774\pm 39\,\text{MeV}%
\end{array}%
\right. &\Rightarrow &\left\langle \Lambda \right\rangle =775\pm 39\,\text{%
MeV} \\
\frac{1}{12m_{Q}^{2}}=\frac{1}{\left\langle \Lambda \right\rangle ^{2}}
&\Rightarrow &m_{Q}=\frac{\left\langle \Lambda \right\rangle }{\sqrt{12}}%
\simeq 224\pm 12\,\text{MeV}
\end{eqnarray*}%
Comparing with the $\mathcal{O}\left( p^{6}\right) $ expression in table 5
gives $m_{Q}=m_{\rho }/\sqrt{12}\simeq 222$ MeV, in excellent agreement with
the empirically derived value. Since we are using the same experimental data
to fix $m_{Q}$ that we use to predict the chiral coefficients, we cannot
compare the two approaches in the $M\rightarrow \gamma \gamma $ case. Using $%
m_{Q}=m_{\rho }/\sqrt{12},$ will of course give the same $C_{22}^{Wr}$
prediction as VMD. Note that no mass parameters appear in (\ref{cqmmgg}).
Unfortunately, in deriving the CQM Lagrangian, some mass terms have been
approximated away, as they do not alter the end result notably. CQM is in
principle able give us predictions for all mass terms. We just have to
remember that the equations derived by comparing with these are subject to
small corrections.

Table 7 shows the CQM $\mathcal{O}\left( p^{6}\right) $ amplitudes, with the
same prefactors as in the ChPT case.

\begin{table}
\begin{center}
\begin{tabular}{lll}
\emph{Process:} & \emph{Amplitude}: & \emph{Lagrangian:} \\ 
$\pi ^{0}\left( \eta \right) \rightarrow \gamma \gamma ^{\star }$ & $\frac{%
k^{\star 2}}{12m_{Q}^{2}}$ & $\Gamma ^{-}\left( a_{3}\right) $ \\ 
$\pi ^{+}\rightarrow \gamma e^{+}\nu $ & $\frac{1}{48m_{Q}^{2}}(-m_{\pi
^{+}}^{2}+2q^{2}+2k^{2})$ & $\Gamma ^{-}\left( a_{3}\right) $ \\ 
$K^{+}\rightarrow \gamma e^{+}\nu $ & $\frac{1}{48m_{Q}^{2}}%
(-m_{K^{+}}^{2}+3q^{2}+3k^{2})$ & $\Gamma ^{-}\left( a_{3}\right) $ \\ 
$\pi ^{0}\rightarrow \gamma \pi ^{+}\pi ^{-}$ & $\frac{k^{2}}{6m_{Q}^{2}}$ & 
$\Gamma ^{-}\left( a_{4}\right) $ \\ 
$\eta \rightarrow \gamma \pi ^{+}\pi ^{-}$ & $\frac{1}{30m_{Q}^{2}}[8m_{\pi
}^{2}-2m_{\eta }^{2}+12p_{+}p_{-}+3k^{2}]$ & $\Gamma ^{-}\left( a_{4}\right) 
$ \\ 
$K^{+}\rightarrow \pi ^{+}\pi ^{-}e^{+}\nu $ & $\frac{1}{30m_{Q}^{2}}%
[m_{K}^{2}-m_{\pi }^{2}-6qp_{+}+3q^{2}]$ & $\Gamma ^{-}\left( a_{4}\right) $%
\end{tabular}%
\end{center}

\caption{
 \emph{CQM }$\mathcal{O}\left( p^{6}\right) $\emph{\
amplitudes}.}
\end{table}

The $\pi ^{+}$ and the $K^{+}$ decays give the same (exact) prediction for $%
C_{22}^{Wr}=1/512m_{Q}^{2}\pi ^{2}\simeq 3.94\times 10^{-9}\unit{MeV}^{-2}.$
It seems that the VMD prediction overestimates the $C_{22}^{Wr}$ value,
whereas CQM underestimates it by roughly the same amount. Both models do
equally well with respect to ChPT.

Comparing the mass terms we get the (inexact) value $C_{7}^{Wr}\simeq
5.10\times 10^{-10}\unit{MeV}^{-2}$, from the $\pi ^{+}$ decay. This can
then be used in the $K^{+}$ amplitude to get $C_{11}^{Wr}\simeq -1.44\times
10^{-12}\unit{MeV}^{-2},$ which is a bit to small to be taken seriously.

Comparing kinematical terms, $\eta \gamma \pi \pi $ \& $K_{e4}$ both give
the same two equations.

\begin{eqnarray}
2C_{14}^{Wr}-C_{13}^{Wr} &=&\frac{2}{5}\frac{1}{128m_{Q}^{2}\pi ^{2}}\simeq
6.\,\allowbreak 31\times 10^{-9}\,\text{MeV}^{-2}  \label{c1} \\
2C_{15}^{Wr}-4C_{14}^{Wr}+C_{13}^{Wr} &=&\frac{4}{5}\frac{1}{128m_{Q}^{2}\pi
^{2}}\simeq 1.\,\allowbreak 26\times 10^{-8}\,\text{MeV}^{-2}  \label{c2}
\end{eqnarray}%
$\gamma 3\pi $ gives one equation that offers no additional constraints.
Equations (\ref{c1}) \& (\ref{c2}) are of the exact the same form as the VMD
predictions (\ref{eta1}) \& (\ref{eta2}), and the predicted value of the RH
side lies in between ($8.0\times 10^{-9}$ MeV$^{-2}$) those of CQM.

Turning to the mass terms, we get one equation from $\gamma 3\pi ,$ which
can be reconstructed with the two equations from $\eta \gamma \pi \pi .$
These are:%
\begin{eqnarray}
-6C_{36}^{Wr}+2C_{14}^{Wr}+C_{13}^{Wr} &=&\frac{4}{15}\frac{1}{%
128m_{Q}^{2}\pi ^{2}}  \label{c3} \\
-3C_{36}^{Wr}+C_{5}^{Wr}+2C_{4}^{Wr}-2C_{14}^{Wr}+C_{13}^{Wr} &=&\frac{8}{15}%
\frac{1}{128m_{Q}^{2}\pi ^{2}}  \label{c4}
\end{eqnarray}%
with $C_{36}^{Wr}\equiv C_{3}^{Wr}-C_{6}^{Wr}.$ Two additional equations
follow from comparing mass terms in $K_{e4}:$%
\begin{eqnarray}
C_{2}^{Wr}+C_{4}^{Wr}-C_{14}^{Wr} &=&\frac{1}{15}\frac{1}{128\pi
^{2}m_{Q}^{2}}  \label{c5} \\
-C_{2}^{Wr}+C_{4}^{Wr}+C_{5}^{Wr}-C_{15}^{Wr} &=&-\frac{1}{15}\frac{1}{%
128\pi ^{2}m_{Q}^{2}}  \label{c6}
\end{eqnarray}%
If we wish to solve for the chiral coefficients using the (inexact)
predictions from the mass terms (eq. (\ref{c3})-(\ref{c6})), we must make
additional assumptions. One approach is to parametrize the solution of the
system of equations (\ref{c1})-(\ref{c6}) in terms of (for example) $%
C_{15}^{Wr}$ \& $C_{4}^{Wr}$, and then vary it freely with the constraint
that the solution should stay as close to the ChPT result as possible, given
the error limits. Table 8 displays the results of applying this procedure on
the full ChPT and extrapolated ChPT\ result, respectively. Looking at figure
8, showing the ChPT along with CQM and VMD predicted coefficients in units
of $10^{-9}$ MeV$^{-2},$ we see that the best result is obtained by
extrapolation$.$The quoted CQM errors arise from the constraints in the
fitting process and from the constituent quark mass error, and are not
intrinsic to the model.

\begin{table}
\begin{center}
\begin{tabular}{||l|c|c||}
\hline\hline
$%
\begin{array}{l}
\emph{Process:}%
\end{array}%
$ & $%
\begin{array}{l}
\mathbf{CQM}%
\end{array}%
$ & $%
\begin{array}{l}
\mathbf{CQM\ }\text{(extrapolation)}%
\end{array}%
$ \\ \hline
$\pi ^{+}\rightarrow \gamma e^{+}\nu $ & \multicolumn{1}{|l|}{$%
\begin{array}{l}
\\ 
C_{7}^{Wr}\simeq 0.51\pm 0.06\smallskip \\ 
C_{22}^{Wr}\simeq 3.94\pm 0.43\smallskip \smallskip \smallskip%
\end{array}%
$} & \multicolumn{1}{|l||}{} \\ \hline
$K^{+}\rightarrow \gamma e^{+}\nu $ & \multicolumn{1}{|l|}{$%
\begin{array}{l}
\\ 
C_{11}^{Wr}\simeq -0.00143\pm 0.03\smallskip \\ 
C_{22}^{Wr}\simeq 3.94\pm 0.43\smallskip \smallskip \smallskip%
\end{array}%
$} & \multicolumn{1}{|l||}{} \\ \hline
$\gamma 3\pi ,$ $\eta \gamma \pi \pi $ \& $K_{e4}$ & \multicolumn{1}{|l|}{$%
\begin{array}{l}
\\ 
C_{2}^{Wr}\simeq 4.96\pm 9.70\smallskip \\ 
C_{36}^{Wr}\simeq 5.07\pm 5.07\smallskip \\ 
C_{4}^{Wr}\simeq 6.32\pm 6.09\smallskip \\ 
C_{5}^{Wr}\simeq 33.05\pm 28.66\smallskip \\ 
C_{13}^{Wr}\simeq 14.15\pm 15.22\smallskip \\ 
C_{14}^{Wr}\simeq 10.23\pm 7.56\smallskip \\ 
C_{15}^{Wr}\simeq 19.70\pm 7.49\smallskip \smallskip \smallskip%
\end{array}%
$} & \multicolumn{1}{|l||}{$%
\begin{array}{l}
\\ 
C_{2}^{Wr}\simeq -0.074\pm 13.3\smallskip \\ 
C_{36}^{Wr}\simeq -2.14\pm 6.54\smallskip \\ 
C_{4}^{Wr}\simeq -0.55\pm 9.05\smallskip \\ 
C_{5}^{Wr}\simeq 34.51\pm 41.13\smallskip \\ 
C_{13}^{Wr}\simeq -7.46\pm 19.62\smallskip \\ 
C_{14}^{Wr}\simeq -0.58\pm 9.77\smallskip \\ 
C_{15}^{Wr}\simeq 8.89\pm 9.72\smallskip \smallskip \smallskip%
\end{array}%
$} \\ \hline\hline
\end{tabular}%
\smallskip
\end{center}
\caption{
\emph{CQM \& chiral predictions in MeV}$^{-2}$\emph{.}}
\end{table}

\begin{figure}
\begin{center}
\includegraphics{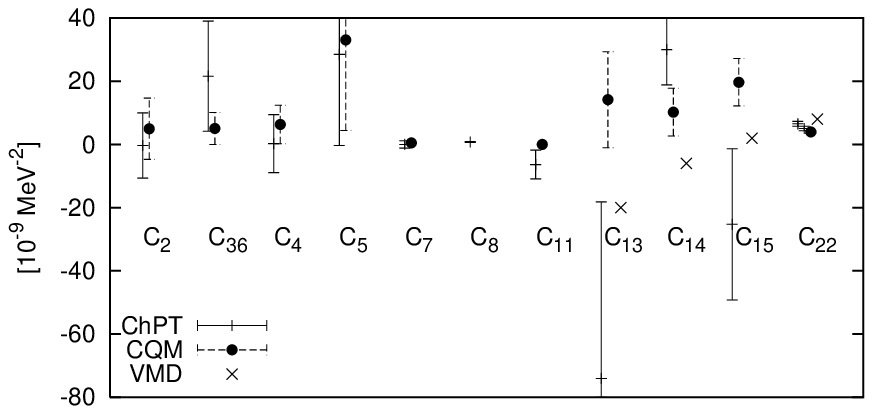}

\includegraphics{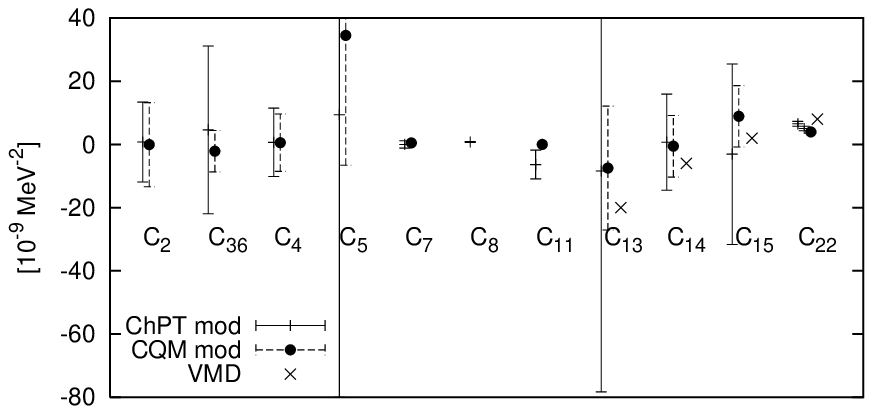}
\end{center}
\caption{
\emph{Chiral coefficients (bottom: using extrapolation)}.}
\end{figure}

\noindent Figure 9 shows the chiral and CQM form factor predictions for $\pi
^{+}\gamma e^{+}\nu $ (top left), $K^{+}\gamma e^{+}\nu $ (top right), $%
\gamma 3\pi $ (bottom left) and $K_{e4}$ (bottom right). For $\pi ^{+}\left(
K^{+}\right) \gamma e^{+}\nu $ CQM does slightly better than VMD, but fails
to predict the mass terms in the $K^{+}\gamma e^{+}\nu $ case. In the $%
\gamma 3\pi $ process, CQM \& VMD fit well with the coefficients obtained by
extrapolating, less so with the full data set. In the $K_{e4}$, both VMD \&
CQM predict roughly the same form factor.

\begin{figure}[t]
\begin{center}
\includegraphics[width=0.49\textwidth]{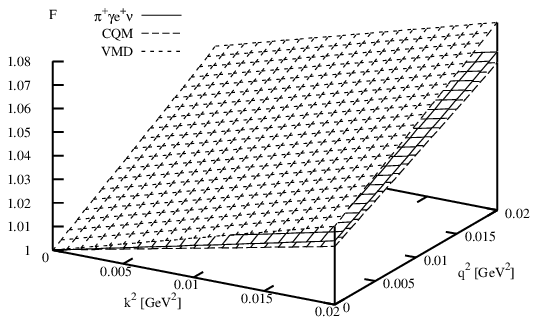}
\includegraphics[width=0.49\textwidth]{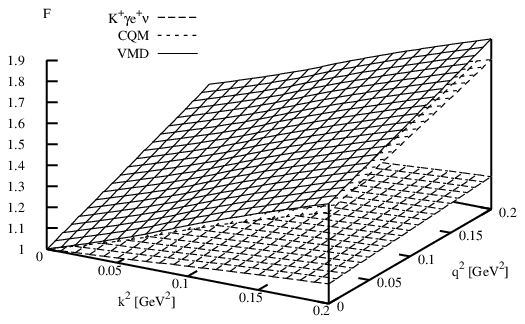}

\includegraphics[width=0.49\textwidth]{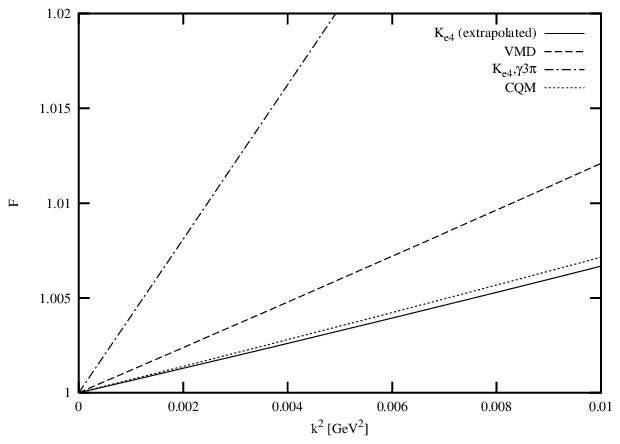}
\includegraphics[width=0.49\textwidth]{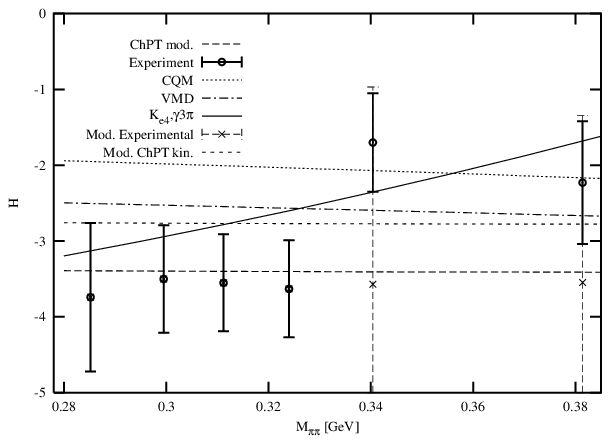}
\end{center}
\caption{
 \emph{Form factors.}}
\end{figure}

\section{Conclusions \& Outlook}

To gain insight into the physics leading to the chiral Lagrangian, it is
vital to chart the anomalous territory. Ultimately, we hope to arrive at a
complete standard model derivation, capable of predicting all the chiral
parameters. Going through the ChPT theory up to the anomalous $\mathcal{O}%
\left( p^{6}\right) $ Lagrangian$,$ we are faced with a large number of
coefficients, that can be empirically determined through comparison with
available experimental data. As much information as possible is extracted
using the amplitudes calculated for a number of different anomalous
processes. The extracted values then allow us to test the predictions of two
other models for the $C_{i}^{Wr}s$ : the vector meson dominance and the
chiral constituent quark model.

VMD does very well in predicting the coefficients that are connected to
kinematical factors, but fails to fully predict those that do not vanish in
the low energy limit. Here the CQM does better, and gives reasonable
predictions for all coefficients.

To improve the accuracy in gauging the validity of these models, and to
reduce the errors of the $C_{i}^{Wr}s$ further, more detailed experiments
must be made, probing the anomalous sector. Many such experiments are in the
planning stages today, e.g. the Primex precision $\pi ^{0}$ lifetime
measurements at Jefferson laboratories. 

\bigskip

\section*{Acknowledgements}

Johan Bijnens - \emph{Thank you} Hans for being the best tutor I could
possibly hope for. I would also like to thank all other experimental and
theoretical physicists, upon who's hard work this thesis is based on. And of
course my mother, for her invaluable moral support.


\appendix

\section{\emph{One-loop Corrections}}

\begin{longtable}{||l|l||}
\hline\hline
\emph{Process:} & \emph{Amplitude:} \\ \hline
$%
\begin{array}{l}
\\ 
\pi ^{0}\rightarrow \gamma e^{+}e^{-}\smallskip \\ 
\smallskip \smallskip \smallskip%
\end{array}%
$ & $%
\begin{array}{l}
\\ 
\frac{1}{4\pi ^{2}}\frac{e^{3}}{F_{\pi }}\varepsilon ^{\mu \nu \alpha \beta
}\varepsilon _{\mu }k_{\nu }\frac{\bar{e}\gamma _{\alpha }^{\star }e}{%
k^{\star 2}}k_{\beta }^{\star }
\,\frac{1}{32\pi ^{2}F^{2}}
\times \smallskip \\ 
\{-\tfrac{1}{3}k^{\star 2}(\log \frac{m_{K}^{2}}{%
\mu ^{2}}+\log \frac{m_{\pi }^{2}}{\mu ^{2}})+\tfrac{10}{9}k^{\star 2}+%
\tfrac{4}{3}[F\left( k^{\star 2},m_{\pi }^{2}\right) +F\left( k^{\star
2},m_{K}^{2}\right) ]\}\smallskip \smallskip \smallskip%
\end{array}%
$\hskip-100pt\\ \hline
$%
\begin{array}{l}
\\ 
\eta \rightarrow \gamma e^{+}e^{-}\smallskip \\ 
\smallskip \smallskip \smallskip%
\end{array}%
$ & $%
\begin{array}{l}
\\ 
\frac{e^{3}}{4\sqrt{3}\pi ^{2}F_{\eta }}\varepsilon ^{\mu \nu \alpha \beta
}\varepsilon _{\mu }k_{\nu }\frac{\bar{e}\gamma _{\alpha }e}{k^{\star 2}}%
k_{\beta }^{\star }
\,\tfrac{1}{32\pi ^{2}F^{2}}
\times \smallskip \\ 
\{-\tfrac{1}{3}k^{\star 2}(\log \tfrac{m_{K}^{2}}{%
\mu ^{2}}+\log \tfrac{m_{\pi }^{2}}{\mu ^{2}})+\tfrac{10}{9}k^{\star 2}+%
\tfrac{4}{3}[F\left( k^{\star 2},m_{\pi }^{2}\right) +F\left( k^{\star
2},m_{K}^{2}\right) ]\}\smallskip \smallskip \smallskip%
\end{array}%
$\hskip-100pt\\ \hline
$%
\begin{array}{l}
\\ 
\pi ^{+}\rightarrow \gamma e^{+}\nu \smallskip \\ 
\smallskip \smallskip \smallskip%
\end{array}%
$ & $%
\begin{array}{l}
\\ 
\frac{eG_{F}\cos \theta }{8\pi ^{2}F_{\pi }}\varepsilon ^{\mu \nu \alpha
\beta }l_{\mu }q_{\nu }\varepsilon _{\alpha }k_{\beta }
\,\frac{1}{32\pi ^{2}F^{2}}
\times \smallskip \\ 
\{-4m_{\pi }^{2}\text{ln}\frac{m_{\pi }^{2}}{\mu
^{2}}-4m_{K}^{2}\text{ln}\frac{m_{K}^{2}}{\mu ^{2}}+4I\left( q^{2},m_{\pi
}^{2},m_{\pi }^{2}\right) +4I\left( k^{2},m_{K}^{2},m_{K}^{2}\right)
\}\smallskip \smallskip \smallskip%
\end{array}%
$\hskip-100pt\\ \hline
$%
\begin{array}{l}
\\ 
K^{+}\rightarrow \gamma e^{+}\nu \smallskip \\ 
\smallskip \\ 
\smallskip \smallskip \smallskip%
\end{array}%
$ & $%
\begin{array}{l}
\\ 
\frac{eG_{F}\sin \theta }{8\pi ^{2}F_{K}}\varepsilon ^{\mu \nu \alpha \beta
}l_{\mu }q_{\alpha }\varepsilon _{\alpha }k_{\beta }\times \smallskip \\ 
\frac{1}{32\pi ^{2}F^{2}}\{-\tfrac{7}{2}m_{\pi }^{2}\text{ln}\frac{m_{\pi
}^{2}}{\mu ^{2}}-3m_{K}^{2}\text{ln}\frac{m_{K}^{2}}{\mu ^{2}}-\tfrac{3}{2}%
m_{\eta }^{2}\text{ln}\frac{m_{\eta }^{2}}{\mu ^{2}}\smallskip \\ 
+4I\left( k^{2},m_{\pi }^{2},m_{\pi }^{2}\right) +2I\left(
q^{2},m_{K}^{2},m_{\pi }^{2}\right) +2I\left( q^{2},m_{K}^{2},m_{\eta
}^{2}\right) \}\smallskip \smallskip \smallskip%
\end{array}%
$ \\ \hline
$%
\begin{array}{l}
\\ 
\gamma \pi ^{0}\pi ^{+}\pi ^{-}\smallskip \\ 
\smallskip \\ 
\smallskip \smallskip \smallskip%
\end{array}%
$ & $%
\begin{array}{l}
\\ 
i\frac{1}{4\pi ^{2}}\frac{e}{F_{\pi }^{3}}\varepsilon ^{\mu \nu \alpha \beta
}\varepsilon _{\mu }p_{\nu }p_{\alpha }p_{\beta }\times \smallskip \\ 
\frac{1}{96\pi ^{2}F^{2}}\{-\left( p_{01}^{2}+p_{02}^{2}+p_{12}^{2}\right)
\log \frac{m_{\pi }^{2}}{\mu ^{2}}+\tfrac{5}{3}\left(
p_{01}^{2}+p_{02}^{2}+p_{12}^{2}\right) \smallskip \\ 
+4[F\left( m_{\pi }^{2},p_{01}^{2}\right) +F\left( m_{\pi
}^{2},p_{02}^{2}\right) +F\left( m_{\pi }^{2},p_{12}^{2}\right) ]\smallskip
\smallskip \smallskip%
\end{array}%
$ \\ \hline
$%
\begin{array}{l}
\\ 
\gamma \eta _{8}\pi ^{+}\pi ^{-}\smallskip \\ 
\smallskip \\ 
\smallskip \smallskip \smallskip%
\end{array}%
$ & $%
\begin{array}{l}
\\ 
-i\frac{1}{4\pi ^{2}\sqrt{3}}\frac{e}{F_{\pi }^{2}F_{\eta _{8}}}\varepsilon
^{\mu \nu \alpha \beta }\varepsilon _{\mu }p_{\nu }^{\eta }p_{\alpha }^{\pi
^{+}}p_{\beta }^{\pi ^{-}}\times \smallskip \\ 
\frac{1}{32\pi ^{2}F^{2}}\{-\left( 4m_{\pi }^{2}+\tfrac{1}{3}%
p_{12}^{2}\right) \log \frac{m_{\pi }^{2}}{\mu ^{2}}+\left( 4m_{K}^{2}-%
\tfrac{2}{3}p_{12}^{2}\right) \log \frac{m_{K}^{2}}{\mu ^{2}}\smallskip \\ 
+\tfrac{5}{3}p_{12}^{2}+\tfrac{4}{3}F\left( m_{\pi }^{2},p_{12}^{2}\right) +%
\tfrac{8}{3}F\left( m_{K}^{2},p_{12}^{2}\right) \}\smallskip \smallskip
\smallskip%
\end{array}%
$ \\ \hline
$%
\begin{array}{l}
\\ 
K^{+}\rightarrow \pi ^{+}\pi ^{-}e^{+}\nu \smallskip \\ 
\smallskip \\ 
\smallskip \\ 
\smallskip \\ 
\smallskip \\ 
\smallskip \smallskip \smallskip%
\end{array}%
$ & $%
\begin{array}{l}
\\ 
\frac{G_{F}\sin \theta }{4\pi ^{2}F_{\pi }^{2}F_{K}}\varepsilon ^{\mu \nu
\alpha \beta }l_{\mu }q_{\nu }p_{\alpha }^{+}p_{\beta }^{-}\times \smallskip
\\ 
\lbrack \frac{1}{32\pi ^{2}F^{2}}\{-\tfrac{11}{2}m_{\pi }^{2}\text{ln}\frac{%
m_{\pi }^{2}}{\mu ^{2}}-5m_{K}^{2}\text{ln}\frac{m_{K}^{2}}{\mu ^{2}}-\tfrac{%
3}{2}m_{\eta }^{2}\text{ln}\frac{m_{\eta }^{2}}{\mu ^{2}}\smallskip \\ 
+2I\left( \left( p^{+}+p^{-}\right) ^{2},m_{\pi }^{2},m_{\pi }^{2}\right)
+I\left( \left( p^{+}+p^{-}\right) ^{2},m_{K}^{2},m_{K}^{2}\right) \smallskip
\\ 
+2I\left( \left( p^{-}+q\right) ^{2},m_{K}^{2},m_{\pi }^{2}\right) +I\left(
\left( p^{-}+q\right) ^{2},m_{K}^{2},m_{\eta }^{2}\right) \smallskip \\ 
+3I\left( q^{2},m_{K}^{2},m_{\pi }^{2}\right) +3I\left(
q^{2},m_{K}^{2},m_{\eta }^{2}\right) \}]\smallskip \smallskip \smallskip%
\end{array}%
$ \\ \hline
$%
\begin{array}{l}
\\ 
K^{+}\rightarrow \pi ^{0}\pi ^{0}e^{+}\nu \smallskip \\ 
\smallskip \\ 
\smallskip \smallskip \smallskip%
\end{array}%
$ & $%
\begin{array}{l}
\\ 
\frac{G_{F}\sin \theta }{4\pi ^{2}F_{K}F_{\pi }^{2}}\varepsilon ^{\mu \nu
\alpha \beta }l_{\mu }q_{\nu }p_{\alpha }p_{\beta }^{\prime }\times
\smallskip \\ 
\lbrack \frac{1}{32\pi ^{2}F^{2}}\{-I\left( \left( q+p\right)
^{2},m_{K}^{2},m_{\pi }^{2}\right) -\tfrac{1}{2}I\left( \left( q+p\right)
^{2},m_{K}^{2},m_{\eta }^{2}\right) \smallskip \\ 
+I\left( \left( q+p^{\prime }\right) ^{2},m_{K}^{2},m_{\pi }^{2}\right) +%
\tfrac{1}{2}I\left( \left( q+p^{\prime }\right) ^{2},m_{K}^{2},m_{\eta
}^{2}\right) ]\smallskip \smallskip \smallskip%
\end{array}%
$ \\ \hline\hline
\end{longtable}
where

\begin{eqnarray}
F\left( m^{2},x\right) &\equiv &m^{2}\left( 1-\frac{x}{4}\right) \sqrt{\frac{%
x-4}{x}}\log \frac{\sqrt{x}-\sqrt{x-4}}{-\sqrt{x}+\sqrt{x-4}}-2m^{2}\quad
;x\equiv \frac{k^{\star 2}}{m^{2}}
\nonumber \\
I\left( k^{2},m_{1}^{2},m_{2}^{2}\right) &\equiv &\int_{0}^{1}dx\left[
m_{1}^{2}-\left( m_{1}^{2}-m_{2}^{2}\right) x-x\left( 1-x\right) k^{2}\right]
\times
\nonumber\\&&
\log \frac{m_{1}^{2}-\left( m_{1}^{2}-m_{2}^{2}\right) x-x\left( 1-x\right)
k^{2}}{\mu ^{2}}
\nonumber
\end{eqnarray}
\begin{eqnarray}
&=&16\pi ^{2}\{\frac{m_{1}^{2}-m_{2}^{2}+k^{2}}{6k^{2}}iA\left(
m_{1}^{2}\right) +\frac{-m_{1}^{2}+m_{2}^{2}+k^{2}}{6k^{2}}iA\left(
m_{2}^{2}\right) 
\nonumber\\
&&-\frac{\left( m_{1}^{2}-m_{2}^{2}-k^{2}\right) ^{2}-4k^{2}m_{2}^{2}}{6k^{2}%
}iB\left( k^{2},m_{1}^{2},m_{2}^{2}\right) -\tfrac{1}{3}\left(
m_{1}^{2}+m_{2}^{2}\right) +\tfrac{1}{9}k^{2}\} \nonumber\\
iA\left( m^{2}\right) &\equiv &\frac{m^{2}}{16\pi ^{2}}\ln \frac{m^{2}}{\mu
^{2}}\nonumber\\
iB\left( k^{2},m_{1}^{2},m_{2}^{2}\right) &\equiv &-\frac{1}{16\pi ^{2}}%
\left( 1-\tfrac{1}{2}\log \frac{m_{1}^{2}m_{2}^{2}}{\mu ^{4}}+\frac{%
m_{2}^{2}-m_{1}^{2}}{2k^{2}}\log \frac{m_{1}^{2}}{m_{2}^{2}}-\frac{1}{k^{2}}%
u_{+}u_{-}\log \frac{u_{+}+u_{-}}{u_{+}-u_{-}}\right) \nonumber\\
u_{\pm } &\equiv &\sqrt{k^{2}-\left( m_{1}^{2}\pm m_{2}^{2}\right) ^{2}}
\nonumber
\end{eqnarray}


\section{ \emph{CQM Lagrangians}}

\emph{5 pseudoscalars}:%
\begin{eqnarray*}
\Gamma ^{-}\left( a_{5}\right) &=&-\frac{N_{c}}{32\pi ^{2}}\int d^{4}x\frac{1%
}{30}\varepsilon _{\mu \nu \alpha \beta } \\
&&\times \text{\textrm{t}}\mathrm{\mathrm{r}}[\left( \Sigma D_{\gamma
}D_{\mu }\Sigma ^{\dagger }-D_{\gamma }D_{\mu }\Sigma \Sigma ^{\dagger
}\right) D_{\nu }(D_{\alpha }\Sigma ^{\dagger }D_{\gamma }\Sigma -D_{\gamma
}\Sigma ^{\dagger }D_{\alpha }\Sigma )D_{\beta }\Sigma ^{\dagger }]
\end{eqnarray*}

\emph{1 external vector field \& 3 pseudoscalars}: 
\begin{eqnarray*}
\Gamma ^{-}\left( a_{4}\right) &=&\frac{N_{c}}{32\pi ^{2}m_{Q}^{2}}\int
d^{4}x\frac{i}{180}\varepsilon _{\mu \nu \alpha \beta } \\
&&\text{t}\mathrm{\mathrm{r}}\{2\left( D_{\gamma }r_{\gamma \mu }+\Sigma
^{\dagger }D_{\gamma }\ell _{\gamma \mu }\Sigma \right) D_{\nu }\Sigma
^{\dagger }D_{\alpha }\Sigma D_{\beta }\Sigma ^{\dagger }\Sigma \\
&&-3\left( D_{\gamma }r_{\mu \nu }+\Sigma ^{\dagger }D_{\gamma }\ell _{\mu
\nu }\Sigma \right) D_{\alpha }\Sigma ^{\dagger }D_{\gamma }\Sigma D_{\beta
}\Sigma ^{\dagger }\Sigma \\
&&+r_{\mu \gamma }[20D_{\nu }\Sigma ^{\dagger }D_{\alpha }\Sigma \Sigma
^{\dagger }D_{\gamma }D_{\beta }\Sigma -20D_{\gamma }D_{\nu }\Sigma
^{\dagger }\Sigma D_{\alpha }\Sigma ^{\dagger }D_{\beta }\Sigma \\
&&-2D_{\nu }\Sigma ^{\dagger }D_{\alpha }\Sigma D_{\gamma }D_{\beta }\Sigma
^{\dagger }\Sigma +2\Sigma ^{\dagger }D_{\gamma }D_{\nu }\Sigma D_{\alpha
}\Sigma ^{\dagger }D_{\beta }\Sigma \\
&&-8D_{\nu }\Sigma ^{\dagger }\left( \Sigma D_{\gamma }D_{\alpha }\Sigma
^{\dagger }-D_{\gamma }D_{\alpha }\Sigma \Sigma ^{\dagger }\right) D_{\beta
}\Sigma ] \\
&&-\ell _{\mu \gamma }[20D_{\nu }\Sigma D_{\alpha }\Sigma ^{\dagger }\Sigma
D_{\gamma }D_{\beta }\Sigma ^{\dagger }-20D_{\gamma }D_{\nu }\Sigma \Sigma
^{\dagger }D_{\alpha }\Sigma D_{\beta }\Sigma ^{\dagger } \\
&&-2D_{\nu }\Sigma D_{\alpha }\Sigma ^{\dagger }D_{\gamma }D_{\beta }\Sigma
\Sigma ^{\dagger }+2\Sigma D_{\gamma }D_{\nu }\Sigma ^{\dagger }D_{\alpha
}\Sigma D_{\beta }\Sigma ^{\dagger } \\
&&-8D_{\nu }\Sigma \left( \Sigma ^{\dagger }D_{\gamma }D_{\alpha }\Sigma
-D_{\gamma }D_{\alpha }\Sigma ^{\dagger }\Sigma \right) D_{\beta }\Sigma
^{\dagger }] \\
&&+3r_{\mu \nu }[D_{\gamma }D_{\alpha }\Sigma ^{\dagger }\Sigma D_{\beta
}\Sigma ^{\dagger }D_{\gamma }\Sigma +D_{\alpha }\Sigma ^{\dagger }D_{\gamma
}D_{\beta }\Sigma \Sigma ^{\dagger }D_{\gamma }\Sigma \\
&&-D_{\gamma }\Sigma ^{\dagger }\Sigma D_{\gamma }D_{\alpha }\Sigma
^{\dagger }D_{\beta }\Sigma -D_{\gamma }\Sigma ^{\dagger }D_{\alpha }\Sigma
\Sigma ^{\dagger }D_{\gamma }D_{\beta }\Sigma ] \\
&&-3\ell _{\mu \nu }[D_{\gamma }D_{\alpha }\Sigma \Sigma ^{\dagger }D_{\beta
}\Sigma D_{\gamma }\Sigma ^{\dagger }+D_{\alpha }\Sigma D_{\gamma }D_{\beta
}\Sigma ^{\dagger }\Sigma D_{\gamma }\Sigma ^{\dagger } \\
&&-D_{\gamma }\Sigma \Sigma ^{\dagger }D_{\gamma }D_{\alpha }\Sigma D_{\beta
}\Sigma ^{\dagger }-D_{\gamma }\Sigma D_{\alpha }\Sigma ^{\dagger }\Sigma
D_{\gamma }D_{\beta }\Sigma ^{\dagger }] \\
&&-2D^{2}D_{\mu }\Sigma \{r_{\nu \alpha }\Sigma ^{\dagger }-\Sigma ^{\dagger
}\ell _{\nu \alpha },\Sigma D_{\beta }\Sigma ^{\dagger }\} \\
&&+2D^{2}D_{\mu }\Sigma ^{\dagger }\{\ell _{\nu \alpha }\Sigma -\Sigma
r_{\nu \alpha },\Sigma ^{\dagger }D_{\beta }\Sigma \} \\
&&+i2D_{\mu }D_{\gamma }D_{\nu }\Sigma \lbrack 2D_{\gamma }D_{\alpha }\Sigma
^{\dagger }\Sigma D_{\beta }\Sigma ^{\dagger }+2D_{\beta }\Sigma ^{\dagger
}\Sigma D_{\gamma }D_{\alpha }\Sigma ^{\dagger } \\
&&+\Sigma ^{\dagger }D_{\gamma }D_{\alpha }\Sigma D_{\beta }\Sigma ^{\dagger
}+D_{\beta }\Sigma ^{\dagger }D_{\gamma }D_{\alpha }\Sigma \Sigma ^{\dagger
}] \\
&&-i2D_{\mu }D_{\gamma }D_{\nu }\Sigma ^{\dagger }[2D_{\gamma }D_{\alpha
}\Sigma \Sigma ^{\dagger }D_{\beta }\Sigma +2D_{\beta }\Sigma \Sigma
^{\dagger }D_{\gamma }D_{\alpha }\Sigma \\
&&+\Sigma D_{\gamma }D_{\alpha }\Sigma ^{\dagger }D_{\beta }\Sigma +D_{\beta
}\Sigma D_{\gamma }D_{\alpha }\Sigma ^{\dagger }\Sigma ]\}
\end{eqnarray*}

\emph{2 external vector fields \& 1 pseudoscalar}:%
\begin{eqnarray*}
\Gamma ^{-}\left( a_{3}\right) &=&-\frac{N_{c}}{32\pi ^{2}m_{Q}^{2}}\int
d^{4}x\frac{1}{120}\varepsilon _{\mu \nu \alpha \beta } \\
&&\text{\textrm{t}}\mathrm{r}\{7\{D_{\gamma }r_{\gamma \mu },r_{\alpha \beta
}\}D_{\nu }\Sigma ^{\dagger }\Sigma \\
&&-7\{D_{\gamma }\ell _{\gamma \mu },\ell _{\alpha \beta }\}D_{\nu }\Sigma
\Sigma ^{\dagger } \\
&&+2D_{\gamma }r_{\gamma \mu }\left( \Sigma ^{\dagger }\ell _{\alpha \beta
}D_{\nu }\Sigma -D_{\nu }\Sigma ^{\dagger }\ell _{\alpha \beta }\Sigma
\right) \\
&&-2D_{\gamma }\ell _{\gamma \mu }\left( \Sigma r_{\alpha \beta }D_{\nu
}\Sigma ^{\dagger }-D_{\nu }\Sigma r_{\alpha \beta }\Sigma ^{\dagger }\right)
\\
&&+3\{r_{\mu \gamma },r_{\alpha \beta }\}\left( \Sigma ^{\dagger }D_{\gamma
}D_{\nu }\Sigma -D_{\gamma }D_{\nu }\Sigma ^{\dagger }\Sigma \right) \\
&&-3\{\ell _{\mu \gamma },\ell _{\alpha \beta }\}\left( \Sigma D_{\gamma
}D_{\nu }\Sigma ^{\dagger }-D_{\gamma }D_{\nu }\Sigma \Sigma ^{\dagger
}\right) \\
&&+4\ell _{\mu \gamma }\left( D_{\gamma }D_{\nu }\Sigma r_{\alpha \beta
}\Sigma ^{\dagger }-\Sigma r_{\alpha \beta }D_{\gamma }D_{\nu }\Sigma
^{\dagger }\right) \\
&&-4r_{\mu \gamma }\left( D_{\gamma }D_{\nu }\Sigma ^{\dagger }\ell _{\alpha
\beta }\Sigma -\Sigma ^{\dagger }\ell _{\alpha \beta }D_{\gamma }D_{\nu
}\Sigma \right) \\
&&+2\ell _{\mu \gamma }\left( D_{\nu }\Sigma D_{\gamma }r_{\alpha \beta
}\Sigma ^{\dagger }-\Sigma D_{\gamma }r_{\alpha \beta }D_{\nu }\Sigma
^{\dagger }\right) \\
&&-2r_{\mu \gamma }\left( D_{\nu }\Sigma ^{\dagger }D_{\gamma }\ell _{\alpha
\beta }\Sigma -\Sigma ^{\dagger }D_{\gamma }\ell _{\alpha \beta }D_{\nu
}\Sigma \right) \\
&&-13\{r_{\mu \gamma },D_{\gamma }r_{\alpha \beta }\}D_{\nu }\Sigma
^{\dagger }\Sigma \\
&&+13\{\ell _{\mu \gamma },D_{\gamma }\ell _{\alpha \beta }\}D_{\nu }\Sigma
\Sigma ^{\dagger } \\
&&+[r_{\alpha \beta },r_{\mu \gamma }]\left( \Sigma ^{\dagger }D_{\gamma
}D_{\nu }\Sigma +D_{\gamma }D_{\nu }\Sigma ^{\dagger }\Sigma \right) \\
&&-\left[ \ell _{\alpha \beta },\ell _{\mu \gamma }\right] \left( \Sigma
D_{\gamma }D_{\nu }\Sigma ^{\dagger }+D_{\gamma }D_{\nu }\Sigma \Sigma
^{\dagger }\right) \}
\end{eqnarray*}


\begin{thebibliography}{99}
\bibitem{WZ} J. Wess and B. Zumino, ''Consequences of the anomalous Ward
identities'',\emph{Phys. Lett.} \textbf{B37} (1971) 95

\bibitem{WZW} E. Witten, ''Global aspects of current algebra'', \emph{Nucl.
Phys.} \textbf{B223} (1983) 422

\bibitem{u1v} J. Bijnens ''Chiral Perturbation Theory and Anomalous
Processes'', \emph{Int.} \emph{J. Mod. Phys.} \textbf{A Vol. 8 No. 18} (1993)

\bibitem{chintref} G.C. Callan, S. Coleman, J. Wess \& B. Zumino, \emph{%
Phys. rev.} \textbf{177} , 2239; 2247 (1969)

\bibitem{chptintro} Gilberto Colangelo \& Gino Isodori, ''An Introduction to
ChPT'', hep-ph/0101264

\bibitem{gasleut} J. Gasser and H. Leutwyler, \emph{Nucl. Phys.} \textbf{B250%
} 465 (1985)

\bibitem{qft1} Michael E. Peskin, Daniel V. Schroeder, ''An Introduction to
Quantum Field Theory'', Westview press (1995), 661

\bibitem{GeOr} M. Gell-Mann, R. Oakes \& B. Renner , ''Behaviour of current
divergences under $SU\left( 3\right) \times SU\left( 3\right) $'', \emph{%
Phys. Rev.} \textbf{175}, 2195 (1968)

\bibitem{dynsm} J. F. Donoghue, E. Golowich, B. R. Holstein, ''Dynamics of
the Standard Model'', Cambridge University Press (1996)

\bibitem{galep4} J. Gasser and H. Leutwyler, \emph{Ann. Phys (N.Y.)} \textbf{%
158}, 142 (1984)

\bibitem{bijp6} J. Bijnens, L. Girlanda and P. Talavera, ''The Anomalous
Chiral Lagrangian of order $p^{6}$'', \emph{Eur. Phys. J. }\textbf{C23, }%
539-544

\bibitem{etamixmod} F. J. Gilman and R. Kauffman ''$\eta $-$\eta ^{\prime }$
mixing angle'' \emph{Phys. Rev. }\textbf{D36}, 2761 (1987)

\bibitem{fuji} K. Fujikawa, ''Path integral measure for gauge invariant
field theories'', \emph{Phys. Rev. }\textbf{D23}, 2262 (1979)

\bibitem{adler} S. L. Adler, \emph{Phys. Rev. }\textbf{177}, 2426 (1969)

\bibitem{belljackiw} J. S. Bell and R. Jackiw, \emph{Nuovo Cim.} \textbf{A60}%
, 47 (1969)

\bibitem{adbardeen} S. L. Adler and W. A. Bardeen, ''Absence of higher order
corrections in the anomalous axial-vector divergence equation'' \emph{Phys.
Rev. }\textbf{182}, 1517 (1969)

\bibitem{steinberger} J. Steinberger, , 1180 (1949)

\bibitem{pdg1} K. Hagiwara et al. (Particle Data Group) \emph{Phys. Rev.} 
\textbf{D66}, 010001 (2002), (URL: http://pdg.lbl.gov)

\bibitem{weinberg} S. Weinberg, ''Phenomenological Lagrangians'', \emph{%
Physica }\textbf{A96}, 327 (1979)

\bibitem{pggloop} J. Bijnens, A. Bramon and F. Cornet \emph{Phys. Rev. Lett. 
}\textbf{61}, 1453 (1988)

\bibitem{semileploop} Ll. Ametller, J. Bijnens, A. Bramon and F. Cornet,
''Semileptonic $\pi $ and $K$ decays and the chiral anomaly at one-loop'', 
\emph{Phys. Lett. }\textbf{B303}, 140 (1993)

\bibitem{bando} M. Bando, T. Kugo and K. Yamawaki, \emph{Phys. Rep. }\textbf{%
164}, 217 (1988)

\bibitem{Ball} R. D. Ball, \emph{Phys. Rep. }\textbf{182}, 1 (1989)

\bibitem{cleo} CLEO Collaboration, ''Measurements of the meson-photon
transition form factors of light pseudoscalar mesons at large momentum
transfer'', CLNS 97/1477, CLEO 97-7, hep-ex/9707031 (1998)

\bibitem{kl4} S. Pislak et al., ''New measurement of $K_{e4}^{+}$ decay and
the s-wave scattering length $a_{0}^{0}$'', \emph{Phys. Rev. Lett}. Vol. 
\textbf{87, No. 22} (2001)

\bibitem{vmd} A. Bramon, J. Bijnens and F. Cornet, ''Chiral Perturbation
Theory for $\gamma PPP$ Processes'', presented at Daphne Workshop 1990;
UAB-FT-263/91, UG-FT-15/91

\bibitem{vtxloop} J. Bijnens, A. Bramon and F. Cornet, \emph{Phys. Rev. Lett.%
} \textbf{61}, 1453 (1988);\newline
J. Donoghue and D. Wyler, \emph{Nucl. Phys }\textbf{B316}, 289 (1989)

\bibitem{area51} Y.N. Antipov et al., \emph{Phys. Rev.} \textbf{D36,} 21
(1987)
\end{thebibliography}
\end{document}